\documentclass[acmsmall]{acmart}

\usepackage[utf8]{inputenc}
\usepackage{wrapfig}
\usepackage{graphicx}
\usepackage{subfig}
\usepackage{multicol}
\usepackage{multirow}
\usepackage{colortbl}
\usepackage{tcolorbox,fancyvrb,xcolor,tikz}
\usepackage{siunitx}
\usepackage{caption}
\usepackage{subcaption}

\usepackage{listings}

\usepackage{algorithm}
\usepackage{algorithmic}

\usepackage{xspace}

\usepackage{url}

\AtBeginDocument{%
  \providecommand\BibTeX{{%
    \normalfont B\kern-0.5em{\scshape i\kern-0.25em b}\kern-0.8em\TeX}}}

\colorlet{punct}{red!60!black}
\definecolor{background}{HTML}{EEEEEE}
\definecolor{delim}{RGB}{20,105,176}
\colorlet{numb}{magenta!60!black}

\newcommand{\response}[1]{#1}

\lstdefinelanguage{json}{
    basicstyle=\normalfont\ttfamily,
    numbers=left,
    numberstyle=\scriptsize,
    stepnumber=1,
    numbersep=8pt,
    showstringspaces=false,
    breaklines=true,
    frame=lines,
    backgroundcolor=\color{background},
    literate=
     *{0}{{{\color{numb}0}}}{1}
      {1}{{{\color{numb}1}}}{1}
      {2}{{{\color{numb}2}}}{1}
      {3}{{{\color{numb}3}}}{1}
      {4}{{{\color{numb}4}}}{1}
      {5}{{{\color{numb}5}}}{1}
      {6}{{{\color{numb}6}}}{1}
      {7}{{{\color{numb}7}}}{1}
      {8}{{{\color{numb}8}}}{1}
      {9}{{{\color{numb}9}}}{1}
      {:}{{{\color{punct}{:}}}}{1}
      {,}{{{\color{punct}{,}}}}{1}
      {\{}{{{\color{delim}{\{}}}}{1}
      {\}}{{{\color{delim}{\}}}}}{1}
      {[}{{{\color{delim}{[}}}}{1}
      {]}{{{\color{delim}{]}}}}{1},
}
\newfloat{Code}{h}{myc}

\title{Sustainable LLM Inference for Edge AI: Evaluating Quantized LLMs for Energy Efficiency, Output Accuracy, and Inference Latency}

\author{Erik Johannes Husom, Arda Goknil, Merve Astekin}
\affiliation{%
  \institution{SINTEF}
  \city{Oslo} 
  \country{Norway} 
}
\email{firstname.lastname@sintef.no}

\author{Lwin Khin Shar}
\affiliation{%
  \institution{Singapore Management University}
  \city{Singapore} 
  \country{Singapore} 
}
\email{lkshar@smu.edu.sg}

\author{Andre Kåsen}
\affiliation{%
  \institution{Oslo Metropolitan University}
  \city{Oslo}
  \country{Norway}
}
\email{ankas9475@oslomet.no}

\author{Sagar Sen, Benedikt Andreas Mithassel}
\affiliation{%
  \institution{SINTEF}
  \city{Oslo} 
  \country{Norway} 
}
\email{firstname.lastname@sintef.no}

\author{Ahmet Soylu}
\email{firstname.lastname@kristiania.no}
\affiliation{%
  \institution{Kristiania University of Applied Sciences}
  \city{Oslo}
  \country{Norway}
}

\begin{document}

\begin{abstract}

Deploying Large Language Models (LLMs) on edge devices presents significant challenges due to computational constraints, memory limitations, inference speed, and energy consumption. Model quantization has emerged as a key technique to enable efficient LLM inference by reducing model size and computational overhead. In this study, we conduct a comprehensive analysis of 28 quantized LLMs from the Ollama library, 
which applies by default Post-Training Quantization (PTQ) and weight-only quantization techniques, deployed on an edge device (Raspberry Pi 4 with 4GB RAM). We evaluate energy efficiency, inference performance, and output accuracy across multiple quantization levels and task types. Models are benchmarked on five standardized datasets (CommonsenseQA, BIG-Bench Hard, TruthfulQA, GSM8K, and HumanEval), and we employ a high-resolution, hardware-based energy measurement tool to capture real-world power consumption. Our findings reveal the trade-offs between energy efficiency, inference speed, and accuracy in different quantization settings, highlighting configurations that optimize LLM deployment for resource-constrained environments. By integrating hardware-level energy profiling with LLM benchmarking, this study provides actionable insights for sustainable AI, bridging a critical gap in existing research on energy-aware LLM deployment.

\end{abstract}

\maketitle

\section{Introduction}
\label{sec:introduction}

\textbf{Context and Motivation.} The rapid advancement of artificial intelligence (AI), particularly in the form of Large Language Models (LLMs), is reshaping our interactions with digital platforms and expanding the capabilities of various applications. LLM architectures, such as GPT (Generative Pre-trained Transformer)~\citep{radford2018improving}, LLaMA (Large Language Model Meta AI)~\citep{touvron2023llama}, and BERT (Bidirectional Encoder Representations from
Transformers)~\cite{devlin2018bert}, exhibit sophisticated abilities to comprehend, generate, and engage in natural language, proving indispensable across diverse use cases—from customer support~\cite{wulf2024exploring} 
and sentiment analysis~\cite{zhong2023can} 
to machine translation~\cite{li2024eliciting} 
and creative content creation~\cite{sudhakaran2024mariogpt}. 
As these models grow in size and complexity, their demand for computational resources and robust cloud infrastructure rises accordingly. However, this dependency on centralized servers introduces challenges related to latency, data privacy, and scalability, particularly as LLM applications become increasingly embedded in personal and mobile devices. To mitigate these issues, researchers and developers are pursuing solutions that bring LLMs closer to the end-user by deploying them on edge devices like IoT gateways (e.g., Raspberry Pi), smartphones, and in-vehicle embedded systems~\cite{friha2024llm, xu2024unleashing, zhou2024survey, qin2024empirical, yu2024edge, cai2024edge, wei2024t, zhang2024edgeshard, haris2024designing, yin2024llm, li2024tpi, khoshsirat2024decentralized}. 
This approach enables real-time processing and reduces reliance on cloud resources. Yet, deploying LLMs on resource-constrained edge devices poses significant challenges due to their high memory, processing, and energy demands, necessitating innovative techniques to make such on-device models feasible and effective~\cite{friha2024llm, liu2024edge, qin2024empirical, qu2024mobile}. Model quantization, a technique reducing model size and computational requirements, has emerged as a promising solution to enhance the performance and efficiency of LLMs on these constrained devices~\cite{xiao2023smoothquant, lin2024awq, park2022lut, yao2022zeroquant, dettmers2022gpt3, frantar2022optq, rahman2023quantized, huang2024billm}. 

Motivated by the dual pressures of energy efficiency and model accuracy, quantization has gained traction as a viable strategy to adapt LLMs for edge environments. Model quantization techniques, including fixed-point, integer-only, and hybrid quantization methods~\cite{lang2024comprehensive}, reduce the model's memory footprint and computational load, making it feasible to deploy on resource-constrained edge devices. However, quantization often affects model accuracy, raising a trade-off between computational efficiency and predictive performance that developers must carefully balance. Additionally, quantization can introduce numerical instability and precision loss, particularly in tasks requiring complex reasoning or arithmetic computations~\cite{gholami2022survey}. The extent of accuracy degradation varies across task types, model architectures, and quantization levels, making it essential to evaluate task-specific performance trade-offs. Moreover, while quantization reduces computational load, its impact on energy consumption remains underexplored, especially in real-world edge deployments where power efficiency is critical.

\textbf{Problem Statement.} Despite the advancements, critical challenges persist in understanding and optimizing the interplay between quantized model deployment, energy consumption, and inference accuracy on edge devices. 
While previous works have investigated quantization methods~\cite{lin2024awq, shen2024agile, shen2024edgeqat, hu2024llm, chai2025flexquant}, accuracy trade-offs~\cite{li2024evaluating, jin2024comprehensive, huang2024empirical}, and other model compression techniques~\cite{lin2024awq, yu2024edge, wang2024model}, there is limited research on the energy consumption of quantized LLMs in real-world edge deployment. Additionally, prior research often focuses on individual models or quantization techniques without systematically comparing multiple quantization techniques or levels across diverse LLM model families and task types~\cite{zhou2024survey}. This gap in research makes it difficult for practitioners and researchers to make informed decisions about which quantization technique, level, and model combination best suits their specific application needs, particularly when balancing energy savings with acceptable levels of model performance for edge deployment scenarios.

Another pertinent issue is the lack of standardized benchmarks and testing environments that enable fair comparisons across LLMs and quantization techniques on edge devices for energy consumption, output accuracy, and inference performance. Several studies undertook evaluations of the performance
of quantized LLMs without considering the trade-off for energy consumption, inference performance (latency), and accuracy on edge deployment scenarios~\cite{liu2024evaluating, liu2023emergent, jaiswal2023compressing, li2024evaluating, huang2024empirical, huang2024good, jin2024comprehensive}. Without a clear, comparative understanding, developers and researchers are often left in a trial-and-error cycle, deploying models that may not be optimal for their edge computing environments. Additionally, the role of dataset and task diversity in affecting energy and accuracy outcomes remains underexplored. Different datasets may stress models in unique ways, potentially impacting the energy-accuracy trade-off and the generalizability of results across use cases.

\textbf{Our Objectives and Experiment Design.} This paper addresses these challenges by presenting a systematic, comparative analysis of quantized LLMs deployed on edge. The primary objective is to measure and evaluate the energy efficiency of different LLMs under different quantization settings and to assess how these configurations impact inference accuracy and latency across multiple datasets and tasks. This analysis will identify and quantify the trade-offs between energy consumption, inference performance, and output accuracy, providing insights into how specific quantization techniques perform on a range of models. By offering this comparative perspective, the paper aims to equip developers with actionable knowledge on selecting quantization techniques and models that align with both energy and accuracy goals for their edge applications.

In our experiments, we utilize 28 quantized versions LLMs from the \textbf{Ollama library}~\cite{ollama-library}, including variants based on Gemma 2 (2B), Lama 3.2 (1B), and Qwen 2.5 (0.5B, 1.5B). 
The models were selected due to their varying parameter scales and architectures, offering a diverse testbed for analyzing the impact of quantization on energy efficiency and accuracy across edge devices. The Ollama library was chosen for its extensive repository of pre-quantized models, ensuring consistency in quantization methods while enabling direct performance comparisons across models. This diversity allows us to capture insights into the trade-offs between model size, energy consumption, latency, and predictive accuracy, providing a comprehensive understanding of LLM behavior in resource-constrained environments.

For our experiments, we utilize datasets from the \textbf{NeurIPS 2024 Challenge: Edge-Device Large Language Model Competition}~\cite{liu2024edge}, which are specifically designed to test the diverse capabilities of LLMs deployed on resource-constrained edge devices. These datasets span critical dimensions such as commonsense knowledge (CommonsenseQA~\cite{commonsenseqa}, TruthfulQA~\cite{truthfulqa}), reasoning (BIG-Bench Hard~\cite{bigbenchhard}), mathematical problem-solving (GSM8K~\cite{gsm8k}), 
and programming (HumanEval~\cite{humaneval}).
Due to the constraints of time and computational resources, we employ a systematic sampling strategy to extract representative subsets from each dataset. This approach ensures a balance between preserving the datasets’ diversity and maintaining a practical evaluation scope, enabling a robust yet efficient assessment of energy efficiency and output accuracy for quantized LLMs on edge.

We measure the energy consumption of quantized LLM inference using \textbf{Joulescope (JS110)}~\cite{joulescope}, a high-precision hardware-based power monitoring tool. The setup consists of a \textbf{Raspberry Pi 4 (4GB RAM)} running the inference workload, with Joulescope capturing real-time voltage and current measurements to compute total energy consumption. Data is recorded on a separate computer, ensuring accurate power profiling without introducing additional computational overhead on the edge device. Joulescope was chosen for its high sampling rate (2 MHz for JS110), fine-grained power measurement capabilities, and ability to provide precise, hardware-based energy profiling, making it ideal for evaluating LLM energy efficiency in real-world edge deployments.

\textbf{Our Findings.} In our experiments, we investigate, based on five datasets, the following Research Questions (RQ)s:

\begin{itemize}

\item \textit{\textbf{RQ1.} How does model quantization affect the energy consumption of LLM inference on an edge device?} This question focuses on quantifying energy savings from different quantization levels while running LLMs on a constrained hardware platform.

\item \textit{\textbf{RQ2.} What are the trade-offs between accuracy and energy efficiency across different quantization levels in LLMs deployed on an edge device?} This question explores how varying degrees of quantization affect inference performance, highlighting the balance between computational savings and potential degradation in output quality.

\item \textit{\textbf{RQ3.} How do different quantization techniques 
impact inference speed
on an edge device?} This research question explores the relationship between quantization levels and inference latency, examining how reduced precision affects token generation rates, model execution time, and overall computational efficiency in a resource-constrained environment.

\end{itemize}

Our findings reveal key tradeoffs between energy efficiency, accuracy, and inference speed in quantized LLMs deployed on edge devices. For RQ1, quantization consistently reduces energy consumption, with q3 and q4 variants cutting energy use by up to 79\% compared to FP16, though extreme quantization can sometimes introduce inefficiencies. For RQ2, accuracy varies significantly across quantization levels and tasks, with some q3 models maintaining competitive accuracy, while others experience substantial degradation, particularly in mathematical reasoning tasks. For RQ3, quantization reduces inference latency by up to 69\%, but the benefits diminish at lower precision due to computational overhead. Smaller models achieve higher throughput, while larger models struggle with latency, particularly in code generation tasks, where longer responses drive higher processing times.

\textbf{Our Contributions.} This paper makes the following key contributions:

\begin{itemize}
\item \textbf{Comprehensive Evaluation of Quantized LLMs on Edge:} We conduct an extensive evaluation of 28 quantized LLMs from the Ollama framework, which by default applies PQT and weight-only quantization techniques, deployed on an edge device. These models span different model families, parameter sizes, and quantization levels. 

\item \textbf{Analysis of the Energy-Accuracy Trade-Off:} We systematically measure and compare the energy consumption and inference accuracy of quantized LLMs using multiple real-world datasets. This analysis highlights the trade-offs between computational efficiency and model performance, providing actionable insights for optimizing LLM deployment in resource-constrained environments.

\item \textbf{Insights into Dataset and Task Influence on Quantized Models:} We investigate the impact of dataset and task characteristics on the energy and accuracy outcomes of quantized LLMs. Our findings offer guidance on selecting datasets that appropriately stress models for specific application scenarios, improving the generalizability of results.

\end{itemize}

The remainder of this paper is as follows. In Section~\ref{sec:background}, we summarize the background on LLMs and energy monitoring. Section~\ref{sec:evaluation} presents our experimental setup and analysis. In Section~\ref{sec:results}, we summarize our findings for each research question. 
Section~\ref{sec:validity} presents threats to the validity of our experiments. In Section~\ref{sec:related}, we review the related work. Finally, Section~\ref{sec:conclusion} concludes the paper. 

\section{Background}
\label{sec:background}

\subsection{Model Quantization Techniques for LLMs}

Model quantization reduces numerical precision to enhance storage efficiency and computational performance while preserving accuracy. It is broadly classified into post-training quantization (PTQ) and quantization-aware training (QAT). Most LLM quantization techniques use PTQ~\cite{zhou2024survey}, as it enables efficient inference without requiring retraining.

\subsubsection{Post-Training Quantization (PTQ)}

Post-training quantization (PTQ) applies quantization to a fully trained model without modifying the original training process. This technique converts high-precision weights (typically 32-bit floating-point, FP32) to lower-bit formats such as \textbf{INT8 (8-bit integers), INT4, or BF16 (brain floating-point 16-bit)}. 

PTQ offers different techniques to reduce memory consumption by replacing high-precision values with low-bit representations for weights, activations, and KV caches:

\begin{enumerate}

\item \textbf{Weight-Only Quantization} only quantizes the weight tensor W of each linear layer. It accelerates memory-bounded General Matrix-Vector Multiply (GEMV) operations during the decoding stage of the LLM inference process~\cite{lin2024awq, park2022lut, frantar2022gptq}. Common weight quantization methods include \textbf{uniform quantization}, mapping floating-point values into a fixed range of integer values, and \textbf{non-uniform Quantization}, using techniques such as logarithmic or k-means clustering to optimize quantization scales.

\item \textbf{Weight-Activation Quantization} quantizes both the input activation X and weight tensor W of each linear layer. It leverages low-precision Tensor Cores in GPUs to optimize compute-bounded General Matrix Multiply (GEMM) operations in the prefill stage of the LLM inference process~\cite{xiao2023smoothquant, wei2022outlier, yao2022zeroquant}.

\item \textbf{KV Cache Quantization} quantizes the key tensor K and value tensor V in each self-attention block. It minimizes memory overhead, improving efficiency for long texts and large batch sizes~\cite{sheng2023flexgen}.

\end{enumerate}

PTQ techniques use different bit-widths to balance accuracy, memory efficiency, and speed. \textbf{FP16 (Half Precision)} reduces 32-bit weights to 16-bit, cutting memory usage. \textbf{INT8 Quantization} maps weights to 8-bit integers, improving speed with minimal accuracy loss. \textbf{INT4 Quantization} quantization further reduces memory and power consumption, maximizing efficiency.

\subsubsection{Quantization-Aware Training (QAT)}

Quantization-aware training (QAT) integrates quantization into the training process, simulating reduced numerical precision during forward and backward passes. Unlike PTQ, QAT enables the model to adjust its weights, minimizing precision loss and improving quantized inference performance. The process involves fake quantization, parameter adaptation via backpropagation, and fine-tuning to retain accuracy while reducing computational demands. QAT is especially useful for deep LLMs, where PTQ alone may cause accuracy degradation, but it requires training data and additional compute resources, making it less practical for large-scale pre-trained models.

\subsubsection{Implications for Edge Deployment and Quantization Techniques in Our Experiments}

Deploying quantized LLMs on edge devices requires careful consideration of the following factors:

\begin{itemize}

\item \textbf{Energy Efficiency vs. Accuracy Trade-offs:} Lower-bit quantization improves energy efficiency but can degrade model accuracy, especially in high-precision tasks.

\item \textbf{Inference Latency:} While quantization reduces memory footprint, extreme quantization levels may introduce non-trivial computational overhead due to dequantization steps.

\item \textbf{Model Adaptability:} QAT and adaptive quantization methods can help maintain accuracy, but they require additional training, which is often infeasible for large LLMs.

\end{itemize}

In this study, we systematically evaluate the quantization techniques implemented in the Ollama framework~\cite{ollama}, which serves as a high-level abstraction of llama.cpp~\cite{llama.cpp}, a highly optimized C++-based inference framework supporting various quantization formats. By default, llama.cpp applies PTQ and weight-only quantization techniques~\cite{llama-doc}. Specifically, it supports multiple precision levels, including q4\_0, q4\_1, q8\_0, q3\_K\_S, q3\_K\_M, q3\_K\_L, q4\_K\_S, and q4\_K\_M~\cite{ollama}. The quantization format follows the pattern \textbf{qX\_K\_Y}, where X denotes bit precision, K indicates K-means clustering (if present), and Y specifies the variant affecting accuracy, memory usage, and speed.

\begin{itemize}
    \item The parameter \( qX \) denotes the \textbf{bit precision for weight quantization}, where \textbf{q4} represents \textbf{4-bit} and \textbf{q8} denotes \textbf{8-bit quantization}.

    \item \( K \) indicates the use of \textbf{K-means clustering}, a \textbf{non-uniform quantization technique}. In contrast, the absence of K signifies \textbf{uniform quantization}, where all values share a fixed scaling factor.

    \item The parameter \( Y \) defines a specific quantization variant that balances \textbf{accuracy, memory usage, and inference speed}:

    \begin{itemize}
       \item \(\mathbf{Y = 0}\) applies \textbf{a single uniform scaling parameter per block of weights}, resulting in a simple but less adaptive compression method.
       \item \(\mathbf{Y = 1}\) utilizes \textbf{two scaling parameters per block of weights}, which ensures that small weight blocks have a minimum scaling factor, reducing the impact of outliers and improving accuracy.
       \item \(\mathbf{Y = S}\) optimizes for \textbf{minimal memory usage} by reducing the number of clusters, making it ideal for \textbf{low-resource edge devices}.
       \item \(\mathbf{Y = M}\) represents a \textbf{balanced quantization strategy}, employing a moderate number of clusters to achieve a compromise between \textbf{memory efficiency and accuracy}, suitable for \textbf{mid-range GPUs and CPUs}.
       \item \(\mathbf{Y = L}\) applies \textbf{high-cluster quantization}, minimizing accuracy loss at the cost of increased memory usage, making it best suited for \textbf{high-end hardware with ample memory}.
    \end{itemize}

\end{itemize}

\subsection{Energy Monitoring Tools and Technologies}

Several energy monitoring tools and technologies have been developed to track and analyze power usage at different levels, including \textit{hardware-level measurements}, \textit{software-based estimation}, and \textit{profiling frameworks tailored for AI workloads}. 

\subsubsection{Hardware-Based Power Measurement Tools}

Hardware-based energy monitoring tools offer the \textbf{most precise power consumption data} by directly measuring electrical power usage at the \textbf{system level}. While essential for accurate energy profiling, they require specialized equipment. Common tools include \textbf{Joulescope}~\cite{joulescope}, a high-precision external power meter for embedded systems and edge devices; \textbf{Monsoon Power Monitor}~\cite{monsoon}, widely used for mobile and embedded device power analysis. 
Though highly accurate, these tools cannot isolate CPU vs. GPU consumption, necessitating software-based analysis for component-level breakdowns.

\subsubsection{Software-Based Energy Estimation Tools}
Software-based energy monitoring tools estimate power consumption using \textbf{pre-calibrated models and system telemetry}, offering non-intrusive, lightweight solutions for large-scale experiments. \textbf{Scaphandre}~\cite{scaphandre} profiles CPU and memory power usage in Linux-based environments, making it ideal for cloud and edge AI deployments. Microsoft’s \textbf{Joulemeter}~\cite{joulemeter} estimates power consumption on Windows using machine learning models and system performance counters. For GPU workloads, \textbf{NVIDIA-SMI}~\cite{nvidia_smi_docs} provides real-time power monitoring widely used for profiling LLM energy consumption on GPU-based devices.

\subsubsection{Profiling Frameworks for AI Energy Analysis}
\textbf{AI-specific profiling frameworks} combine energy tracking with performance evaluation, enabling analysis of trade-offs between inference speed, accuracy, and efficiency. \textbf{PyJoules}~\cite{PyJoules} is a Python-based tool for deep learning workloads, integrating with PyTorch and TensorFlow for per-operation power profiling. \textbf{MELODI}~\cite{husom2024price} monitors LLM inference energy consumption, integrating Scaphandre and NVIDIA-SMI to provide detailed insights into power usage during inference.

\subsubsection{Energy Monitoring Solution used in Our Experiments}

Due to the limitations of software-based estimations and the need for high-precision, real-time monitoring, we selected \textbf{Joulescope}~\cite{joulescope}, an external hardware-based power meter, as our primary energy measurement tool. Unlike software-based tools and AI profiling frameworks, which rely on telemetry data and estimation models, Joulescope provides direct power measurements. Software-based methods are suitable for standardized environments like x86 desktops, data centers, or cloud AI workloads, but they lack reliable telemetry and hardware support for edge devices, making them unsuitable for our study. Both Joulescope and Monsoon Power Monitor offer high-precision, microsecond-level energy measurements. We chose Joulescope for its broader compatibility, ease of use, and higher dynamic range for measuring power consumption in edge devices.

\section{Study Design}
\label{sec:evaluation}

\subsection{Research Questions}
\label{subsec:research-questions}
We investigate, based on five datasets, the following Research Questions (RQ)s:

\begin{itemize}

\item \textit{\textbf{RQ1.} How does model quantization affect the energy consumption of LLM inference on an edge device?}

\item \textit{\textbf{RQ2.} What are the trade-offs between accuracy and energy efficiency across different quantization levels in LLMs deployed on an edge device?}

\item \textit{\textbf{RQ3.} How do different quantization techniques 
impact inference speed on an edge device?}

\end{itemize}

\subsection{Energy Measurement Setup}
\label{subsec:experiment-setup}

\begin{wrapfigure}{h}{0.500\textwidth}
    \vspace*{-1.1em}
    \centerline{\includegraphics[width=0.500\textwidth]{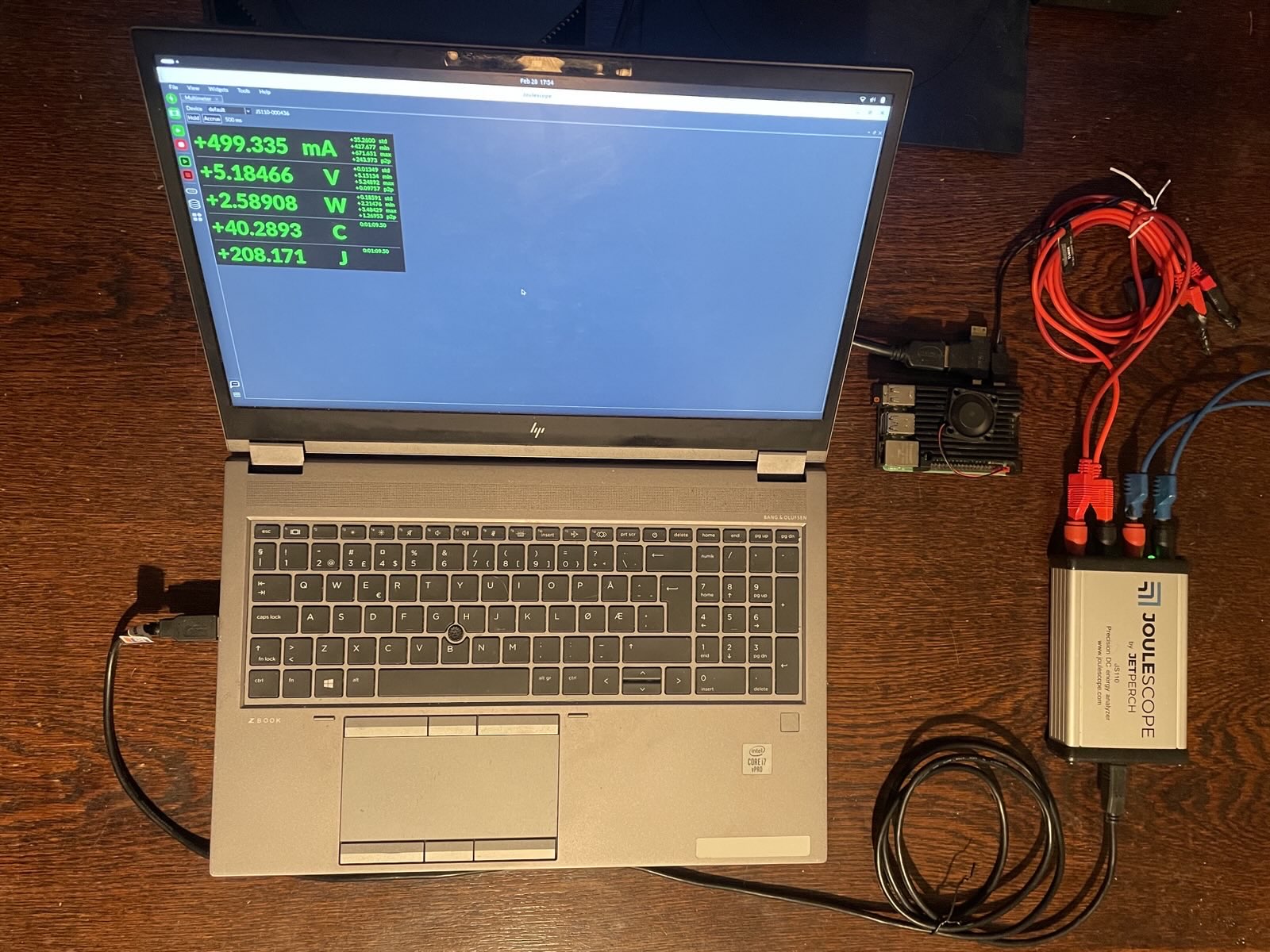}}
    \vspace*{-0.3em}
    \caption{Setup for measuring the energy consumption of LLM inference on an edge device. A Joulescope power meter is between a Raspberry Pi 4 and its power supply, with real-time energy data visualized on a laptop.}
    \vspace*{-0.8em}
    \label{fig:experiment-setup}
    \vspace*{-0.8em}

\end{wrapfigure}

The experiment is designed to precisely measure LLM inference power consumption on a resource-constrained edge device, the Raspberry Pi 4 (4GB RAM). Equipped with a quad-core Cortex-A72 (ARM v8) CPU at 1.5 GHz and running a 64-bit Linux OS, it represents a typical edge computing platform, making it ideal for assessing the feasibility of deploying LLMs under real-world constraints.

\response{To obtain high-resolution power measurements, the amperemeter-ports of a Joulescope precision power analyzer are connected in series between the Raspberry Pi 4 and its power supply, enabling direct measurement the total current (mA) flowing into it. We also measure the voltage (V) from the power supply. This allows us to calculate power (W), and cumulative energy (J) drawn by the Raspberry Pi. The Joulescope operates at a microsecond-level sampling resolution, capturing fine-grained fluctuations in power usage throughout the inference process. Its sampling frequency of the power consumption is 2 MHz, and we recorded the accumulated energy consumption at 2 Hz. The Joulescope software, running on a laptop, records and visualizes the power consumption in real time, providing a detailed view of energy trends across different inference workloads.}

We implemented automated scripts to log LLM-generated responses and corresponding energy consumption values. Responses are stored in structured files for accuracy evaluation, while Joulescope readings are recorded for analyzing quantization effects. Custom scripts process these logs, computing aggregate energy metrics (e.g., average power, total energy per query, peak power) and visualizing energy-performance trade-offs introduced by model quantization.

\subsection{Datasets}
\label{subsec:datasets}

For our experiments, we used datasets from the \textbf{NeurIPS 2024 Challenge: Edge-Device Large Language Model Competition}~\cite{liu2024edge}, designed to evaluate LLMs on resource-constrained edge devices. Table~\ref{tab:dataset_summary} provides the summary of the datasets. These datasets cover commonsense knowledge (CommonsenseQA~\cite{commonsenseqa}, TruthfulQA~\cite{truthfulqa}), reasoning (BIG-Bench Hard~\cite{bigbenchhard}), mathematical problem-solving (GSM8K~\cite{gsm8k}), and programming (HumanEval~\cite{humaneval}). We sampled \textbf{four out of the five datasets—CommonsenseQA, BIG-Bench Hard, TruthfulQA, and GSM8K}—to create a computationally feasible evaluation set for LLM inference with \textbf{28 models on an edge device}. Each dataset is reduced to \textbf{200 tasks} using appropriate sampling strategies while maintaining representativeness. \textbf{HumanEval} is not sampled, as it contains only \textbf{164 tasks}. 

\begin{table}[t]
    \centering
    \scriptsize
    \caption{Summary of the datasets used in our experiments.}
    \vspace*{-1.1em}

    \begin{tabular}{p{2.0cm} p{3.3cm} p{1.3cm} p{1.5cm} p{3.7cm}}
        \toprule
        \textbf{Dataset} & \textbf{Task Type} & \textbf{Full Size} & \textbf{Sampled Size} & \textbf{Sampling Method} \\ 
        \toprule
        \textbf{CommonsenseQA} & Multiple-choice QA & 12,102 & 200 & Uniform random sampling \\ 
        \rowcolor{black!6}
        \textbf{BIG-Bench Hard} & Reasoning & 6,511 & 200 & Hierarchical stratified sampling \\ 
        \textbf{TruthfulQA} & Multiple-choice QA & 817 & 200 & Uniform sampling \\ 
        \rowcolor{black!6}
        \textbf{GSM8K} & Mathematical problem-solving & 8,500 & 200 & Uniform random sampling \\ 
        \textbf{HumanEval} & Code generation & 164 & 164 & No sampling (entire dataset used) \\ 
        \toprule

    \end{tabular}
        \vspace*{-1.3em}

    \label{tab:dataset_summary}
\end{table}

\subsection{Prompts and Augmented Information}
\label{subsec:prompts}

We structured the prompts by combining original dataset questions with explicit instructions to clarify the expected output format and required output data. Following Astekin et al.~\cite{astekin2024comparative}, we instructed the LLMs to enclose answers within <ANS> and </ANS> tags for easy extraction. To reduce post-processing and ensure efficient inference, we explicitly directed the models to omit additional explanations in their responses.

For all datasets except HumanEval, we applied a standard instruction to the original prompts. Since HumanEval focuses on code generation, we adapted the instruction to align with programming tasks, creating a version tailored for this dataset. The instructions in the experiments are as follows:

\begin{itemize}
\item \textbf{\textit{instruction\_general }}= \textit{"Print only the answer surrounded by <ANS> and </ANS>. Never print any extra explanations about how the response was generated."}
\item  \textbf{\textit{instruction\_humaneval}} = \textit{"\# Complete the function implementation based on the provided docstring, and print only the completed function surrounded by <ANS> and </ANS>. Never print any extra explanations about how the code was generated."}
\end{itemize}

\subsection{Accuracy Evaluation Metrics}
\label{subsec:metrics}

To evaluate LLM accuracy across datasets, we selected task-specific metrics suited to multiple-choice, reasoning, math problem-solving, and code generation tasks, ensuring a comprehensive assessment of model performance. The accuracy metrics for each dataset are as follows:

\begin{itemize}

\item \textbf{CommonsenseQA:} \textit{Accuracy} measures the percentage of correctly answered multiple-choice questions, assessing the model's commonsense reasoning ability.

\item \textbf{BIG-Bench Hard:} \textit{Accuracy} evaluates performance across various reasoning challenges by calculating the proportion of correct answers.

\item \textbf{TruthfulQA:} We implemented an \textit{ensemble approach (Exact Match, ROUGE-L, and Cosine Similarity)}. In TruthfulQA, answer choices vary from single words to full sentences, and LLM-generated responses may partially or fully match a choice. To evaluate the accuracy, we first checked for exact matches and then used ROUGE-L and cosine similarity to identify the best matching choice.

    \begin{itemize}
        \item \textit{Exact Match (EM)} checks whether the response exactly matches the ground-truth answer.
        \item \textit{ROUGE-L} evaluates text similarity using the Longest Common Subsequence (LCS), measuring how well word order is preserved without requiring an exact match~\cite{deng2023benchmark, lin2004rouge}.        
        \item \textit{Cosine Similarity} measures the angle between two text vectors in a high-dimensional space. It assesses semantic closeness between the generated and reference answers~\cite{schaik2024field}.
    \end{itemize}

We explored other metrics but found them unsuitable due to limitations. Jaccard Similarity measures word overlap but ignores word order, making it ineffective for partial responses. BLEU Score, designed for full sentences, can misrepresent partial matches. LCS alone lacks the flexibility of ROUGE-L in handling minor variations.

\item \textbf{GSM8K:} \textit{Accuracy} determines correctness by verifying whether the final computed answer matches the target answer in the solution.

\item \textbf{HumanEval:} \textit{pass@k} measures the probability that at least one of the top-k generated code solutions passes all unit tests~\cite{chen2021codex}. In our experiments, we set k = 1 since we request only a single solution from the LLMs.

\end{itemize}

To calculate accuracy, we use the standard formula: \textit{the proportion of correct answers out of the total number of questions or tasks}. 

\subsection{Models}
\label{subsec:models}

\begin{wraptable}[]{t}{0.572\linewidth}
    \centering
    \vspace*{-1.20em}

    \caption{Overview of the quantized LLMs in the experiments.} 
    \scriptsize
    \vspace*{-0.80em}

    \begin{tabular}{|p{0.008\textwidth}|p{0.24\textwidth}|p{0.07\textwidth}|p{0.08\textwidth}| p{0.05\textwidth}|}
        \hline
         \textbf{} & \textbf{\textit{Model}} & \textbf{\textit{Paramet.}} & \textbf{\textit{Size (MB)}} & \textbf{\textit{Quanti.}} \\
        \hline
       \multirow{9}{*}{\rotatebox{90}{\textit{\textbf{Llama\,\,\,\,\,\,\,\,\,\,\,}}}} & \cellcolor{green!1}
       
         \texttt{llama3.2:1b-instruct-fp16} & \cellcolor{green!1}  1.2B & \cellcolor{green!1}  2364.74 MB & \cellcolor{green!1}  \texttt{FP16} \\
        
         & \cellcolor{green!1}  \texttt{llama3.2:1b-instruct-q8\_0} & \cellcolor{green!1} 1.2B & \cellcolor{green!1} 1259.90 MB & \cellcolor{green!1} \texttt{Q8\_0} \\
        
        & \cellcolor{green!1} \texttt{llama3.2:1b-instruct-q4\_1} & \cellcolor{green!1} 1.2B & \cellcolor{green!1} 793.23 MB & \cellcolor{green!1} \texttt{Q4\_1} \\
        
        & \cellcolor{green!1} \texttt{llama3.2:1b-instruct-q4\_K\_M} & \cellcolor{green!1} 1.2B & \cellcolor{green!1} 770.29 MB & \cellcolor{green!1} \texttt{Q4\_K\_M} \\
        
        & \cellcolor{green!1} \texttt{llama3.2:1b-instruct-q4\_0} & \cellcolor{green!1} 1.2B & \cellcolor{green!1} 735.23 MB & \cellcolor{green!1} \texttt{Q4\_0} \\
        
        & \cellcolor{green!1} \texttt{llama3.2:1b-instruct-q4\_K\_S} & \cellcolor{green!1} 1.2B & \cellcolor{green!1} 739.73 MB & \cellcolor{green!1} \texttt{Q4\_K\_S} \\
        
        & \cellcolor{green!1} \texttt{llama3.2:1b-instruct-q3\_K\_L} & \cellcolor{green!1} 1.2B & \cellcolor{green!1} 698.60 MB & \cellcolor{green!1} \texttt{Q3\_K\_L} \\
        
        & \cellcolor{green!1} \texttt{llama3.2:1b-instruct-q3\_K\_M} & \cellcolor{green!1} 1.2B & \cellcolor{green!1} 658.85 MB & \cellcolor{green!1} \texttt{Q3\_K\_M} \\
        
        & \cellcolor{green!1} \texttt{llama3.2:1b-instruct-q3\_K\_S} & \cellcolor{green!1} 1.2B & \cellcolor{green!1} 611.98 MB & \cellcolor{green!1} \texttt{Q3\_K\_S} \\ \hline

        \multirow{16}{*}{\rotatebox{90}{\textit{\textbf{Qwen\,\,\,\,\,\,\,\,\,\,\,\,\,\,\,}}}} & \cellcolor{blue!5}
        
         \texttt{qwen2.5:1.5b-instruct-q4\_1} & \cellcolor{blue!5} 1.5B & \cellcolor{blue!5} 969.75 MB & \cellcolor{blue!5} \texttt{Q4\_1} \\
        
         & \cellcolor{blue!5}  \texttt{qwen2.5:1.5b-instruct-q4\_K\_M} & \cellcolor{blue!5}  1.5B & \cellcolor{blue!5}  940.38 MB & \cellcolor{blue!5}  \texttt{Q4\_K\_M} \\
        
        & \cellcolor{blue!5} \texttt{qwen2.5:1.5b-instruct-q4\_0} & \cellcolor{blue!5} 1.5B & \cellcolor{blue!5} 891.66 MB & \cellcolor{blue!5} \texttt{Q4\_0} \\
        
         & \cellcolor{blue!5}  \texttt{qwen2.5:1.5b-instruct-q4\_K\_S} &  \cellcolor{blue!5} 1.5B & \cellcolor{blue!5}  896.76 MB & \cellcolor{blue!5}  \texttt{Q4\_K\_S} \\
        
        & \cellcolor{blue!5} \texttt{qwen2.5:1.5b-instruct-q3\_K\_L} & \cellcolor{blue!5} 1.5B & \cellcolor{blue!5} 839.40 MB & \cellcolor{blue!5} \texttt{Q3\_K\_L} \\
        
        & \cellcolor{blue!5} \texttt{qwen2.5:1.5b-instruct-q3\_K\_M} & \cellcolor{blue!5} 1.5B & \cellcolor{blue!5} 786.01 MB & \cellcolor{blue!5} \texttt{Q3\_K\_M} \\
        
        & \cellcolor{blue!5} \texttt{qwen2.5:1.5b-instruct-q3\_K\_S} & \cellcolor{blue!5} 1.5B & \cellcolor{blue!5} 725.71 MB & \cellcolor{blue!5} \texttt{Q3\_K\_S} \\

        & \cellcolor{red!5} \texttt{qwen2.5:0.5b-instruct-fp16} & \cellcolor{red!5} 494.03M & \cellcolor{red!5} 948.11 MB & \cellcolor{red!5} \texttt{FP16} \\
        
         & \cellcolor{red!5}  \texttt{qwen2.5:0.5b-instruct-q8\_0} & \cellcolor{red!5}  494.03M & \cellcolor{red!5}  506.48 MB & \cellcolor{red!5}  \texttt{Q8\_0} \\
        
        & \cellcolor{red!5} \texttt{qwen2.5:0.5b-instruct-q4\_1} & \cellcolor{red!5} 494.03M & \cellcolor{red!5} 357.18 MB & \cellcolor{red!5} \texttt{Q4\_1} \\
        
         &  \cellcolor{red!5} \texttt{qwen2.5:0.5b-instruct-q4\_K\_M} &  \cellcolor{red!5} 494.03M & \cellcolor{red!5}  379.39 MB & \cellcolor{red!5}  \texttt{Q4\_K\_M} \\
        
        & \cellcolor{red!5} \texttt{qwen2.5:0.5b-instruct-q4\_0} & \cellcolor{red!5} 494.03M & \cellcolor{red!5} 335.85 MB & \cellcolor{red!5} \texttt{Q4\_0} \\
        
         & \cellcolor{red!5} \texttt{qwen2.5:0.5b-instruct-q4\_K\_S} &  \cellcolor{red!5} 494.03M &  \cellcolor{red!5} 367.63 MB &  \cellcolor{red!5} \texttt{Q4\_K\_S} \\
        
        & \cellcolor{red!5} \texttt{qwen2.5:0.5b-instruct-q3\_K\_L} & \cellcolor{red!5} 494.03M & \cellcolor{red!5} 352.26 MB & \cellcolor{red!5} \texttt{Q3\_K\_L} \\
        
        & \cellcolor{red!5} \texttt{qwen2.5:0.5b-instruct-q3\_K\_M} & \cellcolor{red!5} 494.03M & \cellcolor{red!5} 339.01 MB & \cellcolor{red!5} \texttt{Q3\_K\_M} \\
        
        & \cellcolor{red!5} \texttt{qwen2.5:0.5b-instruct-q3\_K\_S} & \cellcolor{red!5} 494.03M & \cellcolor{red!5} 322.61 MB & \cellcolor{red!5} \texttt{Q3\_K\_S} \\
        \hline

        \multirow{3}{*}{\rotatebox{90}{\textit{\textbf{Gemma}}}} & \cellcolor{yellow!5}
        \texttt{gemma2:2b-instruct-q3\_K\_S} & \cellcolor{yellow!5} 2.6B & \cellcolor{yellow!5} 1297.64 MB & \cellcolor{yellow!5} \texttt{Q3\_K\_S} \\
        
        & \cellcolor{yellow!5} \texttt{gemma2:2b-instruct-q3\_K\_M} & \cellcolor{yellow!5} 2.6B & \cellcolor{yellow!5} 1393.96 MB & \cellcolor{yellow!5} \texttt{Q3\_K\_M} \\
        
        & \cellcolor{yellow!5} \texttt{gemma2:2b-instruct-q3\_K\_L} & \cellcolor{yellow!5} 2.6B & \cellcolor{yellow!5} 1478.62 MB & \cellcolor{yellow!5} \texttt{Q3\_K\_L} \\
        \hline
    \end{tabular}
    \vspace*{-0.70em}

    \label{tab:model_info}
\end{wraptable}

Table~\ref{tab:model_info} presents an \textbf{overview of the quantized LLMs used in our experiments}, detailing their model names, parameter sizes, storage requirements, and quantization formats. The models are categorized into three main \textit{model families}: \textbf{LLaMA 3.2, Qwen 2.5, and Gemma 2}, with parameter sizes ranging from 494 million to 2.6 billion. We have in total four \textit{base models}: LLaMA 3.2:1b, Qwen 2.5:0.5b, Qwen 2.5:1.5b, and Gemma 2:2b. 
To evaluate the trade-offs between model size, inference accuracy, and energy consumption on an edge device, we selected models that have been quantized using different precision levels, giving us multiple \textit{model variants} for each base model.

The table consists of four columns: \textbf{Model, Parameter Size, Size (MB), and Quantization Type}. The Model column lists the model and quantization format applied. The Parameter Size column indicates the total number of model parameters, ranging from \textbf{494M (million) to 2.6B (billion)}. The Size (MB) column represents the storage footprint of each model, which varies based on the quantization level. Finally, the Quantization Type column specifies the applied quantization format, directly influencing model accuracy, memory footprint, and inference efficiency.

The \textbf{LLaMA 3.2 family} consists of 1.2B parameter models employing PTQ and weight-only quantization techniques, including FP16 (half precision), 8-bit (q8\_0), 4-bit (q4\_0, q4\_1, q4\_K\_M, q4\_K\_S), and 3-bit (q3\_K\_L, q3\_K\_M, q3\_K\_S). The \textbf{Qwen 2.5 family} includes models with 1.5B and 494M parameters, supporting 8-bit, 4-bit, and 3-bit quantization, with formats such as FP16, q8\_0, q4\_1, q4\_0, q4\_K\_M, q4\_K\_S, q3\_K\_L, and q3\_K\_S. These models provide a balance between memory efficiency and accuracy, making them well-suited for various edge-computing environments. The \textbf{Gemma 2 family} comprises 2.6B parameter models utilizing 3-bit quantization (q3\_K\_S, q3\_K\_M, q3\_K\_L), which are the largest models used in our experiments. While these models require more memory, they offer enhanced performance for complex inference tasks.

Quantization significantly reduces model size, making LLM deployment feasible on resource-constrained hardware. The \textbf{FP16 models} are the largest, with \textbf{LLaMA 3.2 FP16 requiring 2364.74MB and Qwen 2.5 FP16 using 948.11MB}. In contrast, \textbf{lower-bit models (q3\_K\_S, q3\_K\_M, q3\_K\_L) require substantially less memory}, making them ideal for \textbf{edge applications}. However, this reduction in size comes with \textbf{trade-offs in accuracy}, as higher-bit quantization formats (e.g., q8\_0, q4\_1, q4\_0) retain more precision, whereas lower-bit formats (q3\_K\_S, q3\_K\_L) prioritize efficiency at the cost of accuracy. Additionally, different quantization formats optimize models for \textbf{various hardware constraints}. Weight-only quantization formats (\textbf{qX\_K\_Y}) help balance \textbf{memory efficiency and computational performance}, with higher K values (q3\_K\_L, q4\_K\_L) reducing accuracy loss but demanding more memory.

By evaluating these quantized models on an edge device, we analyzed how \textbf{reduced precision affects inference energy consumption} while maintaining usable accuracy. This selection of models allows us to systematically compare:

\begin{itemize}

\item Higher precision vs. lower precision models (e.g., FP16 vs. INT8 vs. INT4 vs. INT3).

\item Different quantization methods (e.g., uniform quantization (Q4\_0) vs. balanced, non-uniform K-means quantization (Q4\_K\_M)).

\item Model family impact on energy efficiency and inference latency (e.g., LLaMA3.2 vs. Qwen2.5 vs. Gemma2).

\end{itemize}

\subsection{Variables}
\label{subsec:variables}

In our experiment, we analyzed the \textbf{energy efficiency, inference performance, and output accuracy} of quantized LLMs deployed on an edge device. These aspects are influenced by various factors, which can be categorized as \textbf{independent} and \textbf{dependent} variables.

\subsubsection{Independent Variables (Factors We Manipulate)}
\label{subsubsec:independent-variables}

These are the variables we control or modify to observe their effects on the dependent variables:

\begin{enumerate}
\item \textbf{LLM family} – The specific language model family tested (e.g., LLaMA3.2, Qwen2.5, Gemma2) 

\item \textbf{Model size in number of parameters} - The number of parameters for a given model, which is independent of the quantization technique.

\item \textbf{Model size in bytes} - The file size of the model, which varies according to the quantization type and level.

\item \textbf{Model Quantization Type and Level} – The bit-width of the quantized models (e.g., 8-bit). 

\item \textbf{Task Type} – The nature of the benchmark datasets (e.g., multiple-choice reasoning, mathematical problem-solving, code generation).

\item \textbf{Dataset Used} – The dataset from which inference tasks are drawn (CommonsenseQA, BIG-Bench Hard, TruthfulQA, GSM8K, HumanEval).

\end{enumerate}

\subsubsection{Dependent Variables (Measured Outcomes)}
\label{subsubsec:dependent-procedure}

These are the variables we observe and measure to evaluate the impact of the independent variables:

\begin{enumerate}

\item \textbf{Energy Consumption} – The total power usage (measured in joules) during inference, captured using the hardware-based energy monitoring tool (i.e., Joulescope).

\item \textbf{Inference Latency} – The time taken for the model to generate a response. 

\item \textbf{Token Generation Throughput} – The number of tokens generated per second. 

\item \textbf{Output Accuracy} – The correctness of LLM responses, benchmarked against ground truth labels for each dataset.


\end{enumerate}

By analyzing the impact of independent variables, 
we aimed to gain insights into optimal configurations for deploying LLMs in energy-constrained environments.

\subsection{Experiment Execution}
\label{subsec:experiment-execution}

The experiment setup includes a Raspberry Pi 4 for inference, a Joulescope power meter for energy measurement, and a computer for data recording (see Section~\ref{subsec:experiment-setup}). A systematic procedure ensures precise power monitoring across all models and datasets. The five sampled datasets and accompanying scripts are available in our repository~\cite{experiment-repo}.

To ensure accurate energy measurements, we followed a structured execution procedure. The system was initialized by powering on the computer and Joulescope, supplying power to the Raspberry Pi 4 (RPi). We verified the Ollama server and ensured all models fit within the RPi’s memory to prevent inference failures. The experiment script (exp.sh) was configured to specify models and datasets, with session sizes adjusted based on available disk space. To align power measurements with inference execution, we synchronized clocks between the RPi and the recording computer. The inference.py script included a 20-minute timeout to prevent indefinite hangs. Joulescope-UI recorded raw signal samples and statistical data, ensuring precise energy profiling.

During inference, models were evaluated sequentially per dataset, with continuous power logging. To ensure consistency, we completed all model evaluations on one dataset before proceeding to the next. After each session, power logs and outputs were backed up, and the RPi was restarted. For datasets with long responses (e.g., HumanEval), a batching strategy was used to manage storage constraints. This systematic approach ensured repeatability, precision, and reliable energy-performance analysis of quantized LLMs.

\subsection{Data Analysis}
\label{subsec:data-analysis}

\textbf{Data Preparation.} Our raw data includes energy consumption, token counts, processing time, and accuracy metrics per inference operation. To ensure statistical robustness, we aggregated measurements across multiple runs for each model variant. The data is categorized by model architecture, quantization level, and dataset, with analysis focused on the dependent variables outlined in Section~\ref{subsubsec:dependent-procedure}.

The recorded energy consumption includes the entire device's usage, including background processes. To isolate LLM inference energy, we measured the Raspberry Pi's idle power consumption and used the average value to subtract the baseline energy consumption from our measurements. As shown in Table~\ref{tab:idle_consumption}, the device's average idle power was 2.85W (energy consumption of 2.85J/s), which was deducted from the energy consumption values recorded during inference.

\begin{wraptable}[]{t}{0.500\textwidth}
\vspace*{-1.20em}
\caption{Idle power consumption (in Watts) of the Raspberry Pi 4, measured over a period of 104 minutes.}
\label{tab:idle_consumption}
\scriptsize
\vspace*{-1.20em}
\begin{tabular}{lrrrrrrr}
\toprule
 & mean & std & min & 25\% & 50\% & 75\% & max \\
\midrule
power (W) & 2.85 & 0.17 & 2.81 & 2.81 & 2.81 & 2.82 & 5.63 \\
\bottomrule
\end{tabular}
\vspace*{-1.20em}
\end{wraptable}

\textbf{Data Post-processing.} 
The first step in data post-processing involves extracting the generated answer enclosed within <ANS> and </ANS> tags. Since LLMs may not strictly follow the instructed format, responses often require cleaning and extraction from unstructured or semi-structured text. Missing opening or closing tags, extra parentheses, or misplaced characters are corrected to isolate the target answer. Invalid cases, such as empty responses or improperly formatted tags, are filtered out. Additionally, some dataset target answers require post-processing to extract the final value. For example, in GSM8K, solutions include detailed explanations, requiring scripts to identify and extract the final numerical or textual answer. We developed scripts to extract the actual target answers from the dataset.

\textbf{Statistical Analysis.} For each model variant and quantization level, we computed mean, median, and variance of the collected metrics to support visualization and analysis of the results. 
To identify optimal model configurations balancing energy efficiency and task performance, we applied Pareto efficiency analysis. A configuration is Pareto-optimal if no other variant achieves higher accuracy with the same or lower energy consumption or equal accuracy with lower energy usage. Formally, a model variant $P_1(energy_1, accuracy_1)$ dominates another configuration $P_2(energy_2, accuracy_2)$ if:

\begin{enumerate}
    \item $energy_1 \leq energy_2$ (equal or less energy consumption)
    \item $accuracy_1 \geq accuracy_2$ (equal or higher accuracy)
    \item Either $(energy_1 < energy_2)$ or $(accuracy_1 > accuracy_2)$ (strictly better in at least one dimension)
\end{enumerate}

The Pareto-optimal models represent the best-performing configurations, enabling practitioners to choose the most suitable balance between task performance and energy efficiency.

\textbf{Scaling Analysis.} To assess the impact of model quantization on performance and efficiency, we analyzed the relationship between model size (bytes), energy consumption, and accuracy. Model size was used instead of parameter count as it directly reflects memory footprint differences across quantization levels. For each model-dataset pair, we applied linear regression (scikit-learn's \texttt{LinearRegression}) to quantify scaling trends across quantization levels.

\textbf{Visualization.} We employed multiple visualization techniques to represent our analysis:
\begin{enumerate}
    \item Box plots: To visualize the distribution of energy consumption metrics across different models and quantization levels. Each box plot depicts the median (horizontal line splitting the main box), interquartile range (IQR, main box), and whiskers extending to 1.5 times the IQR, effectively summarizing central tendency and variability.
    \item Scatter plots: To examine relationships between energy consumption and accuracy, with points representing different model configurations. Data points are colored and styled according to which model and dataset they represent. For certain figures we plot the fitted regression lines indicating trend lines.
    \item Pareto frontier curves: Connecting Pareto-optimal configurations to visualize the trade-off boundary between energy efficiency and accuracy.
\end{enumerate}

\section{Experiment Results}
\label{sec:results}


\subsection{How does model quantization affect the energy consumption of LLM inference on an edge device? (RQ1)}
\label{subsec:RQ1}


To address RQ1, we computed the average energy consumption (Joules per token) for all model variants across the five benchmark datasets. Results are shown in Table~\ref{tab:energy} and Figure~\ref{fig:energy_distribution_per_token_log_all}.

\begin{wraptable}[]{t}{0.280\textwidth}
\vspace*{-1.20em}
\scriptsize
\caption{Energy per token for model families averaged across the datasets.}
\label{tab:energy_per_token}
\vspace*{-1.30em}
\begin{tabular}{lr}
\toprule
 Base Model  & \begin{tabular}[l]{@{}r@{}}Mean Energy\\ per Token (J/tok)\end{tabular} \\
\midrule
qwen25\_0.5b & $2.61 \pm 1.49$ \\
qwen25\_1.5b & $7.57 \pm 4.54$ \\
llama3.2\_1b & $8.40 \pm 5.36$ \\
gemma2\_2b &  $9.35 \pm 3.30$ \\
\bottomrule
\end{tabular}
\vspace*{-1.30em}
\end{wraptable}

\subsubsection{Baseline Energy Consumption Across Base Models}

\begin{figure}
    \centering
    \includegraphics[width=0.8\linewidth]{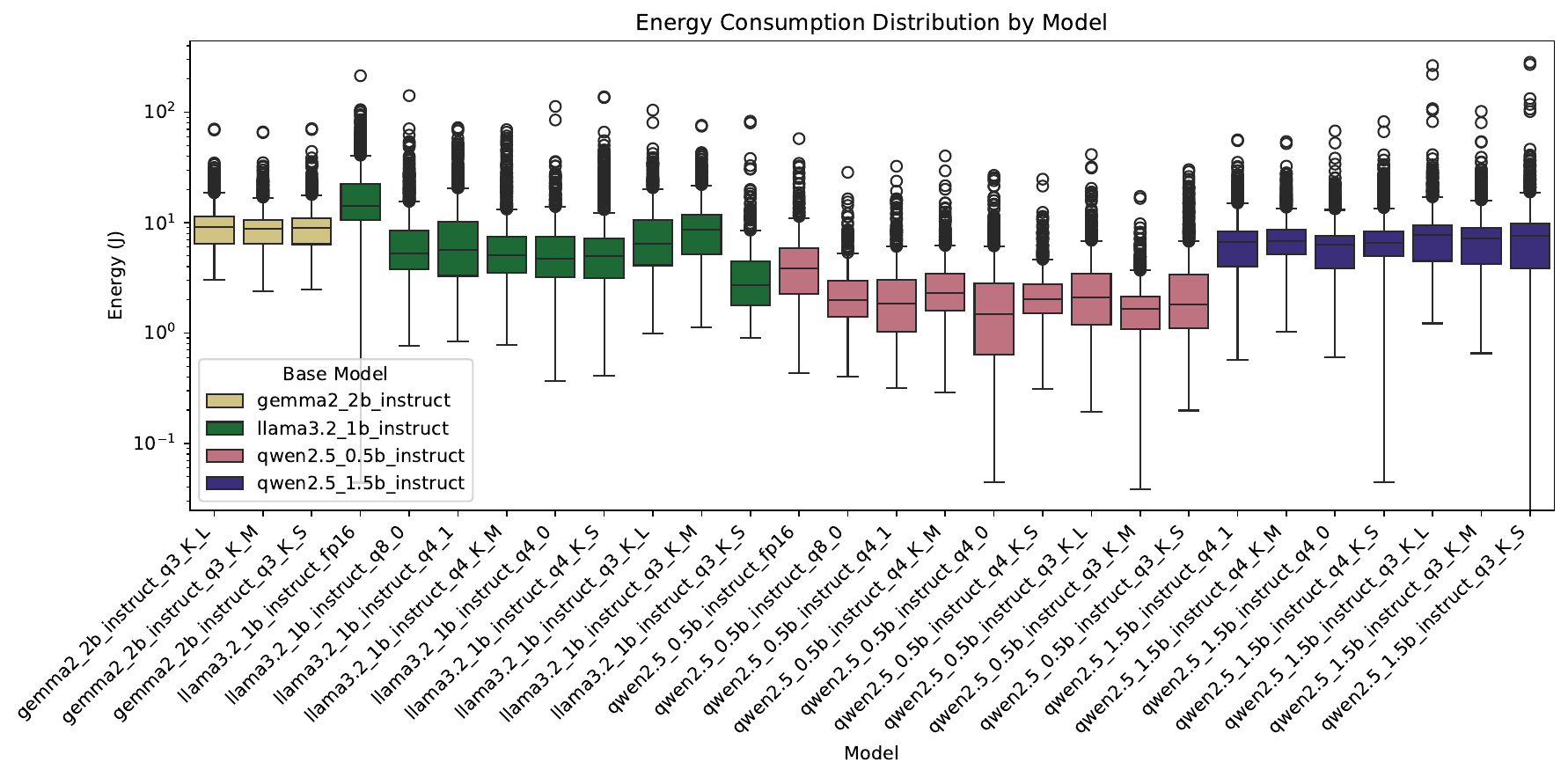}
    \vspace*{-1.20em}
    \caption{Distribution of energy consumption per token for all models, all datasets included.}
    \label{fig:energy_distribution_per_token_log_all}
\end{figure}

As shown in Figure~\ref{fig:energy_distribution_per_token_log_all}, energy consumption per token varies significantly across base models. Table 4 summarizes these differences, with qwen2.5\_0.5b being the most efficient at 2.61 J/token, approximately 3–3.5 times more efficient than other model families due to its smaller parameter size. Despite having more parameters, qwen2.5\_1.5b (7.57 J/token) is slightly more efficient than llaMA3.2\_1b (8.40 J/token), and also outperforms Gemma2\_2b (9.35 J/token).
Standard deviations indicate high variability, particularly in llama3.2\_1b (±5.36 J/token), which shows the greatest range between its full-precision and most quantized variants.


\begin{wraptable}[]{t}{0.670\textwidth}
\vspace*{-1.20em}
\scriptsize
\caption{Average energy consumption per response (Joules) by base model and quantization level. For comparison, the average cost of prompting Gemma2\_2b equals powering a 5W light bulb for 51 seconds.}
\vspace*{-1.20em}
\label{tab:avg_energy_per_response}
\begin{tabular}{lllll|r}
\toprule
 & gemma2\_2b & llama3.2\_1b & qwen25\_0.5b & qwen25\_1.5b & Average \\
\hline
fp16 & N/A & 159.42 & 91.59 & N/A & 125.51 \\
q8\_0 & N/A & 75.85 & 42.34 & N/A & 59.10 \\
q4 variants (avg) & N/A & 83.95 & 49.75 & 107.00 & 80.23 \\
q3 variants (avg) & 255.79 & 101.02 & 51.65 & 113.72 & 130.54 \\
\hline
Average & 255.79 & 105.06 & 58.83 & 110.36 & \\
\bottomrule
\end{tabular}
\vspace*{-1.20em}
\end{wraptable}

Examining energy consumption per response (Table~\ref{tab:avg_energy_per_response}) reveals patterns similar to per-token measurements. \\Qwen2.5\_0.5b is the most efficient at 58.83 J/response, while Gemma2\_2b consumes the most at 255.79 J/response. Quantization significantly reduces energy usage, with savings of 52\% in LLaMA3.2\_1b and 54\% in qwen2.5\_0.5b when comparing FP16 to q8\_0. Notably, q4 quantization is generally more efficient than q3, suggesting that aggressive quantization may introduce overhead that offsets energy gains.


\begin{table}
\scriptsize
\caption{Mean and standard deviation energy consumption of models on various datasets.}
\vspace*{-1.0em}
\label{tab:energy}
\begin{tabular}{lllllll}
\toprule
 & bigbenchhard & commonsenseqa & gsm8k & humaneval & truthfulqa & Average \\
Model &  &  &  &  &  &  \\
\midrule
gemma2\_2b\_instruct\_q3\_K\_L & 14.40 ± 9.02 & 10.38 ± 1.65 & 10.34 ± 2.20 & 4.69 ± 1.15 & 7.98 ± 2.59 & 9.56 ± 3.32 \\
gemma2\_2b\_instruct\_q3\_K\_M & 13.86 ± 8.66 & 9.56 ± 1.46 & 10.26 ± 1.92 & 4.70 ± 1.20 & 8.00 ± 2.66 & 9.28 ± 3.18 \\
gemma2\_2b\_instruct\_q3\_K\_S & 14.38 ± 9.69 & 9.67 ± 1.43 & 10.48 ± 2.49 & 3.14 ± 0.71 & 8.40 ± 2.67 & 9.22 ± 3.40 \\
\hline
llama3.2\_1b\_instruct\_fp16 & 26.89 ± 25.11 & 15.01 ± 7.34 & 25.44 ± 16.74 & 5.93 ± 2.22 & 14.74 ± 7.09 & 17.60 ± 11.70 \\
llama3.2\_1b\_instruct\_q8\_0 & 16.74 ± 15.31 & 5.81 ± 2.41 & 9.59 ± 6.22 & 2.40 ± 0.88 & 4.83 ± 2.02 & 7.87 ± 5.37 \\
llama3.2\_1b\_instruct\_q4\_1 & 20.66 ± 14.02 & 6.69 ± 2.06 & 5.74 ± 3.42 & 1.67 ± 0.46 & 7.36 ± 4.13 & 8.42 ± 4.82 \\
llama3.2\_1b\_instruct\_q4\_K\_M & 16.70 ± 13.19 & 5.21 ± 0.94 & 6.58 ± 3.73 & 2.25 ± 0.79 & 4.82 ± 2.21 & 7.11 ± 4.17 \\
llama3.2\_1b\_instruct\_q4\_0 & 12.84 ± 11.27 & 5.00 ± 1.60 & 6.21 ± 2.53 & 1.95 ± 1.02 & 4.56 ± 2.50 & 6.11 ± 3.78 \\
llama3.2\_1b\_instruct\_q4\_K\_S & 18.78 ± 16.65 & 5.21 ± 1.19 & 5.90 ± 2.50 & 1.91 ± 0.59 & 4.43 ± 1.93 & 7.25 ± 4.57 \\
llama3.2\_1b\_instruct\_q3\_K\_L & 13.95 ± 12.71 & 8.84 ± 3.72 & 9.20 ± 4.53 & 2.23 ± 1.10 & 7.02 ± 4.70 & 8.25 ± 5.35 \\
llama3.2\_1b\_instruct\_q3\_K\_M & 13.82 ± 10.75 & 10.93 ± 3.88 & 7.90 ± 3.53 & 2.33 ± 0.99 & 11.22 ± 5.64 & 9.24 ± 4.96 \\
llama3.2\_1b\_instruct\_q3\_K\_S & 6.03 ± 9.46 & 4.00 ± 2.30 & 3.02 ± 1.70 & 2.33 ± 1.23 & 3.38 ± 2.76 & 3.75 ± 3.49 \\
\hline
qwen2.5\_0.5b\_instruct\_fp16 & 7.25 ± 6.87 & 6.74 ± 2.27 & 3.68 ± 1.16 & 1.01 ± 0.10 & 3.43 ± 1.47 & 4.42 ± 2.37 \\
qwen2.5\_0.5b\_instruct\_q8\_0 & 3.40 ± 3.08 & 3.14 ± 1.40 & 1.79 ± 0.59 & 0.58 ± 0.06 & 1.79 ± 0.80 & 2.14 ± 1.19 \\
qwen2.5\_0.5b\_instruct\_q4\_1 & 5.00 ± 4.56 & 3.19 ± 0.88 & 1.60 ± 0.87 & 0.58 ± 0.10 & 1.82 ± 1.01 & 2.44 ± 1.48 \\
qwen2.5\_0.5b\_instruct\_q4\_K\_M & 4.43 ± 4.59 & 3.48 ± 1.49 & 2.17 ± 1.00 & 0.62 ± 0.09 & 2.08 ± 0.94 & 2.56 ± 1.62 \\
qwen2.5\_0.5b\_instruct\_q4\_0 & 5.51 ± 4.78 & 3.06 ± 1.01 & 1.01 ± 0.61 & 0.52 ± 0.08 & 1.42 ± 0.55 & 2.30 ± 1.41 \\
qwen2.5\_0.5b\_instruct\_q4\_K\_S & 3.80 ± 3.19 & 2.34 ± 1.01 & 2.09 ± 0.67 & 0.58 ± 0.08 & 2.10 ± 1.03 & 2.18 ± 1.20 \\
qwen2.5\_0.5b\_instruct\_q3\_K\_L & 5.52 ± 5.36 & 3.68 ± 0.75 & 2.07 ± 0.69 & 0.54 ± 0.06 & 1.77 ± 1.34 & 2.72 ± 1.64 \\
qwen2.5\_0.5b\_instruct\_q3\_K\_M & 3.03 ± 2.32 & 1.87 ± 0.47 & 1.90 ± 0.51 & 0.51 ± 0.05 & 1.61 ± 0.78 & 1.78 ± 0.83 \\
qwen2.5\_0.5b\_instruct\_q3\_K\_S & 7.04 ± 6.01 & 3.51 ± 0.85 & 1.54 ± 0.66 & 0.91 ± 0.30 & 1.51 ± 0.57 & 2.90 ± 1.68 \\
\hline
qwen2.5\_1.5b\_instruct\_q4\_1 & 12.07 ± 8.02 & 7.86 ± 1.34 & 8.13 ± 4.02 & 1.84 ± 0.42 & 4.99 ± 1.50 & 6.98 ± 3.06 \\
qwen2.5\_1.5b\_instruct\_q4\_K\_M & 12.44 ± 7.76 & 7.44 ± 1.77 & 7.24 ± 1.97 & 2.28 ± 0.94 & 5.50 ± 2.02 & 6.98 ± 2.89 \\
qwen2.5\_1.5b\_instruct\_q4\_0 & 11.51 ± 8.18 & 7.35 ± 1.20 & 6.77 ± 1.55 & 2.11 ± 0.95 & 4.88 ± 1.83 & 6.52 ± 2.74 \\
qwen2.5\_1.5b\_instruct\_q4\_K\_S & 15.23 ± 10.19 & 7.15 ± 1.57 & 6.79 ± 1.86 & 1.96 ± 0.61 & 5.22 ± 2.11 & 7.27 ± 3.27 \\
qwen2.5\_1.5b\_instruct\_q3\_K\_L & 16.58 ± 26.36 & 8.81 ± 1.37 & 9.87 ± 4.96 & 2.37 ± 0.65 & 6.04 ± 2.44 & 8.73 ± 7.16 \\
qwen2.5\_1.5b\_instruct\_q3\_K\_M & 13.28 ± 11.44 & 8.51 ± 1.51 & 7.56 ± 3.94 & 2.09 ± 0.54 & 6.23 ± 2.77 & 7.53 ± 4.04 \\
qwen2.5\_1.5b\_instruct\_q3\_K\_S & 18.62 ± 31.40 & 8.58 ± 1.82 & 11.18 ± 6.84 & 0.25 ± 0.60 & 6.18 ± 2.44 & 8.96 ± 8.62 \\
\hline
Average & 11.92 ± 10.71 & 6.54 ± 1.81 & 6.64 ± 2.98 & 1.94 ± 0.64 & 5.12 ± 2.30 & 6.43 \\
\bottomrule
\end{tabular}
\vspace*{-1.00em}
\end{table}

\begin{figure}
    \centering
    \includegraphics[width=1.0\linewidth]{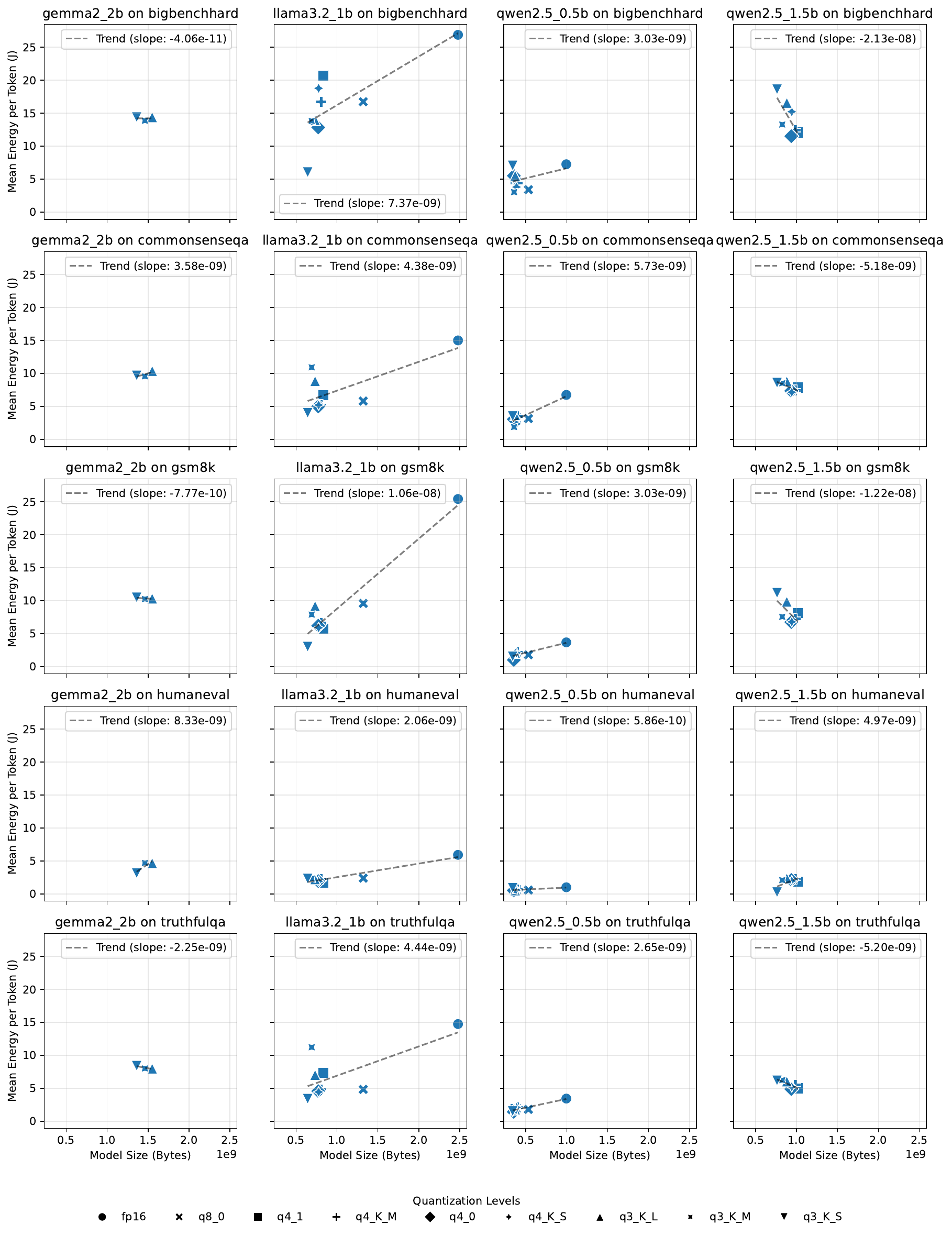}
    \vspace*{-1.40em}
    \caption{Relationship between model size (in bytes) and energy consumption per token across different model families and benchmark datasets. Each subplot represents a specific model-dataset combination with fitted linear regression trend lines showing the scaling relationship between model size and energy consumption. Marker shapes indicate different quantization techniques.}
    \label{fig:model_size_vs_energy}
\end{figure}

\subsubsection{Impact of Quantization on Energy Consumption}


Quantization significantly reduces energy consumption compared to higher-precision models. Llama3.2\_1b\_instruct\_fp16 consumes 17.60 J/token, while its most efficient quantized variant (q3\_K\_S) requires only 3.75 J/token, achieving a 79\% reduction. Similarly, for qwen2.5\_0.5b\_instruct, energy consumption drops from 4.42 J/token (FP16) to 1.78 J/token (q3\_K\_M), a 60\% reduction. The impact of quantization levels below FP16 on energy consumption is non-linear, and more aggressive quantization does not always yield greater efficiency. For example, in qwen2.5\_0.5b, q3\_K\_M (1.78 J/token) is 62\% more efficient than q3\_K\_S (2.90 J/token) despite having the same bit precision. Similarly, in qwen2.5\_1.5b, all q3 variants consume more energy per token than q4 variants, emphasizing that quantization method selection is as crucial as reducing bit width.



Table~\ref{tab:energy} shows high variability in energy consumption, particularly in highly quantized models. Llama3.2\_1b\_instruct\_q3\_K\_S exhibits extremely high variability (±3.49 J/token on average), sometimes exceeding the mean. This variance suggests that energy consumption in highly quantized models may be less predictable and more influenced by input characteristics. In contrast, qwen2.5\_0.5b\_instruct\_q3\_K\_M is more stable (±0.83 J/token).

Figure~\ref{fig:model_size_vs_energy} examines how energy consumption per token scales with model size (bytes) across five benchmark datasets. Model size correlates with quantization level, where higher precision (e.g., FP16, Q8) increases size, while aggressive quantization (e.g., Q3) reduces it. This enables a relative comparison of quantization methods across different precision levels. With four base models and five datasets, we analyzed twenty configurations, though some quantization levels were excluded due to memory limitations on the Raspberry Pi. For each configuration, we fitted linear regression trend lines, revealing distinct scaling patterns across models and tasks. A positive trend indicates that energy consumption rises with model size, meaning aggressive quantization reduces energy use. Conversely, a negative trend suggests that larger models may consume less energy per token than their more aggressively quantized counterparts.


For llama3.2\_1b and qwen2.5\_0.5b, positive scaling coefficients across all benchmarks (0.625 to 1.15) indicate that energy consumption increases with model size, driven by the efficiency gains of quantization beyond FP16. However, other models show unexpected patterns. Qwen2.5\_1.5b exhibits negative scaling coefficients (-0.572 to -1.27) on several benchmarks, suggesting that some larger quantized variants may be more energy-efficient than smaller ones. Gemma2\_2b displays mixed behavior, with coefficients ranging from -0.403 to 3.16 across tasks. These diverse trends highlight the complex interplay between model architecture, quantization strategy, and task characteristics. 

\subsubsection{Task-Specific Energy Consumption Patterns}

Task characteristics greatly impact energy consumption. Table~\ref{tab:energy} shows HumanEval has the lowest energy use (1.94 J/token), while BigBenchHard has the highest (11.92 J/token)—a 6× difference. Table~\ref{tab:response_length} reveals that HumanEval responses are significantly longer (162.49 tokens on average) compared to 8.00–21.63 tokens for other tasks. Since HumanEval requires code generation, longer outputs distribute the fixed inference overhead across more tokens, reducing energy per token. The quantization impact varies by task. In GSM8K (mathematical reasoning), llama3.2\_1b shows high variability, with energy consumption ranging from 3.02 J/token (q3\_K\_S) to 25.44 J/token (FP16)—an 8.4× difference. In contrast, for HumanEval (code generation), the range is narrower (2.33 to 5.93 J/token, 2.5× difference). This suggests quantization provides greater efficiency gains for mathematical reasoning than for code generation.



\subsubsection{Correlation Between Response Length and Energy Consumption}

\begin{wrapfigure}{h}{0.590\textwidth}

    \centering
    \includegraphics[width=0.590\textwidth]{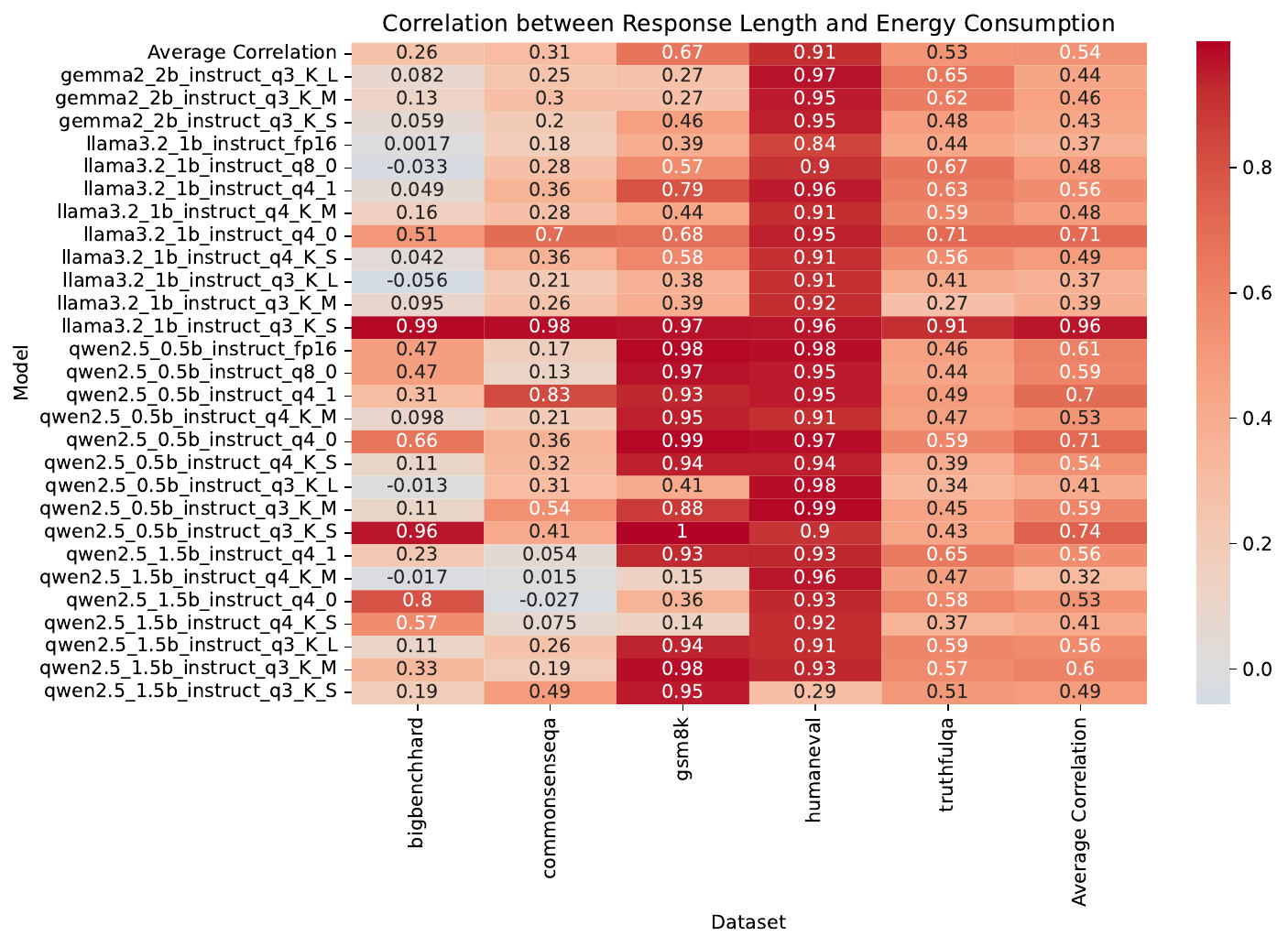}

    \vspace*{-1.40em}

    \caption{Correlation between response length and energy consumption per response.}
        \vspace*{-1.00em}

    \label{fig:response_length_correlation}
\end{wrapfigure}

Due to varying response lengths across datasets (Table~\ref{tab:response_length}), we analyzed their correlation with energy consumption. Figure~\ref{fig:response_length_correlation} shows the Pearson correlation coefficients for each model variant and benchmark, highlighting this relationship. The average correlation across all models and benchmarks is 0.54, indicating that response length significantly impacts energy consumption but is not the sole factor. This analysis reveals several key patterns:


The correlation strength varies across datasets, with HumanEval showing the highest (0.91 on average). This is likely due to its longer, variable-length responses (162.49 tokens on average) in code generation tasks. The strong correlation suggests response length can be a reliable proxy for estimating energy consumption in such tasks.

BigBenchHard and CommonsenseQA show weaker correlations (0.26 and 0.31), suggesting for shorter responses, factors beyond token count play a larger role in energy consumption.

Energy consumption varies across model families, with the variant llama3.2\_1b\_instruct\_q3\_K\_S showing a strong correlation (0.96) between response length and energy usage across benchmarks. Its higher average response length (74.54 vs. overall 44.49, Table~\ref{tab:response_length}.
For models with longer responses, energy consumption becomes more predictable based on output length because fixed computational overhead becomes proportionally smaller. In contrast, models generating shorter responses show more variable energy-output relationships, as initialization and prompt processing costs dominate.


These findings are crucial for energy estimation in real-world deployments. In tasks with longer responses (e.g., code generation), response length reliably predicts energy consumption. However, for shorter responses, energy estimates must consider additional factors beyond length.

\begin{tcolorbox}[boxsep=2pt,left=2pt,right=2pt,top=2pt,bottom=2pt]
\paragraph{\textbf{RQ1 Conclusion.}} 
Our analysis reveals that quantization significantly reduces energy consumption, with more aggressive quantization (e.g., q3\_K\_M vs. FP16) yielding up to 79\% savings in some models. However, the relationship between quantization level and energy efficiency is non-linear, as extreme quantization sometimes increases energy usage, possibly due to overhead or inefficiencies. Additionally, task characteristics influence energy consumption, with longer response tasks (e.g., HumanEval) showing lower energy per token due to fixed inference overhead distribution, while short-response tasks exhibit greater variability. These highlight the importance of selecting quantization strategies based on task type and deployment constraints.


\end{tcolorbox}

\subsection{What are the trade-offs between accuracy and energy efficiency across different quantization levels in LLMs deployed on an edge device? (RQ2)}
\label{subsec:RQ2}


To address RQ2, we analyzed the accuracy-energy tradeoff across quantization levels and model families. Table~\ref{tab:accuracy} reports accuracy scores for all models on the five benchmark datasets, while Figure~\ref{fig:accuracy_benchmarks} visualizes accuracy comparisons across variants.

\subsubsection{Impact of Quantization on Accuracy}



\begin{table}
\scriptsize
\caption{Average response length for models on various datasets, with standard deviation.}
\label{tab:response_length}
\vspace*{-1.20em}

\begin{tabular}{lllllll}
\toprule
 & bigbenchhard & commonsenseqa & gsm8k & humaneval & truthfulqa & Average \\
Model &  &  &  &  &  &  \\
\midrule
gemma2\_2b\_instruct\_q3\_K\_L & 14.07 ± 8.25 & 11.12 ± 1.53 & 12.28 ± 3.32 & 190.56 ± 106.81 & 19.54 ± 10.20 & 49.51 ± 26.02 \\
gemma2\_2b\_instruct\_q3\_K\_M & 14.37 ± 11.39 & 11.53 ± 1.60 & 11.89 ± 1.70 & 168.29 ± 89.30 & 19.18 ± 9.68 & 45.05 ± 22.73 \\
gemma2\_2b\_instruct\_q3\_K\_S & 14.90 ± 10.00 & 12.19 ± 1.34 & 13.40 ± 5.87 & 196.04 ± 87.90 & 18.64 ± 5.59 & 51.03 ± 22.14 \\
\hline
llama3.2\_1b\_instruct\_fp16 & 9.27 ± 8.92 & 7.41 ± 2.13 & 6.49 ± 5.32 & 52.72 ± 38.65 & 12.19 ± 6.64 & 17.61 ± 12.33 \\
llama3.2\_1b\_instruct\_q8\_0 & 9.16 ± 8.90 & 7.68 ± 1.94 & 6.68 ± 4.94 & 64.20 ± 40.02 & 12.59 ± 7.58 & 20.06 ± 12.67 \\
llama3.2\_1b\_instruct\_q4\_1 & 10.98 ± 13.78 & 6.62 ± 2.05 & 15.59 ± 18.46 & 124.67 ± 83.34 & 9.86 ± 8.24 & 33.54 ± 25.17 \\
llama3.2\_1b\_instruct\_q4\_K\_M & 10.32 ± 17.00 & 7.95 ± 1.08 & 9.14 ± 5.15 & 68.99 ± 46.95 & 12.93 ± 7.88 & 21.87 ± 15.61 \\
llama3.2\_1b\_instruct\_q4\_0 & 13.26 ± 32.87 & 9.10 ± 7.88 & 9.62 ± 8.00 & 103.50 ± 85.60 & 14.43 ± 14.30 & 29.98 ± 29.73 \\
llama3.2\_1b\_instruct\_q4\_K\_S & 10.57 ± 12.84 & 7.99 ± 2.70 & 9.33 ± 5.96 & 81.93 ± 50.05 & 13.31 ± 7.38 & 24.63 ± 15.79 \\
llama3.2\_1b\_instruct\_q3\_K\_L & 9.13 ± 10.72 & 5.88 ± 2.15 & 6.95 ± 4.03 & 84.43 ± 66.48 & 10.97 ± 6.24 & 23.47 ± 17.93 \\
llama3.2\_1b\_instruct\_q3\_K\_M & 7.18 ± 10.24 & 4.36 ± 2.12 & 7.30 ± 3.18 & 110.13 ± 82.76 & 5.48 ± 3.28 & 26.89 ± 20.32 \\
llama3.2\_1b\_instruct\_q3\_K\_S & 80.35 ± 316.80 & 27.71 ± 33.27 & 47.45 ± 57.44 & 179.57 ± 193.26 & 37.62 ± 35.56 & 74.54 ± 127.27 \\
\hline
qwen25\_0.5b\_instruct\_fp16 & 13.29 ± 21.25 & 4.64 ± 0.96 & 24.72 ± 60.38 & 271.23 ± 109.90 & 14.47 ± 5.63 & 65.67 ± 39.62 \\
qwen25\_0.5b\_instruct\_q8\_0 & 11.96 ± 18.42 & 4.65 ± 1.02 & 20.57 ± 53.84 & 209.98 ± 69.21 & 14.54 ± 5.42 & 52.34 ± 29.58 \\
qwen25\_0.5b\_instruct\_q4\_1 & 11.15 ± 21.93 & 6.07 ± 8.71 & 66.37 ± 119.06 & 233.70 ± 107.01 & 14.07 ± 6.70 & 66.27 ± 52.68 \\
qwen25\_0.5b\_instruct\_q4\_K\_M & 12.33 ± 12.43 & 5.58 ± 2.80 & 31.31 ± 64.23 & 205.92 ± 78.68 & 14.90 ± 6.97 & 54.01 ± 33.02 \\
qwen25\_0.5b\_instruct\_q4\_0 & 18.09 ± 45.29 & 5.25 ± 2.25 & 121.16 ± 142.43 & 236.14 ± 114.56 & 16.91 ± 7.96 & 79.51 ± 62.50 \\
qwen25\_0.5b\_instruct\_q4\_K\_S & 12.42 ± 10.26 & 8.56 ± 2.93 & 16.88 ± 43.78 & 201.92 ± 81.86 & 14.57 ± 6.23 & 50.87 ± 29.01 \\
qwen25\_0.5b\_instruct\_q3\_K\_L & 10.97 ± 10.36 & 4.45 ± 0.85 & 10.92 ± 4.43 & 258.34 ± 104.98 & 15.45 ± 6.28 & 60.03 ± 25.38 \\
qwen25\_0.5b\_instruct\_q3\_K\_M & 12.73 ± 12.66 & 9.55 ± 3.00 & 13.85 ± 21.87 & 259.72 ± 111.43 & 15.94 ± 8.07 & 62.36 ± 31.41 \\
qwen25\_0.5b\_instruct\_q3\_K\_S & 27.30 ± 218.86 & 4.82 ± 2.31 & 45.29 ± 300.07 & 206.67 ± 105.84 & 16.43 ± 4.93 & 60.10 ± 126.40 \\
\hline
qwen25\_1.5b\_instruct\_q4\_1 & 10.60 ± 18.21 & 7.08 ± 0.48 & 13.14 ± 36.77 & 158.28 ± 85.26 & 16.14 ± 5.13 & 41.05 ± 29.17 \\
qwen25\_1.5b\_instruct\_q4\_K\_M & 8.36 ± 6.13 & 7.08 ± 0.86 & 9.27 ± 1.32 & 125.84 ± 96.84 & 15.39 ± 4.86 & 33.19 ± 22.00 \\
qwen25\_1.5b\_instruct\_q4\_0 & 14.55 ± 72.90 & 7.01 ± 0.14 & 10.16 ± 5.02 & 130.84 ± 83.80 & 16.02 ± 8.20 & 35.71 ± 34.01 \\
qwen25\_1.5b\_instruct\_q4\_K\_S & 11.04 ± 40.66 & 7.03 ± 0.17 & 9.43 ± 1.23 & 133.92 ± 91.46 & 15.38 ± 4.80 & 35.36 ± 27.66 \\
qwen25\_1.5b\_instruct\_q3\_K\_L & 10.67 ± 15.45 & 7.26 ± 0.83 & 16.48 ± 38.77 & 144.63 ± 79.20 & 16.19 ± 8.63 & 39.05 ± 28.57 \\
qwen25\_1.5b\_instruct\_q3\_K\_M & 12.60 ± 23.88 & 7.24 ± 0.93 & 26.62 ± 66.98 & 151.29 ± 96.29 & 15.19 ± 8.56 & 42.59 ± 39.33 \\
qwen25\_1.5b\_instruct\_q3\_K\_S & 13.32 ± 20.97 & 8.22 ± 4.75 & 13.29 ± 35.52 & 196.29 ± 106.73 & 16.02 ± 7.71 & 49.43 ± 35.14 \\
\hline
Average & 14.82 ± 36.83 & 8.00 ± 3.31 & 21.63 ± 39.97 & 162.49 ± 89.08 & 15.51 ± 8.17 & 44.49 \\
\bottomrule
\end{tabular}
\vspace*{-0.20em}
\end{table}

\begin{table}
\scriptsize
\caption{Average accuracy of models.}
\label{tab:accuracy}
\vspace*{-1.20em}
\begin{tabular}{lrrrrr|r}
\toprule
Dataset & bigbenchhard & commonsenseqa & gsm8k & humaneval & truthfulqa & Average \\
Model &  &  &  &  &  &  \\
\hline
gemma2\_2b\_instruct\_q3\_K\_L & 0.35 ± 0.48 & 0.67 ± 0.47 & 0.07 ± 0.26 & 0.25 ± 0.43 & 0.37 ± 0.48 & 0.34 ± 0.43 \\
gemma2\_2b\_instruct\_q3\_K\_M & 0.36 ± 0.48 & 0.67 ± 0.47 & 0.07 ± 0.26 & 0.77 ± 0.42 & 0.36 ± 0.48 & 0.45 ± 0.42 \\
gemma2\_2b\_instruct\_q3\_K\_S & 0.34 ± 0.47 & 0.70 ± 0.46 & 0.04 ± 0.21 & 0.02 ± 0.13 & 0.40 ± 0.49 & 0.30 ± 0.35 \\
\hline
llama3.2\_1b\_instruct\_fp16 & 0.16 ± 0.37 & 0.35 ± 0.48 & 0.03 ± 0.16 & 0.62 ± 0.49 & 0.31 ± 0.46 & 0.29 ± 0.39 \\
llama3.2\_1b\_instruct\_q8\_0 & 0.17 ± 0.37 & 0.38 ± 0.49 & 0.04 ± 0.18 & 0.85 ± 0.36 & 0.34 ± 0.48 & 0.35 ± 0.38 \\
llama3.2\_1b\_instruct\_q4\_1 & 0.15 ± 0.35 & 0.23 ± 0.43 & 0.04 ± 0.18 & 0.84 ± 0.37 & 0.27 ± 0.44 & 0.30 ± 0.35 \\
llama3.2\_1b\_instruct\_q4\_K\_M & 0.23 ± 0.43 & 0.43 ± 0.50 & 0.06 ± 0.24 & 0.87 ± 0.34 & 0.34 ± 0.47 & 0.39 ± 0.39 \\
llama3.2\_1b\_instruct\_q4\_0 & 0.20 ± 0.40 & 0.27 ± 0.44 & 0.05 ± 0.22 & 0.79 ± 0.41 & 0.39 ± 0.49 & 0.34 ± 0.39 \\
llama3.2\_1b\_instruct\_q4\_K\_S & 0.27 ± 0.44 & 0.43 ± 0.50 & 0.04 ± 0.18 & 0.84 ± 0.37 & 0.39 ± 0.49 & 0.39 ± 0.40 \\
llama3.2\_1b\_instruct\_q3\_K\_L & 0.14 ± 0.35 & 0.34 ± 0.47 & 0.03 ± 0.17 & 0.34 ± 0.48 & 0.14 ± 0.35 & 0.20 ± 0.36 \\
llama3.2\_1b\_instruct\_q3\_K\_M & 0.10 ± 0.30 & 0.13 ± 0.34 & 0.03 ± 0.17 & 0.54 ± 0.50 & 0.03 ± 0.17 & 0.17 ± 0.30 \\
llama3.2\_1b\_instruct\_q3\_K\_S & 0.08 ± 0.26 & 0.26 ± 0.44 & 0.02 ± 0.14 & 0.49 ± 0.50 & 0.07 ± 0.26 & 0.18 ± 0.32 \\
\hline
qwen2.5\_0.5b\_instruct\_fp16 & 0.27 ± 0.45 & 0.32 ± 0.47 & 0.04 ± 0.21 & 0.74 ± 0.44 & 0.23 ± 0.42 & 0.32 ± 0.40 \\
qwen2.5\_0.5b\_instruct\_q8\_0 & 0.29 ± 0.45 & 0.32 ± 0.47 & 0.04 ± 0.18 & 0.54 ± 0.50 & 0.22 ± 0.42 & 0.28 ± 0.40 \\
qwen2.5\_0.5b\_instruct\_q4\_1 & 0.28 ± 0.45 & 0.36 ± 0.48 & 0.07 ± 0.26 & 0.75 ± 0.44 & 0.29 ± 0.46 & 0.35 ± 0.42 \\
qwen2.5\_0.5b\_instruct\_q4\_K\_M & 0.25 ± 0.43 & 0.32 ± 0.47 & 0.06 ± 0.24 & 0.56 ± 0.50 & 0.26 ± 0.44 & 0.29 ± 0.42 \\
qwen2.5\_0.5b\_instruct\_q4\_0 & 0.22 ± 0.42 & 0.28 ± 0.45 & 0.05 ± 0.22 & 0.72 ± 0.45 & 0.24 ± 0.43 & 0.30 ± 0.39 \\
qwen2.5\_0.5b\_instruct\_q4\_K\_S & 0.25 ± 0.43 & 0.36 ± 0.48 & 0.08 ± 0.27 & 0.52 ± 0.50 & 0.26 ± 0.44 & 0.29 ± 0.43 \\
qwen2.5\_0.5b\_instruct\_q3\_K\_L & 0.28 ± 0.45 & 0.30 ± 0.46 & 0.03 ± 0.17 & 0.72 ± 0.45 & 0.31 ± 0.46 & 0.33 ± 0.40 \\
qwen2.5\_0.5b\_instruct\_q3\_K\_M & 0.26 ± 0.44 & 0.37 ± 0.48 & 0.07 ± 0.25 & 0.70 ± 0.46 & 0.24 ± 0.43 & 0.33 ± 0.41 \\
qwen2.5\_0.5b\_instruct\_q3\_K\_S & 0.20 ± 0.40 & 0.26 ± 0.44 & 0.01 ± 0.10 & 0.77 ± 0.42 & 0.31 ± 0.46 & 0.31 ± 0.36 \\
\hline
qwen2.5\_1.5b\_instruct\_q4\_1 & 0.33 ± 0.47 & 0.67 ± 0.47 & 0.09 ± 0.28 & 0.08 ± 0.27 & 0.29 ± 0.46 & 0.29 ± 0.39 \\
qwen2.5\_1.5b\_instruct\_q4\_K\_M & 0.32 ± 0.47 & 0.62 ± 0.49 & 0.12 ± 0.33 & 0.20 ± 0.40 & 0.24 ± 0.43 & 0.30 ± 0.42 \\
qwen2.5\_1.5b\_instruct\_q4\_0 & 0.33 ± 0.47 & 0.66 ± 0.48 & 0.09 ± 0.29 & 0.27 ± 0.44 & 0.24 ± 0.43 & 0.32 ± 0.42 \\
qwen2.5\_1.5b\_instruct\_q4\_K\_S & 0.33 ± 0.47 & 0.57 ± 0.50 & 0.13 ± 0.34 & 0.18 ± 0.39 & 0.23 ± 0.42 & 0.29 ± 0.42 \\
qwen2.5\_1.5b\_instruct\_q3\_K\_L & 0.29 ± 0.45 & 0.45 ± 0.50 & 0.09 ± 0.28 & 0.03 ± 0.17 & 0.32 ± 0.47 & 0.23 ± 0.37 \\
qwen2.5\_1.5b\_instruct\_q3\_K\_M & 0.26 ± 0.44 & 0.40 ± 0.49 & 0.07 ± 0.25 & 0.02 ± 0.13 & 0.24 ± 0.43 & 0.20 ± 0.35 \\
qwen2.5\_1.5b\_instruct\_q3\_K\_S & 0.26 ± 0.44 & 0.57 ± 0.50 & 0.03 ± 0.17 & 0.17 ± 0.37 & 0.18 ± 0.39 & 0.24 ± 0.37 \\
\hline
Average Accuracy & 0.25 ± 0.42 & 0.42 ± 0.47 & 0.06 ± 0.22 & 0.51 ± 0.40 & 0.27 ± 0.43 & 0.30 \\
\bottomrule
\end{tabular}
\end{table}

\begin{wrapfigure}{h}{0.580\textwidth}
\vspace*{-1.30em}

    \centering
    \includegraphics[width=0.580\textwidth]{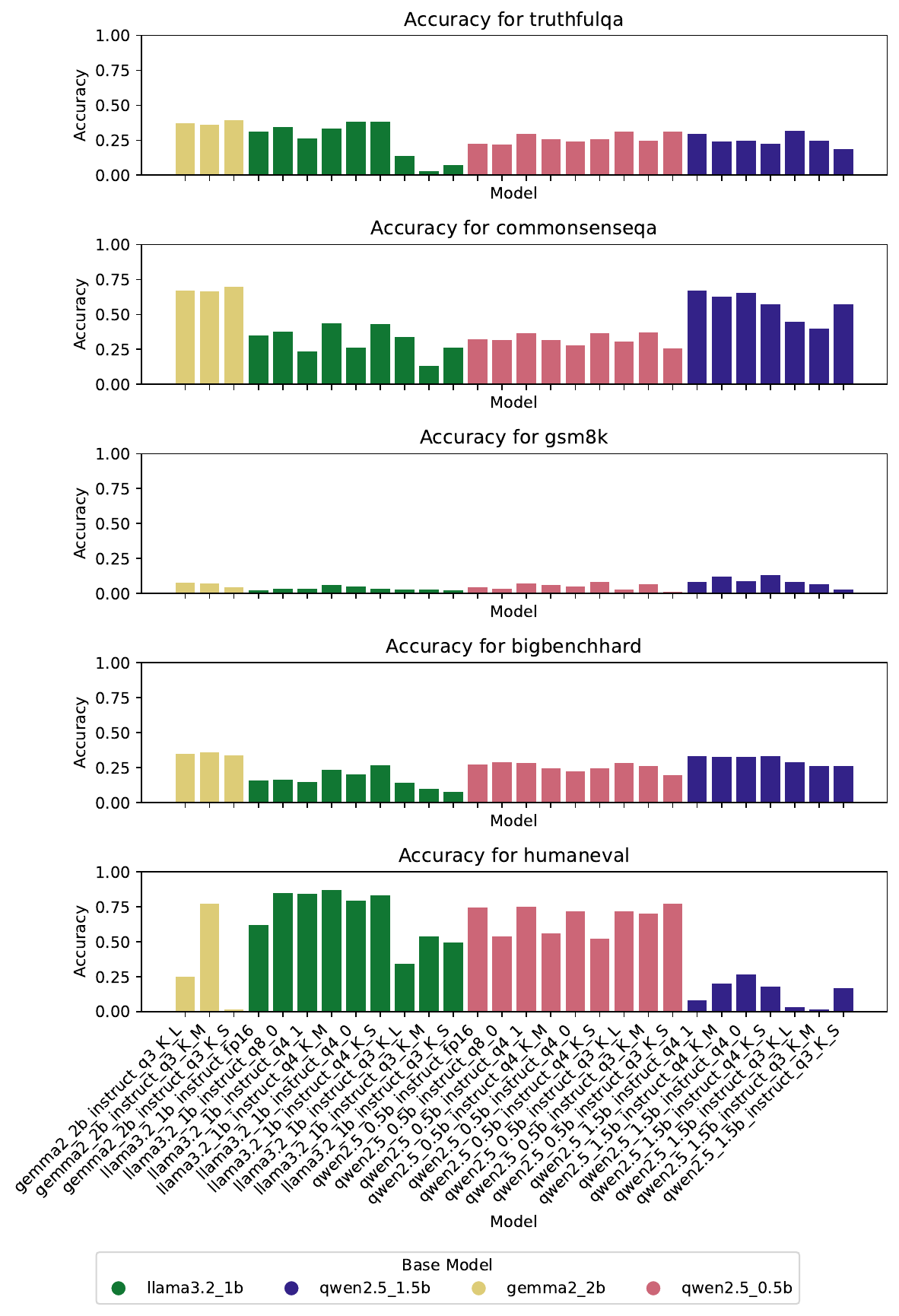}
    \vspace*{-1.20em}

    \caption{Accuracy comparison across datasets for all model variants, grouped by model family, highlighting task-specific sensitivities to model architecture and quantization.}
    \label{fig:accuracy_benchmarks}
    \vspace*{-1.20em}

\end{wrapfigure}

Figure~\ref{fig:accuracy_benchmarks} shows that quantization impacts accuracy differently across model families. In gemma2\_2b base model, accuracy ranges from 0.30 ± 0.35 to 0.45 ± 0.42, with q3\_K\_M (0.45 ± 0.42) outperforming q3\_K\_L despite more aggressive quantization. This suggests that some quantization techniques can preserve or even enhance model performance for specific architectures. The llama3.2\_1b base model shows the widest accuracy variation (0.17 ± 0.30 to 0.39 ± 0.40), with greater quantization leading to clear degradation. While q4\_K variants retain competitive accuracy (0.39 ± 0.40), q3\_K variants drop to 0.17–0.20, losing about 40\% accuracy compared to higher-precision models.



The qwen2.5\_0.5b base model remains stable across quantization levels, with accuracy ranging from 0.28 ± 0.40 to 0.35 ± 0.42 and no clear link between quantization aggressiveness and accuracy. Even the most aggressively quantized variant (q3\_K\_S) achieves 0.31 ± 0.36, only slightly below the full-precision model (0.32 ± 0.40). In contrast, qwen2.5\_1.5b shows moderate accuracy loss with stronger quantization, dropping from 0.32 ± 0.42 (q4\_0) to 0.20 ± 0.35 (q3\_K\_M), suggesting greater sensitivity to precision reduction than its smaller counterpart.

\subsubsection{Task-Specific Accuracy Patterns}


Table~\ref{tab:accuracy} highlights task-specific accuracy variations and quantization sensitivity. HumanEval (code generation) has the highest average accuracy (0.51) but exhibits high variability. For example, gemma2\_2b ranges from 0.02 to 0.77, while qwen2.5\_1.5b varies from 0.02 to 0.27, showing inconsistent responses to quantization.

GSM8K (mathematical reasoning) is challenging across all models and quantization levels, with an average accuracy of 0.06. Notably, aggressive quantization does not significantly degrade performance, suggesting that higher precision offers little benefit for models struggling with complex reasoning. 



CommonsenseQA shows a strong correlation between model size and accuracy, with larger models (gemma2\_2b, qwen2.5\_1.5b) outperforming smaller ones, regardless of quantization level. Gemma2\_2b achieves 0.67–0.70 accuracy, while qwen2.5\_1.5b reaches 0.57–0.67, compared to 0.26–0.43 for smaller models. 
For commonsense reasoning, model scale remains crucial, even under quantization constraints.


High standard deviations (0.35–0.45) across tasks indicate significant instance-level variability in model performance. This suggests that overall accuracy metrics may not fully capture quantization effects, as some instances are highly impacted while others remain stable.

\subsubsection{Energy-Accuracy Tradeoffs}

\begin{wrapfigure}{h}{0.560\textwidth}
\vspace*{-1.30em}

    \centering
    \includegraphics[width=0.560\textwidth]{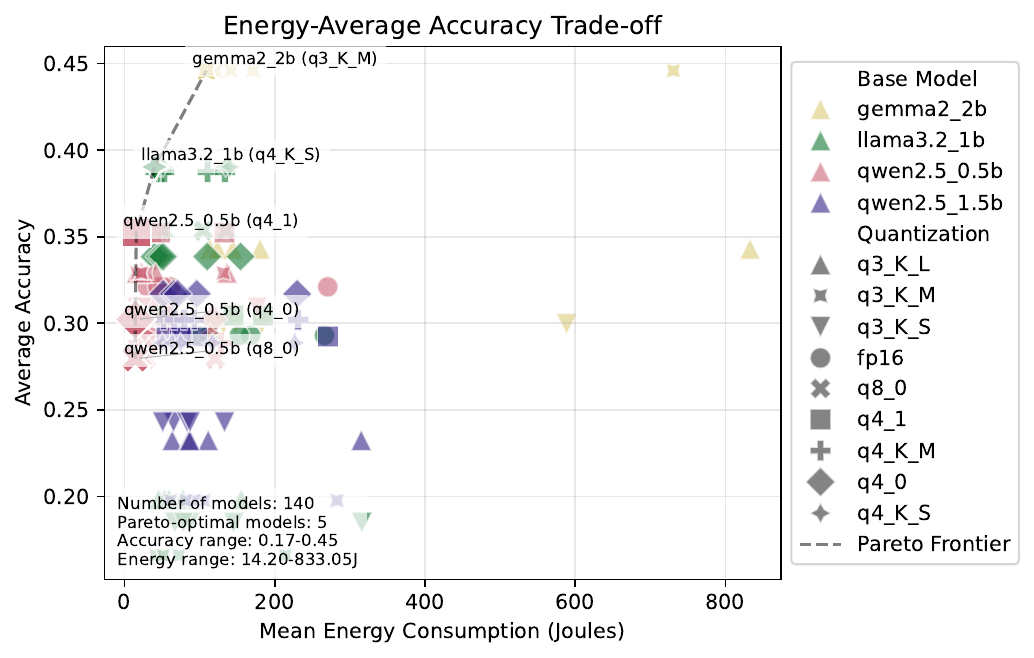}
    \vspace*{-1.90em}

    \caption{Energy-accuracy tradeoff across all datasets. The Pareto frontier (dashed line) highlights optimal models, with points colored by the dataset and shaped by the quantization technique. Pareto-optimal models are annotated.}
    
    \label{fig:energy_accuracy_all}
    \vspace*{-1.20em}

\end{wrapfigure}

Figure~\ref{fig:energy_accuracy_all} highlights the tradeoff between energy consumption and average accuracy across all datasets. The Pareto frontier (dashed line) marks models that achieve optimal accuracy-energy efficiency balances. It includes only five models, showing that few configurations achieve truly optimal energy-accuracy tradeoffs across tasks. These Pareto-optimal models vary in model family and quantization technique.


\begin{enumerate}
    \item Gemma2\_2b\_instruct\_q3\_K\_M achieves the highest overall accuracy (0.45) despite aggressive q3 quantization, showing that effective quantization can maintain performance at low precision. It consumes 109 Joules/response on average, placing it mid-range in energy use.

    \item Llama3.2\_1b\_instruct\_q4\_K\_S balances accuracy (0.39) and energy efficiency (40 Joules/response), making it ideal for performance-conscious, energy-constrained deployments.
    
    \item The qwen2.5\_0.5b base model dominates the low-energy Pareto frontier, with q4\_1, q4\_0, and q8\_0 variants achieving 0.28–0.35 accuracy while consuming just 14–16 Joules. These configurations are ideal for energy-constrained environments.

\end{enumerate}

No qwen2.5\_1.5b variants reach the Pareto frontier despite their larger parameter count. While strong in CommonsenseQA, their energy-efficiency ratio is lower than other models across diverse tasks. The Pareto frontier shows diminishing returns—improving from qwen2.5\_0.5b (0.35 accuracy) to llama3.2\_1b (0.39 accuracy) nearly doubles energy use, while reaching gemma2\_2b (0.45 accuracy) requires 40\% more energy. This nonlinear tradeoff is critical for deployment decisions, highlighting the energy cost of incremental accuracy gains.


\begin{wrapfigure}{h}{0.560\textwidth}
\vspace*{-1.30em}
    \centering
    \includegraphics[width=0.560\textwidth]{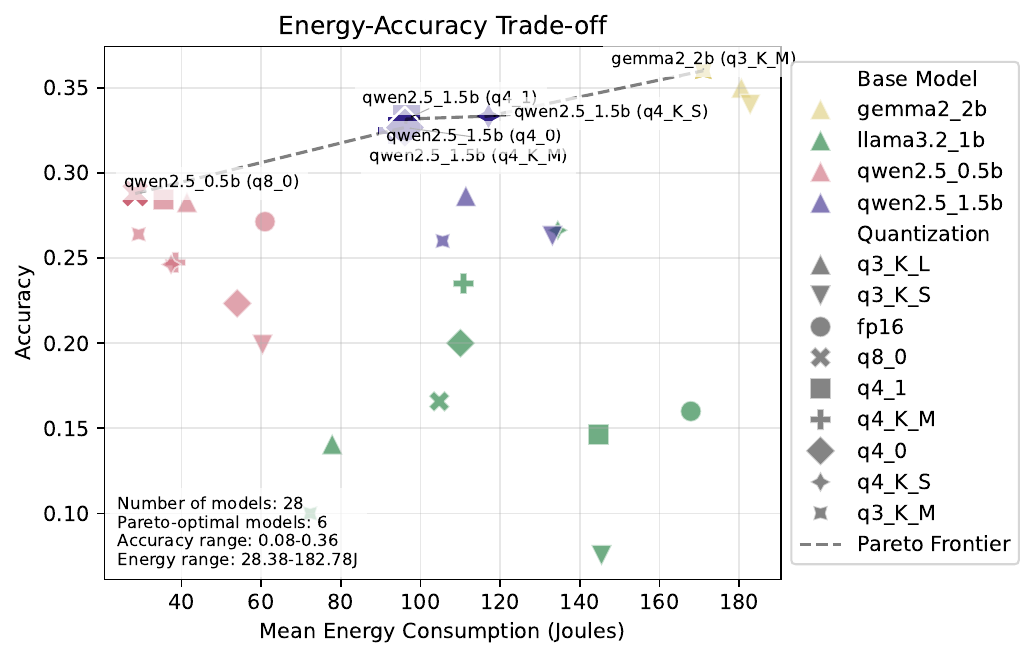}
    \caption{Energy-accuracy tradeoff for BigBenchHard. The Pareto frontier highlights six optimal models, with qwen2.5\_0.5b and qwen2.5\_1.5b variants achieving the best balance for this complex reasoning task.}
    
    \label{fig:energy_bigbenchhard}
\end{wrapfigure}

\begin{wrapfigure}{h}{0.560\textwidth}
\vspace*{-0.10em}
    \centering
    \includegraphics[width=0.560\textwidth]{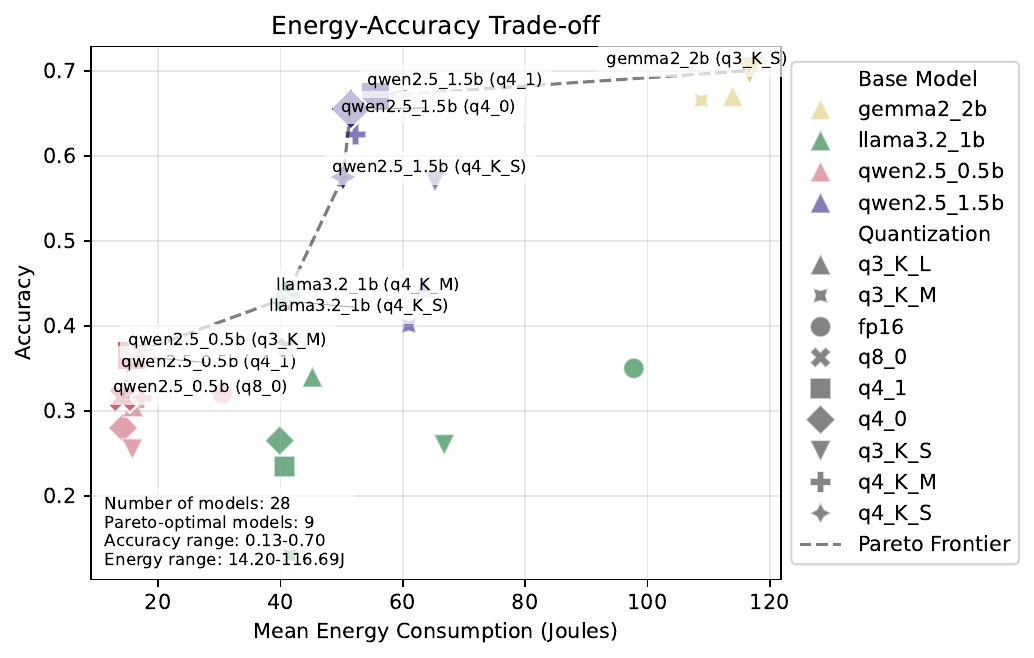}
    \vspace*{-1.90em}

    \caption{Energy-accuracy tradeoff for CommonsenseQA. The Pareto frontier identifies nine optimal models, with gemma2\_2b variants achieving the highest accuracy (0.70) at higher energy costs.}
    
    \label{fig:energy_commonsenseqa}
        \vspace*{-1.10em}

\end{wrapfigure}
\begin{wrapfigure}{h}{0.560\textwidth}
\vspace*{-1.10em}
    \centering

    \includegraphics[width=0.560\textwidth]{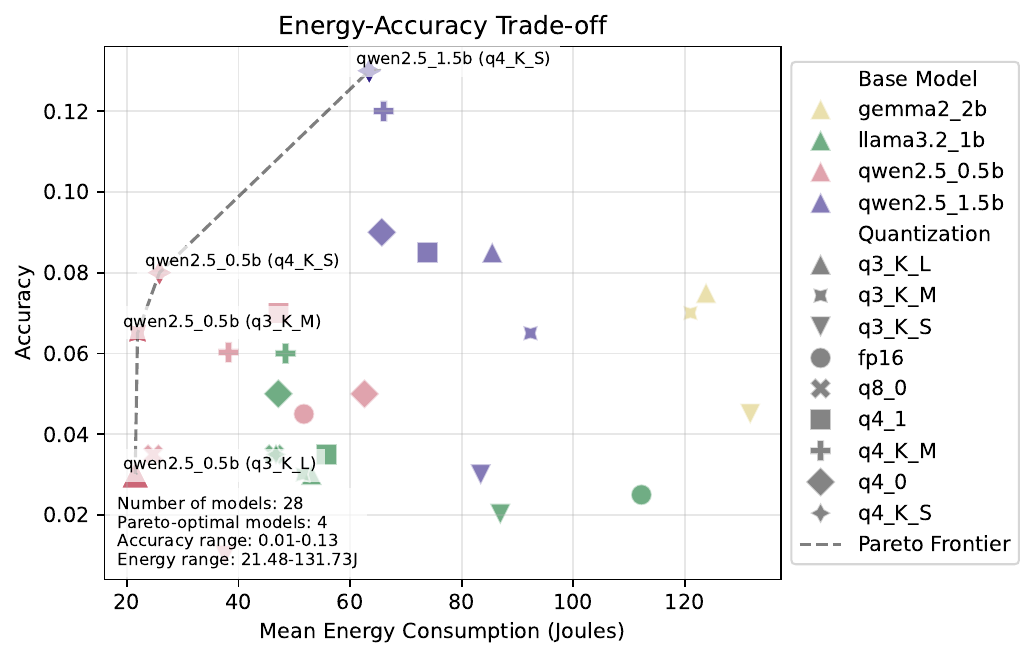}
    \vspace*{-1.90em}

    \caption{Energy-accuracy tradeoff for GSM8K. The Pareto frontier includes four models, with qwen2.5\_1.5b and qwen2.5\_0.5b achieving the highest accuracy for mathematical reasoning despite low overall performance across models.}
    
    \label{fig:energy_gsm8k}
    \vspace*{-1.70em}

\end{wrapfigure}
\begin{figure}
    \centering
    \begin{minipage}{0.45\textwidth}
        \centering
        \includegraphics[width=\textwidth]{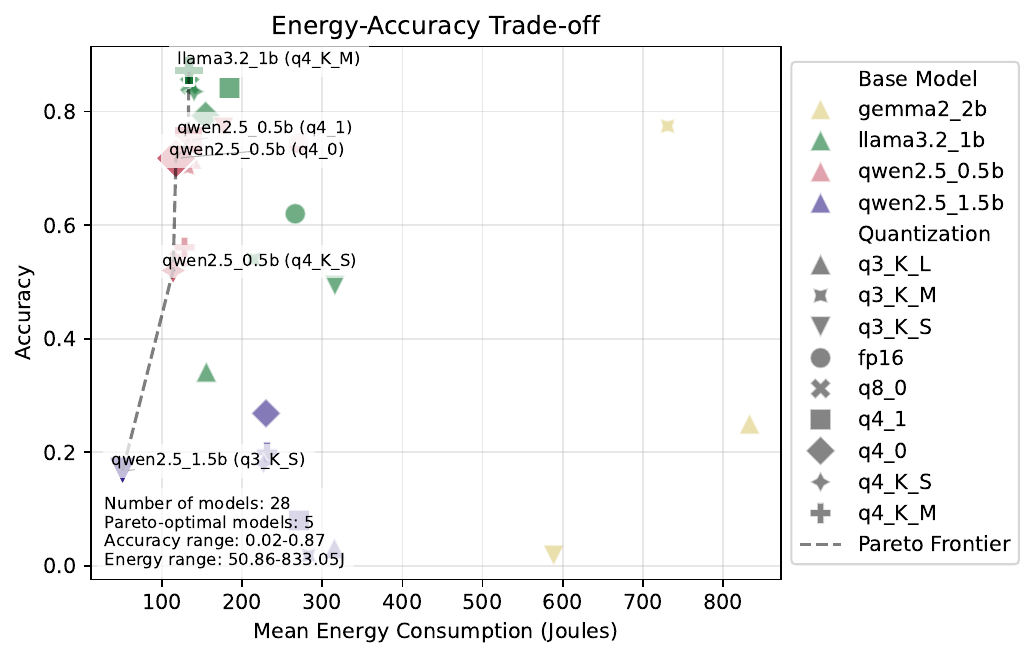}
            \vspace*{-2.10em}

        \caption{Energy-accuracy tradeoff for HumanEval. The Pareto frontier highlights five optimal configurations, with llama3.2\_1b\_q4\_K\_M achieving the highest accuracy (0.87) for code generation while maintaining a moderate energy cost.}
            \vspace*{-1.60em}

        \label{fig:energy_humaneval}
    \end{minipage}
    \hfill
    \begin{minipage}{0.45\textwidth}
        \centering
        \includegraphics[width=\textwidth]{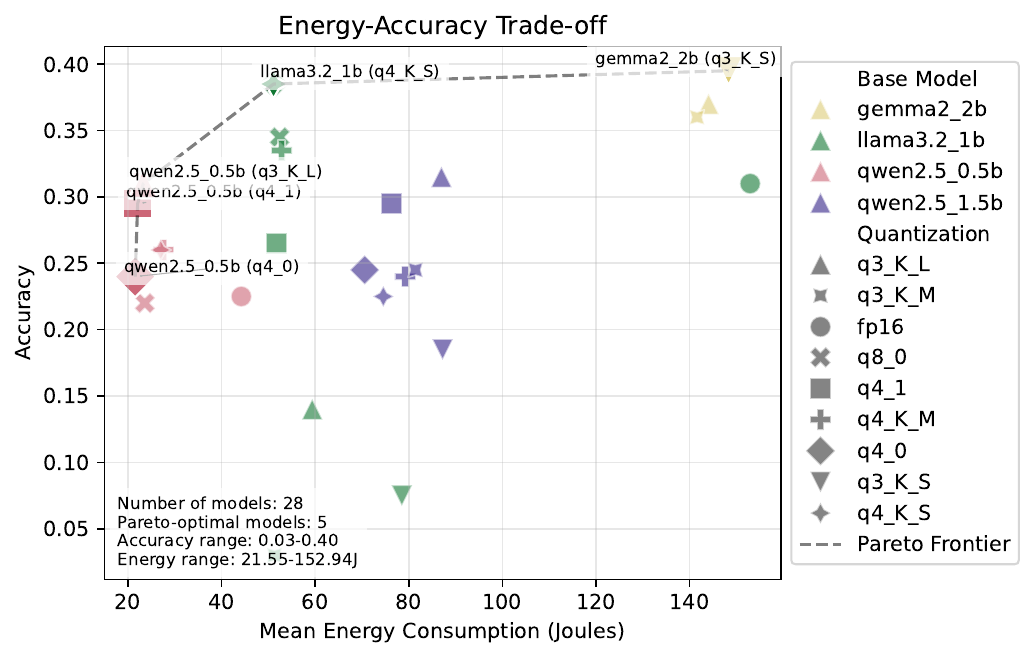}
            \vspace*{-2.10em}

        \caption{Energy-accuracy tradeoff for TruthfulQA. The Pareto frontier highlights optimal configurations for factual reasoning, with qwen2.5\_1.5b\_q3\_K\_S achieving 0.18 accuracy at the lowest energy cost.}
            \vspace*{-1.60em}

        \label{fig:energy_truthfulqa}
    \end{minipage}
\end{figure}

Comparing the average Pareto frontier (Figure~\ref{fig:energy_accuracy_all}) with task-specific frontiers (Figures \ref{fig:energy_accuracy_all}–\ref{fig:energy_truthfulqa}) reveals key differences. Models optimal on average may not be best for specific tasks, and vice versa. For example, qwen2.5\_1.5b, absent from the average frontier, dominates in BigBenchHard and GSM8K. This underscores the need for task-specific model selection rather than relying solely on average performance.


In BigBenchHard (Figure~\ref{fig:energy_bigbenchhard}), qwen2.5\_1.5b and gemma2\_2b achieve the highest accuracy (0.33–0.36), but qwen2.5\_1.5b is more energy-efficient. At the lower end of the Pareto frontier, qwen2.5\_0.5b\_q8\_0 offers 0.29 accuracy with minimal energy use (28 Joules/response), making it a strong low-power option.


In CommonsenseQA (Figure~\ref{fig:energy_commonsenseqa}), larger models (gemma2\_2b, qwen2.5\_1.5b) dominate in accuracy (0.57–0.70), emphasizing the role of model scale in commonsense reasoning. However, smaller models, llama3.2\_1b\_q4\_K, achieve lower accuracy (0.43) but offer greater energy efficiency, appearing on the Pareto frontier.

GSM8K (Figure~\ref{fig:energy_gsm8k}) highlights the challenges of mathematical reasoning, with low accuracy across all models. Qwen2.5\_0.5b\_q4\_K\_S offers the best efficiency-accuracy balance (0.08 accuracy, 26 Joules/response), while qwen2.5\_1.5b\_q4\_K\_S achieves higher accuracy (0.13) but at 2× the energy cost (63 Joules/response) for accuracy-critical deployments.


HumanEval (Figure~\ref{fig:energy_humaneval}) presents sharp energy-accuracy tradeoffs, with model llama3.2\_1b\_q4\_K\_M achieving high accuracy (0.87) in code generation at moderate energy use (134 Joules/response). Notably, qwen2.5\_0.5b variants with aggressive quantization also perform well, reaching 0.70–0.77 accuracy with slightly lower energy consumption (114–133 Joules/response).

TruthfulQA (Figure~\ref{fig:energy_truthfulqa}) features a distinct Pareto frontier with five optimal models. Gemma2\_2b\_q3\_K\_S (0.40 accuracy, 148 Joules/response) and model llama3.2\_1b\_q4\_K\_S (0.39 accuracy, 51 Joules/response) achieve high accuracy but differ in energy use. Meanwhile, three qwen2.5\_0.5b variants (0.24–0.31 accuracy, 21–23 Joules/response) offer 60\% lower energy consumption than llama3.2\_1b, trading some accuracy for efficiency. Notably, qwen2.5\_1.5b variants are absent, suggesting that larger models don’t always excel in factual reasoning under resource constraints.


Comparing task-specific Pareto frontiers reveals distinct energy-accuracy tradeoff patterns. HumanEval and CommonsenseQA show steep accuracy gains at high energy costs, while GSM8K exhibits diminishing returns, as more powerful models offer little improvement in this task. BigBenchHard and TruthfulQA follow a moderate tradeoff, where small energy increases yield notable accuracy gains. These task-dependent patterns are crucial for deployment decisions based on performance constraints.

\subsubsection{Model Size and Accuracy Scaling}

\begin{figure}
    \centering
    \includegraphics[width=1.0\linewidth]{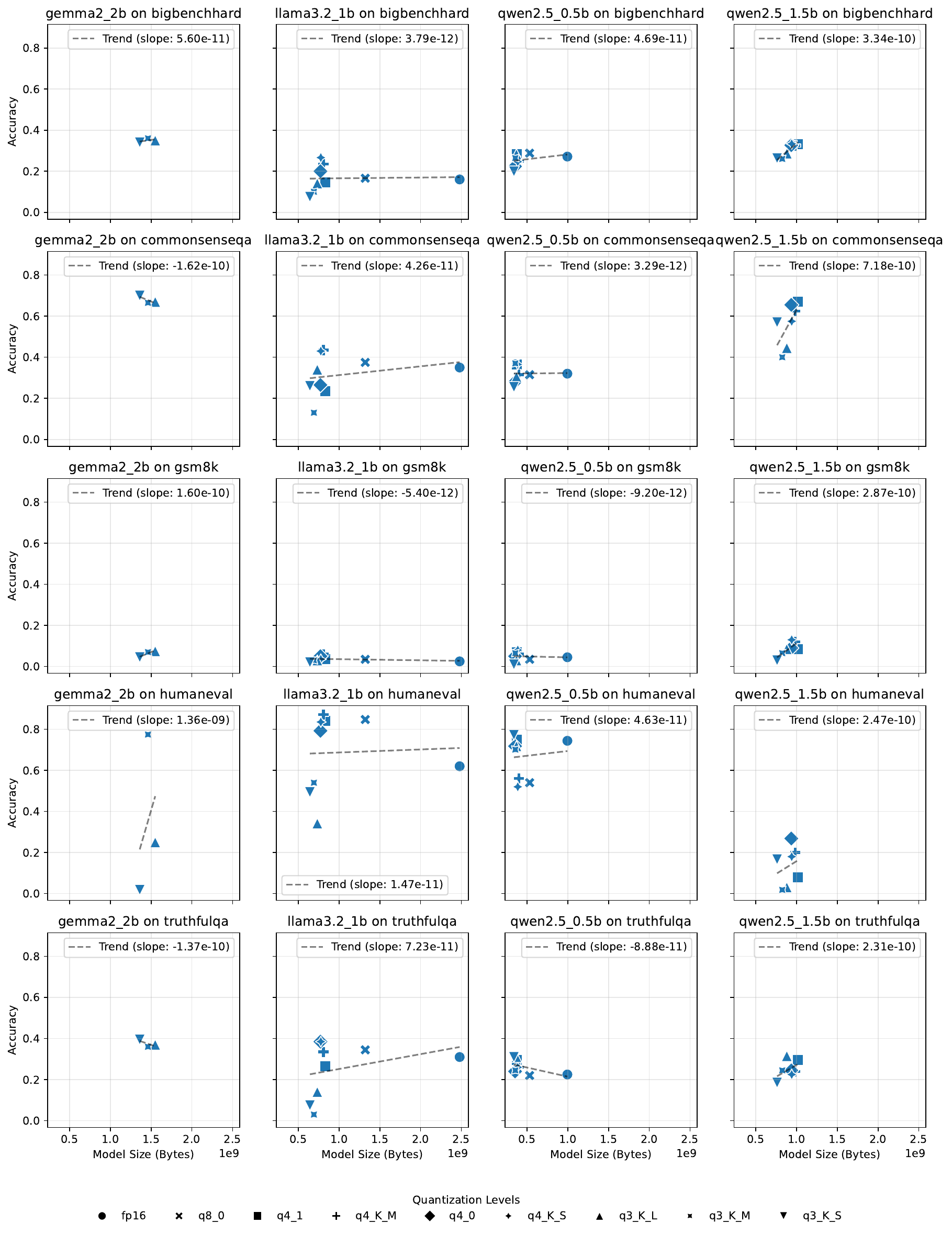}
    \caption{Relationship between model size (in bytes) and accuracy across different base models and datasets. Trend lines show how accuracy scales with increasing or decreasing model size resulting from different quantization techniques. Positive slopes indicate that higher precision (larger model sizes) generally correlates with better accuracy.}
    \label{fig:model_size_accuracy}
\end{figure}

Figure~\ref{fig:model_size_accuracy} explores the scaling of accuracy with model size across quantization techniques, revealing complex, task-dependent patterns.

%

\begin{itemize}

\item For qwen2.5\_1.5b, accuracy increases with model size across most tasks (slopes: 0.21 to 0.61), suggesting higher-precision variants better retain model capabilities.
\item Gemma2\_2b exhibits mixed scaling patterns. In CommonsenseQA (-0.24) and TruthfulQA (-0.20), aggressive quantization can maintain or enhance accuracy. However, HumanEval (2.08) shows a strong positive correlation, benefiting significantly from higher precision.

\item Llama3.2\_1b shows modest positive scaling (0.01 to 0.13) across tasks, except for GSM8K (-0.005), where quantization has little effect on its low performance.


\item Qwen2.5\_0.5b exhibits task-dependent scaling, with positive slopes in BigBenchHard (0.03), negative in TruthfulQA (-0.06), and near-zero in HumanEval (0.001), indicating the minimal impact of quantization across tasks.

\end{itemize}

These scaling patterns highlight the complex interplay between model architecture, quantization technique, and task-specific requirements when optimizing for both accuracy and energy efficiency.

\begin{tcolorbox}[boxsep=2pt,left=2pt,right=2pt,top=2pt,bottom=2pt]
\paragraph{\textbf{RQ2 Conclusion.}} 
Our analysis reveals that quantization impacts accuracy differently across models and tasks. While higher-precision variants generally retain better accuracy, some aggressive quantization techniques perform comparably, particularly in CommonsenseQA and TruthfulQA. Gemma2\_2b benefits from higher precision in code generation, whereas qwen2.5\_0.5b shows minimal sensitivity to quantization across tasks. The Pareto frontier analysis highlights that optimal energy-accuracy tradeoffs are highly task-dependent, emphasizing the need for task-specific model selection rather than relying solely on average performance metrics.

\end{tcolorbox}

\subsection{How do different quantization techniques impact inference speed on an edge device? (RQ3)}
\label{subsec:RQ3}


\begin{wrapfigure}{h}{0.430\textwidth}
\vspace*{-1.80em}
    \centering
    \includegraphics[width=0.430\textwidth]{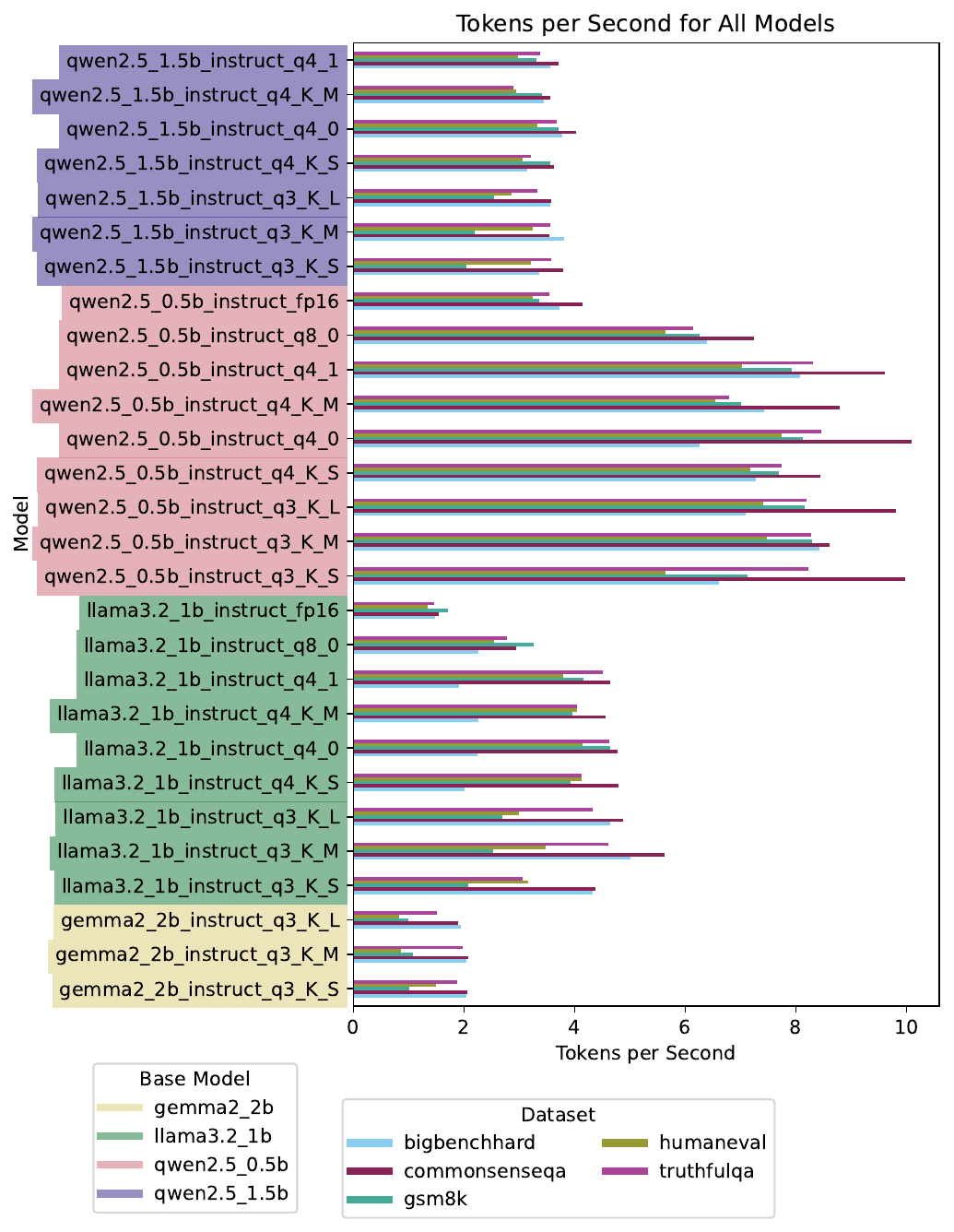}
    \vspace*{-1.90em}
    \caption{Token generation throughput (tokens per second) for all models across the datasets.} 
    \label{fig:tokens_per_second}
        \vspace*{-1.90em}

\end{wrapfigure}

\begin{figure}
    \centering
    \includegraphics[width=0.8\linewidth]{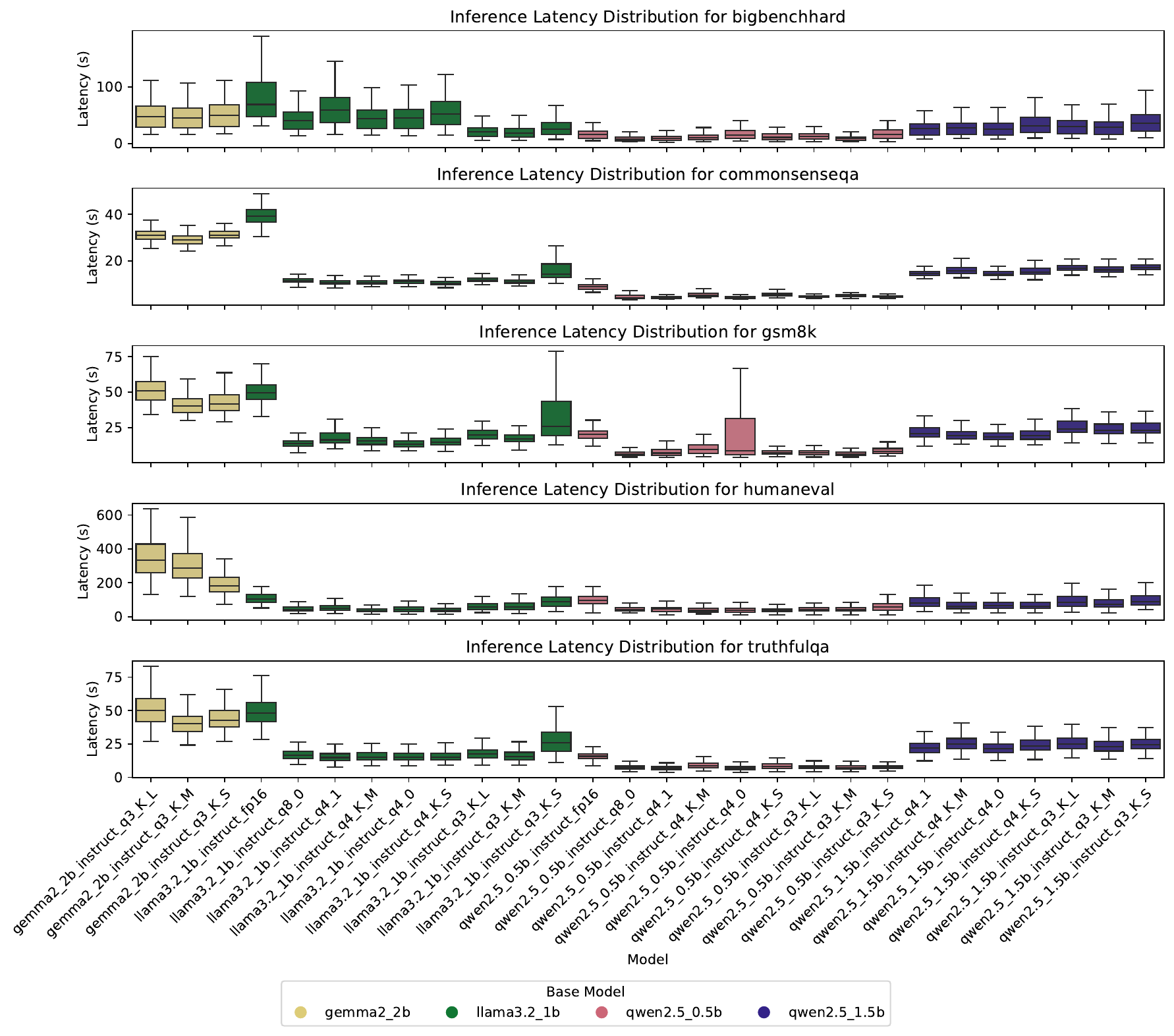}
    \vspace*{-1.20em}
    \caption{Inference latency distribution (in seconds) for all models across the five benchmark datasets. Lower values indicate faster response times. Note the significant differences in scale across benchmarks, particularly for HumanEval which requires longer responses.}
    \label{fig:inference_latency}
    \vspace*{-1.20em}
\end{figure}

To address RQ3, we analyzed latency (time from prompt request to response completion) and token generation throughput (tokens generated per second).

\subsubsection{Token Generation Throughput}

Figure~\ref{fig:tokens_per_second} displays token generation throughput across all models and datasets, showing significant variation by model family and quantization technique. Higher values indicate faster generation. Model variants are grouped by base model and the color of the bars represent different dataset. Qwen2.5\_0.5b consistently achieves the highest throughput (3.2–10.1 tokens/sec), with quantized variants outperforming fp16 (3.2–4.1 tokens/sec) by 2–3×, particularly in q4 and q3 formats.


Llama3.2\_1b exhibits variable throughput, heavily influenced by the quantization technique. Q4 variants (1.9–4.8 tokens/sec) generally outperform FP16 (1.4–1.7 tokens/sec) and q8\_0 (2.3–3.3 tokens/sec), suggesting that aggressive quantization can enhance performance. 
Larger models (gemma2\_2b, qwen2.5\_1.5b) exhibit lower throughput, with gemma2\_2b limited to 0.8–2.1 tokens/sec and qwen2.5\_1.5b performing slightly better at 2.0–4.0 tokens/sec. Unlike smaller models, quantization has less impact on throughput in these larger architectures.


Throughput varies by task, with CommonsenseQA achieving the highest rates due to its short, simple responses, while GSM8K (math reasoning) and HumanEval (code) show lower throughput.

\subsubsection{Inference Latency}

\begin{wraptable}[]{t}{0.700\textwidth}
\vspace*{-1.20em}
\scriptsize
\caption{Average inference latency (seconds) by base model and quantization level. The N/A values are ignored when computing averages.}
\label{tab:avg_latency}
\vspace*{-1.20em}

\begin{tabular}{lllll|r}
\toprule
 & gemma2\_2b & llama3.2\_1b & qwen2.5\_0.5b & qwen2.5\_1.5b & Average \\
\hline
fp16 & N/A & 59.66 & 31.40 & N/A & 45.53 \\
q8\_0 & N/A & 26.32 & 9.84 & N/A & 18.08 \\
q4 variants (avg) & N/A & 27.97 & 14.09 & 29.42 & 23.83 \\
q3 variants (avg) & 86.43 & 31.39 & 16.19 & 38.28 & 43.07 \\
\hline
Average & 86.43 & 36.33 & 17.88 & 33.85 & \\
\bottomrule
\end{tabular}
\vspace*{-1.20em}

\end{wraptable}



Figure~\ref{fig:inference_latency} shows inference latency across models and benchmarks, varying by architecture, quantization, and task. Table~\ref{tab:avg_latency} summarizes average latency, with qwen2.5\_0.5b being the fastest (17.88s avg.), particularly q8\_0 (9.84s). Llama3.2\_1b has moderate latency (36.33s avg.), where quantized variants (26–31s) outperform fp16 (59.66s). Larger models show higher latency, with gemma2\_2b (86.43s avg.) and qwen2.5\_1.5b (33.85s avg.).


Quantization significantly reduces latency, with FP16 to quantized formats improving performance by 56\% (llama3.2\_1b) and 69\% (qwen2.5\_0.5b) when comparing against the model variants with the lowest latency. However, differences among q8, q4, and q3 are less consistent. In llama3.2\_1b, q8 (26.32s) and q4 (27.97s) perform similarly, while q3 (31.39s) is slightly slower. In qwen2.5\_0.5b, q8\_0 (9.84s) is fastest, outperforming q4 (14.09s) and q3 (16.19s). This indicates that despite reducing model size, more aggressive quantization may introduce computational overhead, limiting latency gains on resource-constrained edge devices.

\begin{wraptable}[]{t}{0.320\textwidth}
\vspace*{-1.40em}
\scriptsize
\caption{Average inference latency (seconds) by the datasets.}
\label{tab:avg_latency_dataset}
\vspace*{-1.40em}
\begin{tabular}{lr}
\toprule
Dataset & Average Latency (s) \\
\midrule
bigbenchhard & 33.19 \\
commonsenseqa & 13.96 \\
gsm8k & 22.52 \\
humaneval & 94.84 \\
truthfulqa & 21.07 \\
\midrule
Average & 37.12 \\
\bottomrule
\end{tabular}
\end{wraptable}


Number of parameters alone does not determine latency. Despite having 1.5B parameters, qwen2.5\_1.5b (33.85s avg.) outperforms llama3.2\_1b (36.33s avg.), which has only 1B parameters. This suggests that architecture and optimization play a greater role in inference speed than parameter count.


Table~\ref{tab:avg_latency_dataset} shows HumanEval has the highest latency (94.84s avg.), far exceeding other benchmarks (13.96–33.19s). This aligns with its longer responses (Table~\ref{tab:response_length}) and a strong correlation (>0.9) between response length and latency across most models (as derived from Figure \ref{fig:response_length_correlation}).

\begin{tcolorbox}[boxsep=2pt,left=2pt,right=2pt,top=2pt,bottom=2pt]
\paragraph{\textbf{RQ3 Conclusion.}} 

Our analysis highlights the impact of quantization, model architecture, and task complexity on inference latency and token throughput. Quantized models significantly reduce latency, with fp16 to quantized formats improving performance by up to 69\%. However, more aggressive quantization (e.g., q3) does not always yield additional speed gains due to computational overhead. Smaller models (qwen2.5\_0.5b) consistently achieve the highest throughput, while larger models (gemma2\_2b) exhibit the slowest performance. Task complexity also plays a major role, with HumanEval showing the highest latency (94.84s avg.) due to longer response lengths, strongly correlating with latency increases across models.

\end{tcolorbox}

\section{Threats to Validity}
\label{sec:validity}

\subsection{Internal Validity} 
Internal validity concerns factors that could affect \textbf{the accuracy of our measurements} and results. One primary threat is \textbf{energy measurement variability} due to hardware or software fluctuations. While we use Joulescope for precise power monitoring, external factors such as background processes on the Raspberry Pi (RPi), thermal throttling, and power fluctuations could introduce noise into the energy readings. Additionally, model inference times may vary due to system load and memory management, potentially influencing the energy efficiency results. To mitigate these issues, we ensured \textbf{consistent experimental conditions}, active and passive cooling of the Raspberry Pi, synchronized system clocks, and repeated measurements for verification.

Another potential internal validity threat is \textbf{inference timeout handling}. The inference timeout parameter prevents excessive runtime, but an improperly set threshold could \textbf{prematurely terminate} models that require more time for complex tasks. While we used a \textbf{default timeout of 20 minutes}, tuning this parameter for different models may affect the results. We mitigated this by \textbf{manually verifying model behavior} and adjusting timeouts when necessary.

\subsection{External Validity} 
External validity relates to how well our findings \textbf{generalize to other settings, hardware platforms, and real-world applications}. Our experiments were conducted on a single edge device (Raspberry Pi 4 with 4GB RAM), which may not be representative of other edge AI hardware such as Jetson Nano, Coral TPU, or mobile AI accelerators. Differences in hardware architectures, power management strategies, and optimization techniques could lead to varying energy consumption trends. While our results provide insights into LLM quantization trade-offs on constrained devices, further experiments on diverse edge platforms are necessary for broader generalization.

Additionally, our evaluation focuses on a specific set of benchmark datasets (CommonsenseQA, BBH, TruthfulQA, GSM8K, and HumanEval). While these datasets cover diverse reasoning, mathematical, and programming tasks, they may not fully represent the entire spectrum of real-world LLM use cases. Performance and energy consumption could vary in interactive applications, dialogue-based systems, and multilingual environments. Future work should explore a wider range of tasks and real-world deployment scenarios.

\subsection{Construct Validity} 
Construct validity refers to the degree to which our chosen \textbf{metrics and evaluation methods} accurately measure what we intend to study. Our primary focus is energy efficiency, measured via total energy consumption during inference. While this provides a direct measure of power usage, other efficiency-related factors, such as latency, throughput, and cost per inference, are not explicitly analyzed. In real-world deployment, energy efficiency should be evaluated alongside latency constraints and task performance requirements.

Furthermore, we evaluate quantized models in terms of accuracy and energy consumption, but we do not explicitly measure \textbf{quantization-induced accuracy degradation} beyond standard benchmark results. Some quantization techniques may introduce specific types of errors, such as precision loss in arithmetic tasks or sensitivity to specific linguistic structures. Expanding the evaluation to quantization-aware robustness assessments could improve construct validity.

\subsection{Future Considerations}

We ensure that our findings provide valuable insights into energy-efficient LLM inference on edge devices, while acknowledging the need for further validation across different environments. Future work should (i) extend the evaluation to multiple edge devices (e.g., Jetson Nano, Edge TPUs) to assess hardware-dependent energy efficiency trends, (ii) analyze a broader range of real-world AI tasks, including conversational AI and multilingual inference, 
and (iii) incorporate additional efficiency metrics, such as inference latency, throughput, memory usage, and task-specific accuracy degradation due to quantization.

\section{Related Work}
\label{sec:related}

\subsection{Energy Efficiency in Large Language Models}

\textbf{Energy Consumption in LLM Inference.} 
Recent studies have analyzed the impact of model architecture, token processing speed, and parallelization on LLM energy efficiency. Argerich and Patiño-Martínez~\cite{argerich2024measuring} identified model size, layer count, and quantization levels as key factors influencing power consumption, showing that quantized LLMs significantly reduce energy usage. 
Doo et al.~\cite{doo2024optimal} examined accuracy-energy trade-offs in medical AI, finding that smaller fine-tuned models (e.g., Vicuna 1.5, 7B/13B) were both more energy-efficient and accurate than larger models like LLaMA 2 (70B), which consumed over seven times more power than its 7B counterpart. These findings underscore the importance of choosing the right model architecture to optimize both predictive performance and energy efficiency. Husom et al.~\cite{husom2024price} examined LLM energy consumption during inference, finding that response length and duration strongly correlate with power usage, while prompt complexity 
has minimal impact. Their results suggest that controlling response generation is more effective for reducing energy consumption than simplifying prompts.

\textbf{Reducing the Carbon Footprint of LLMs.} Various strategies have been proposed to reduce the carbon footprint of LLM training and deployment. Shi et al.~\cite{shi2024greening} introduced Avatar, an optimization framework reducing model size, inference latency, and energy consumption while maintaining accuracy, achieving up to 184× lower energy usage and 157× lower carbon emissions through multi-objective tuning. Bai et al.~\cite{bai2024beyond} categorized energy-efficient LLM techniques, highlighting architecture design, training efficiency, and system-level optimizations. Their study underscores the role of energy-aware fine-tuning and adaptive precision scaling in minimizing carbon footprint.

\textbf{Energy-Efficient LLM Deployment on Edge.} Deploying LLMs on edge devices is challenging due to limited battery life and computational resources, requiring energy-efficient solutions. Yuan et al.~\cite{yuan2024generative} proposed intelligent offloading strategies, dynamically balancing computations between mobile devices and edge servers, minimizing energy consumption.
Tian et al.~\cite{tian2024greenllm} introduced GreenLLM, a framework optimizing memory, power, and computational efficiency through pruning, achieving a 34.1\% energy and 33.5\% latency reduction within acceptable performance loss.

While prior studies have analyzed \textbf{LLM energy consumption, carbon footprint, and edge deployment in a limited scope or even without any edge focus} (e.g., estimating power usage through software-based profiling, evaluating only select quantization methods, or focusing on theoretical energy models rather than real-world edge deployments), our work extends this research by \textbf{conducting real-world energy measurements using hardware-based profiling, benchmarking 28 quantized LLMs from the Ollama framework (applying by default PTQ and weight-only quantization) on an edge device, and analyzing energy-performance trade-offs across quantization levels and task types}. By integrating hardware-level energy profiling with LLM benchmarking, we provide a comprehensive evaluation of energy-efficient deployment strategies.

\subsection{Quantized Language Models on Edge Devices}

Several studies~\cite{shen2024agile, shen2024edgeqat, mao2024compressibility, rahman2023quantized, tan2024mobilequant, hasan2024optimizing, ccoplu2023performance} have investigated quantized LLMs in edge environments to address limited computational resources, memory constraints, and energy efficiency requirements

\textbf{Weight and Activation Quantization for LLMs on Edge.} Shen et al.~\cite{shen2024agile} introduced AgileQuant, an activation-guided quantization framework that simultaneously quantizes weights and activations, achieving up to 2.55× speedup on edge devices over FP16 models. Unlike weight-only quantization, it enhances inference efficiency by optimizing activations. Similarly, Tan et al.~\cite{tan2024mobilequant} proposed MobileQuant, a post-training quantization method for on-device language models. The approach jointly optimizes weight transformation and activation range parameters, reducing latency and energy consumption by 20\%-50\% compared to other on-device quantization strategies. 
While weight and activation quantization improves inference efficiency, latency, and energy consumption, it comes with trade-offs. Weight-only quantization retains higher computational efficiency and accuracy, as activations remain in full or higher precision. In contrast, weight and activation quantization provides greater memory savings, making it more suitable where both model and activations must fit within strict memory limits.

\textbf{Optimizing Memory and Latency for Quantized LLMs.} Rahman et al.~\cite{rahman2023quantized} evaluated quantized transformer models, MobileBERT~\cite{sun2020mobilebert}, on edge devices, showing that the converted and quantized MobileBERT models have 160× smaller footprints for a 4.1\% drop in accuracy. 
Shen et al.~\cite{shen2024edgeqat} proposed EdgeQAT, a quantization-aware training framework  for the optimization of lightweight LLMs to achieve inference acceleration on Edge devices. Using entropy and distribution guided techniques, it achieves an on-device speedup of 2.37× compared with its FP16 counterparts. 

Existing studies focus on reducing memory footprint, improving inference speed, and preserving accuracy in quantized LLMs but \textbf{often overlook energy consumption trade-offs across different quantization techniques}. Different than these studies, our work conducts \textbf{real-world energy measurements using hardware-based profiling, benchmarking quantized LLMs from the Ollama framework on an edge device, and analyzing energy-performance-accuracy trade-offs across quantization levels and task types.} 

\subsection{Existing Benchmarks for Evaluating LLM Performance and Efficiency on Edge}

Several benchmarking frameworks have been developed to assess the performance, efficiency, and resource footprint of LLMs deployed on edge devices. Traditional LLM benchmarks often focus on accuracy and latency, but emerging frameworks are now incorporating memory consumption and energy efficiency as critical evaluation metrics. \textbf{PalmBench}~\cite{li2024palmbench} is a comprehensive benchmark that evaluates quantized LLMs on mobile platforms, analyzing memory usage, GPU execution time, and power consumption. \textbf{The Edge-Device Large Language Model Competition (NeurIPS 2024)}~\cite{liu2024edge} provides a platform for benchmarking LLM inference efficiency on edge devices, with a focus on model size, throughput, and latency. \textbf{MLPerf Tiny}~\cite{banbury2021mlperf}, \textbf{MLPerf Power}~\cite{tschand2024mlperf}, \textbf{MLPerf Inference}~\cite{reddi2020mlperf}, and \textbf{Ai benchmark}~\cite{ignatov2019ai} offer standardized AI benchmarking tools that include power-aware evaluations, making them valuable for assessing energy-efficient AI models.

Task-specific datasets like CommonsenseQA and HumanEval
serve as standardized benchmarks for evaluating LLM reasoning, mathematical problem-solving, and programming under computational constraints. Our work builds upon these datasets by introducing energy consumption as a key performance metric. While previous evaluations have focused primarily on memory and latency, our study provides a comprehensive energy-efficiency and LLM output accuracy analysis of quantized LLMs deployed on edge, leveraging high-resolution hardware-based energy monitoring.

\section{Conclusion}
\label{sec:conclusion}

Our study provides a comprehensive evaluation of quantized LLMs on edge devices, analyzing the trade-offs between energy efficiency, accuracy, and inference speed. We find that quantization significantly reduces energy consumption and latency, but the impact on accuracy varies across tasks and model families. While q3 and q4 quantization achieve substantial energy savings, extreme compression can degrade performance in complex reasoning tasks. Our findings emphasize the importance of selecting the right quantization strategy based on task requirements and deployment constraints. Future work should explore adaptive quantization techniques to further optimize efficiency and performance for real-world edge AI applications.

\begin{acks}
\begin{wrapfigure}{h}{0.080\textwidth}
  \vspace*{-1.5em}
    \centerline{\includegraphics[width=0.100\textwidth]{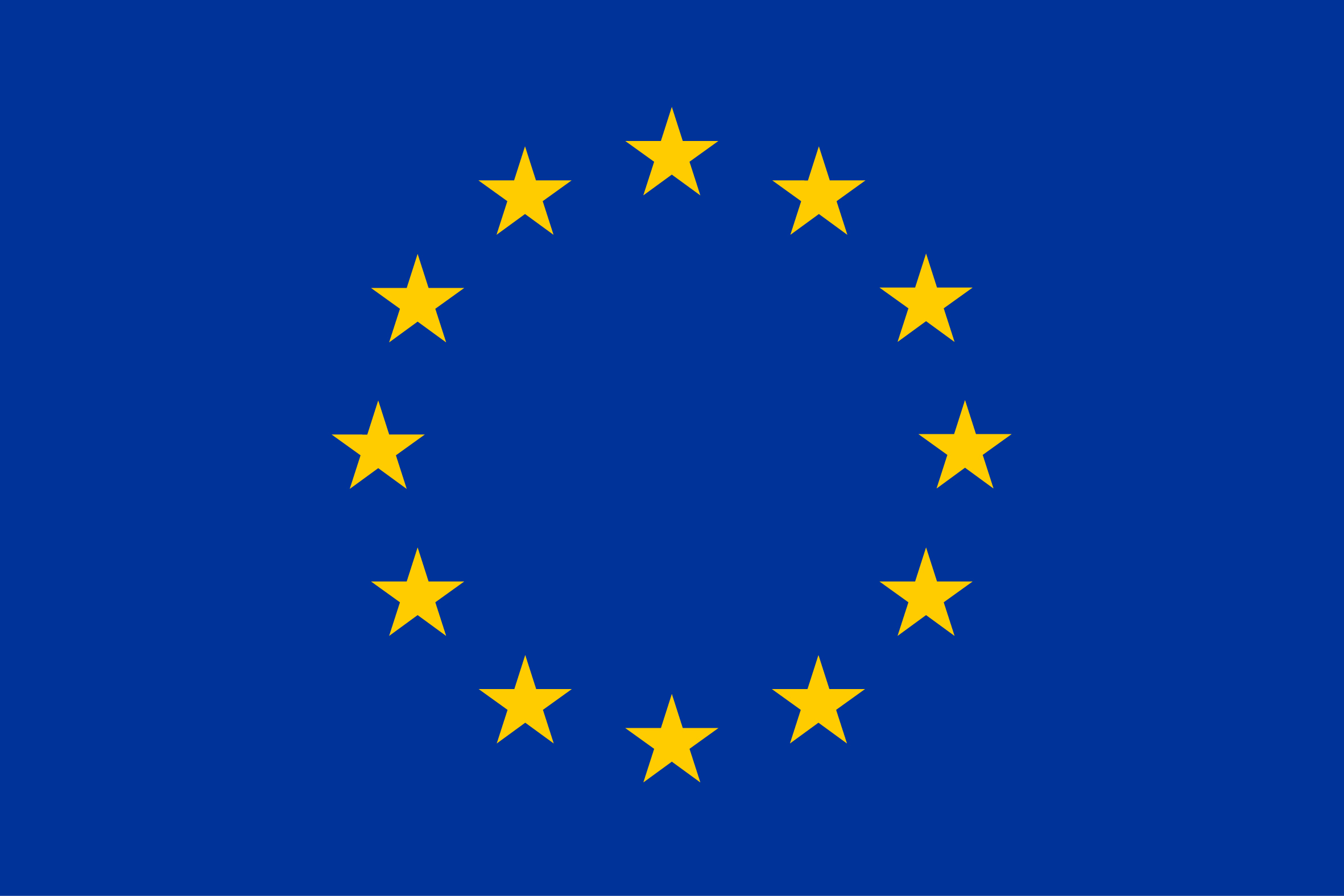}}
    \vspace*{-0.4em}
    \vspace*{-1.1em}
\end{wrapfigure}

The work has been conducted as part of the ENFIELD project (101120657) funded by the European Commission within the HEU Programme, and the National Research Foundation of Korea (NRF) grant RS-2023-00268071.

\end{acks}

\bibliography{references}


\begin{thebibliography}{84}


\ifx \showCODEN    \undefined \def \showCODEN     #1{\unskip}     \fi
\ifx \showDOI      \undefined \def \showDOI       #1{#1}\fi
\ifx \showISBNx    \undefined \def \showISBNx     #1{\unskip}     \fi
\ifx \showISBNxiii \undefined \def \showISBNxiii  #1{\unskip}     \fi
\ifx \showISSN     \undefined \def \showISSN      #1{\unskip}     \fi
\ifx \showLCCN     \undefined \def \showLCCN      #1{\unskip}     \fi
\ifx \shownote     \undefined \def \shownote      #1{#1}          \fi
\ifx \showarticletitle \undefined \def \showarticletitle #1{#1}   \fi
\ifx \showURL      \undefined \def \showURL       {\relax}        \fi
\providecommand\bibfield[2]{#2}
\providecommand\bibinfo[2]{#2}
\providecommand\natexlab[1]{#1}
\providecommand\showeprint[2][]{arXiv:#2}

\bibitem[Argerich and Pati{\~n}o-Mart{\'\i}nez(2024)]%
        {argerich2024measuring}
\bibfield{author}{\bibinfo{person}{Mauricio~Fadel Argerich} {and} \bibinfo{person}{Marta Pati{\~n}o-Mart{\'\i}nez}.} \bibinfo{year}{2024}\natexlab{}.
\newblock \showarticletitle{Measuring and Improving the Energy Efficiency of Large Language Models Inference}.
\newblock \bibinfo{journal}{\emph{IEEE Access}} (\bibinfo{year}{2024}).
\newblock


\bibitem[Astekin et~al\mbox{.}(2024)]%
        {astekin2024comparative}
\bibfield{author}{\bibinfo{person}{Merve Astekin}, \bibinfo{person}{Max Hort}, {and} \bibinfo{person}{Leon Moonen}.} \bibinfo{year}{2024}\natexlab{}.
\newblock \showarticletitle{A Comparative Study on Large Language Models for Log Parsing}. In \bibinfo{booktitle}{\emph{ESEM'24}}. \bibinfo{pages}{234–244}.
\newblock


\bibitem[Bai et~al\mbox{.}(2024)]%
        {bai2024beyond}
\bibfield{author}{\bibinfo{person}{Guangji Bai}, \bibinfo{person}{Zheng Chai}, \bibinfo{person}{Chen Ling}, \bibinfo{person}{Shiyu Wang}, \bibinfo{person}{Jiaying Lu}, \bibinfo{person}{Nan Zhang}, \bibinfo{person}{Tingwei Shi}, \bibinfo{person}{Ziyang Yu}, \bibinfo{person}{Mengdan Zhu}, \bibinfo{person}{Yifei Zhang}, {et~al\mbox{.}}} \bibinfo{year}{2024}\natexlab{}.
\newblock \showarticletitle{Beyond efficiency: A systematic survey of resource-efficient large language models}.
\newblock \bibinfo{journal}{\emph{arXiv preprint arXiv:2401.00625}} (\bibinfo{year}{2024}).
\newblock


\bibitem[Banbury et~al\mbox{.}(2021)]%
        {banbury2021mlperf}
\bibfield{author}{\bibinfo{person}{Colby Banbury}, \bibinfo{person}{Vijay~Janapa Reddi}, \bibinfo{person}{Peter Torelli}, \bibinfo{person}{Jeremy Holleman}, \bibinfo{person}{Nat Jeffries}, \bibinfo{person}{Csaba Kiraly}, \bibinfo{person}{Pietro Montino}, \bibinfo{person}{David Kanter}, \bibinfo{person}{Sebastian Ahmed}, \bibinfo{person}{Danilo Pau}, {et~al\mbox{.}}} \bibinfo{year}{2021}\natexlab{}.
\newblock \showarticletitle{Mlperf tiny benchmark}.
\newblock \bibinfo{journal}{\emph{arXiv preprint arXiv:2106.07597}} (\bibinfo{year}{2021}).
\newblock


\bibitem[{BIG-Bench-Hard}(2024)]%
        {bigbenchhard}
\bibfield{author}{\bibinfo{person}{{BIG-Bench-Hard}}.} \bibinfo{year}{Visited in 2024}\natexlab{}.
\newblock \bibinfo{howpublished}{https://github.com/suzgunmirac/BIG-Bench-Hard}.
\newblock


\bibitem[Cai et~al\mbox{.}(2024)]%
        {cai2024edge}
\bibfield{author}{\bibinfo{person}{Fenglong Cai}, \bibinfo{person}{Dong Yuan}, \bibinfo{person}{Zhe Yang}, {and} \bibinfo{person}{Lizhen Cui}.} \bibinfo{year}{2024}\natexlab{}.
\newblock \showarticletitle{Edge-LLM: A Collaborative Framework for Large Language Model Serving in Edge Computing}. In \bibinfo{booktitle}{\emph{ICWS'24}}. \bibinfo{pages}{799--809}.
\newblock


\bibitem[Chai et~al\mbox{.}(2025)]%
        {chai2025flexquant}
\bibfield{author}{\bibinfo{person}{Yuji Chai}, \bibinfo{person}{Mujin Kwen}, \bibinfo{person}{David Brooks}, {and} \bibinfo{person}{Gu-Yeon Wei}.} \bibinfo{year}{2025}\natexlab{}.
\newblock \showarticletitle{FlexQuant: Elastic Quantization Framework for Locally Hosted LLM on Edge Devices}.
\newblock \bibinfo{journal}{\emph{arXiv preprint arXiv:2501.07139}} (\bibinfo{year}{2025}).
\newblock


\bibitem[Chen et~al\mbox{.}(2021)]%
        {chen2021codex}
\bibfield{author}{\bibinfo{person}{Mark Chen}, \bibinfo{person}{Jerry Tworek}, {and} \bibinfo{person}{etc}.} \bibinfo{year}{2021}\natexlab{}.
\newblock \showarticletitle{Evaluating Large Language Models Trained on Code}.
\newblock  (\bibinfo{year}{2021}).
\newblock
\showeprint[arxiv]{2107.03374}~[cs.LG]


\bibitem[{CommonsenseQA}(2024)]%
        {commonsenseqa}
\bibfield{author}{\bibinfo{person}{{CommonsenseQA}}.} \bibinfo{year}{Visited in 2024}\natexlab{}.
\newblock \bibinfo{howpublished}{https://www.tau-nlp.sites.tau.ac.il/commonsenseqa}.
\newblock


\bibitem[{\c{C}}{\"o}pl{\"u} et~al\mbox{.}(2023)]%
        {ccoplu2023performance}
\bibfield{author}{\bibinfo{person}{Tolga {\c{C}}{\"o}pl{\"u}}, \bibinfo{person}{Marc Loedi}, \bibinfo{person}{Arto Bendiken}, \bibinfo{person}{Mykhailo Makohin}, \bibinfo{person}{Joshua~J Bouw}, {and} \bibinfo{person}{Stephen Cobb}.} \bibinfo{year}{2023}\natexlab{}.
\newblock \showarticletitle{A performance evaluation of a quantized large language model on various smartphones}.
\newblock \bibinfo{journal}{\emph{arXiv preprint arXiv:2312.12472}} (\bibinfo{year}{2023}).
\newblock


\bibitem[Corporation(2023)]%
        {nvidia_smi_docs}
\bibfield{author}{\bibinfo{person}{NVIDIA Corporation}.} \bibinfo{year}{2023}\natexlab{}.
\newblock \bibinfo{title}{NVIDIA System Management Interface}.
\newblock
\newblock
\urldef\tempurl%
\url{https://docs.nvidia.com/deploy/nvidia-smi/index.html}
\showURL{%
\tempurl}
\newblock
\shownote{2024-05-03}.


\bibitem[Deng et~al\mbox{.}(2023)]%
        {deng2023benchmark}
\bibfield{author}{\bibinfo{person}{Chunyuan Deng}, \bibinfo{person}{Yilun Zhao}, \bibinfo{person}{Xiangru Tang}, \bibinfo{person}{Mark Gerstein}, {and} \bibinfo{person}{Arman Cohan}.} \bibinfo{year}{2023}\natexlab{}.
\newblock \showarticletitle{Benchmark probing: Investigating data leakage in large language models}. In \bibinfo{booktitle}{\emph{NeurIPS 2023 Workshop on Backdoors in Deep Learning-The Good, the Bad, and the Ugly}}.
\newblock


\bibitem[Dettmers et~al\mbox{.}(2022)]%
        {dettmers2022gpt3}
\bibfield{author}{\bibinfo{person}{Tim Dettmers}, \bibinfo{person}{Mike Lewis}, \bibinfo{person}{Younes Belkada}, {and} \bibinfo{person}{Luke Zettlemoyer}.} \bibinfo{year}{2022}\natexlab{}.
\newblock \showarticletitle{Gpt3. int8 (): 8-bit matrix multiplication for transformers at scale}.
\newblock \bibinfo{journal}{\emph{Advances in Neural Information Processing Systems}}  \bibinfo{volume}{35} (\bibinfo{year}{2022}), \bibinfo{pages}{30318--30332}.
\newblock


\bibitem[Devlin et~al\mbox{.}(2018)]%
        {devlin2018bert}
\bibfield{author}{\bibinfo{person}{Jacob Devlin}, \bibinfo{person}{Ming-Wei Chang}, \bibinfo{person}{Kenton Lee}, {and} \bibinfo{person}{Kristina Toutanova}.} \bibinfo{year}{2018}\natexlab{}.
\newblock \showarticletitle{Bert: Pre-training of deep bidirectional transformers for language understanding}.
\newblock \bibinfo{journal}{\emph{arXiv preprint arXiv:1810.04805}} (\bibinfo{year}{2018}).
\newblock


\bibitem[Doo et~al\mbox{.}(2024)]%
        {doo2024optimal}
\bibfield{author}{\bibinfo{person}{Florence~X Doo}, \bibinfo{person}{Dharmam Savani}, \bibinfo{person}{Adway Kanhere}, \bibinfo{person}{Ruth~C Carlos}, \bibinfo{person}{Anupam Joshi}, \bibinfo{person}{Paul~H Yi}, {and} \bibinfo{person}{Vishwa~S Parekh}.} \bibinfo{year}{2024}\natexlab{}.
\newblock \showarticletitle{Optimal large language model characteristics to balance accuracy and energy use for sustainable medical applications}.
\newblock \bibinfo{journal}{\emph{Radiology}} \bibinfo{volume}{312}, \bibinfo{number}{2} (\bibinfo{year}{2024}), \bibinfo{pages}{e240320}.
\newblock


\bibitem[{Experiment Repo}(2025)]%
        {experiment-repo}
\bibfield{author}{\bibinfo{person}{{Experiment Repo}}.} \bibinfo{year}{Visited in 2025}\natexlab{}.
\newblock \bibinfo{howpublished}{https://github.com/ejhusom/neurips-edge-llm-challenge-sampled/}.
\newblock


\bibitem[Frantar et~al\mbox{.}(2022a)]%
        {frantar2022gptq}
\bibfield{author}{\bibinfo{person}{Elias Frantar}, \bibinfo{person}{Saleh Ashkboos}, \bibinfo{person}{Torsten Hoefler}, {and} \bibinfo{person}{Dan Alistarh}.} \bibinfo{year}{2022}\natexlab{a}.
\newblock \showarticletitle{Gptq: Accurate post-training quantization for generative pre-trained transformers}.
\newblock \bibinfo{journal}{\emph{arXiv preprint arXiv:2210.17323}} (\bibinfo{year}{2022}).
\newblock


\bibitem[Frantar et~al\mbox{.}(2022b)]%
        {frantar2022optq}
\bibfield{author}{\bibinfo{person}{Elias Frantar}, \bibinfo{person}{Saleh Ashkboos}, \bibinfo{person}{Torsten Hoefler}, {and} \bibinfo{person}{Dan Alistarh}.} \bibinfo{year}{2022}\natexlab{b}.
\newblock \showarticletitle{OPTQ: Accurate quantization for generative pre-trained transformers}. In \bibinfo{booktitle}{\emph{The Eleventh International Conference on Learning Representations}}.
\newblock


\bibitem[Friha et~al\mbox{.}(2024)]%
        {friha2024llm}
\bibfield{author}{\bibinfo{person}{Othmane Friha}, \bibinfo{person}{Mohamed~Amine Ferrag}, \bibinfo{person}{Burak Kantarci}, \bibinfo{person}{Burak Cakmak}, \bibinfo{person}{Arda Ozgun}, {and} \bibinfo{person}{Nassira Ghoualmi-Zine}.} \bibinfo{year}{2024}\natexlab{}.
\newblock \showarticletitle{Llm-based edge intelligence: A comprehensive survey on architectures, applications, security and trustworthiness}.
\newblock \bibinfo{journal}{\emph{IEEE Open Journal of the Communications Society}} (\bibinfo{year}{2024}).
\newblock


\bibitem[{ggml-org}(2025)]%
        {llama-doc}
\bibfield{author}{\bibinfo{person}{{ggml-org}}.} \bibinfo{year}{Visited in 2025}\natexlab{}.
\newblock \bibinfo{title}{llama.cpp}.
\newblock \bibinfo{howpublished}{https://github.com/ggml-org/llama.cpp}.
\newblock


\bibitem[Gholami et~al\mbox{.}(2022)]%
        {gholami2022survey}
\bibfield{author}{\bibinfo{person}{Amir Gholami}, \bibinfo{person}{Sehoon Kim}, \bibinfo{person}{Zhen Dong}, \bibinfo{person}{Zhewei Yao}, \bibinfo{person}{Michael~W Mahoney}, {and} \bibinfo{person}{Kurt Keutzer}.} \bibinfo{year}{2022}\natexlab{}.
\newblock \showarticletitle{A survey of quantization methods for efficient neural network inference}.
\newblock In \bibinfo{booktitle}{\emph{Low-power computer vision}}. \bibinfo{publisher}{Chapman and Hall/CRC}, \bibinfo{pages}{291--326}.
\newblock


\bibitem[{GSM8K}(2024)]%
        {gsm8k}
\bibfield{author}{\bibinfo{person}{{GSM8K}}.} \bibinfo{year}{Visited in 2024}\natexlab{}.
\newblock \bibinfo{howpublished}{https://github.com/openai/grade-school-math}.
\newblock


\bibitem[Haris et~al\mbox{.}(2024)]%
        {haris2024designing}
\bibfield{author}{\bibinfo{person}{Jude Haris}, \bibinfo{person}{Rappy Saha}, \bibinfo{person}{Wenhao Hu}, {and} \bibinfo{person}{Jos{\'e} Cano}.} \bibinfo{year}{2024}\natexlab{}.
\newblock \showarticletitle{Designing Efficient LLM Accelerators for Edge Devices}.
\newblock \bibinfo{journal}{\emph{arXiv preprint arXiv:2408.00462}} (\bibinfo{year}{2024}).
\newblock


\bibitem[Hasan(2024)]%
        {hasan2024optimizing}
\bibfield{author}{\bibinfo{person}{Jahid Hasan}.} \bibinfo{year}{2024}\natexlab{}.
\newblock \showarticletitle{Optimizing Large Language Models through Quantization: A Comparative Analysis of PTQ and QAT Techniques}.
\newblock \bibinfo{journal}{\emph{arXiv preprint arXiv:2411.06084}} (\bibinfo{year}{2024}).
\newblock


\bibitem[Hu et~al\mbox{.}(2024)]%
        {hu2024llm}
\bibfield{author}{\bibinfo{person}{Xing Hu}, \bibinfo{person}{Yuan Chen}, \bibinfo{person}{Dawei Yang}, \bibinfo{person}{Sifan Zhou}, \bibinfo{person}{Zhihang Yuan}, \bibinfo{person}{Jiangyong Yu}, {and} \bibinfo{person}{Chen Xu}.} \bibinfo{year}{2024}\natexlab{}.
\newblock \showarticletitle{I-LLM: Efficient Integer-Only Inference for Fully-Quantized Low-Bit Large Language Models}.
\newblock \bibinfo{journal}{\emph{arXiv preprint arXiv:2405.17849}} (\bibinfo{year}{2024}).
\newblock


\bibitem[Huang et~al\mbox{.}(2024a)]%
        {huang2024billm}
\bibfield{author}{\bibinfo{person}{Wei Huang}, \bibinfo{person}{Yangdong Liu}, \bibinfo{person}{Haotong Qin}, \bibinfo{person}{Ying Li}, \bibinfo{person}{Shiming Zhang}, \bibinfo{person}{Xianglong Liu}, \bibinfo{person}{Michele Magno}, {and} \bibinfo{person}{Xiaojuan Qi}.} \bibinfo{year}{2024}\natexlab{a}.
\newblock \showarticletitle{Billm: Pushing the limit of post-training quantization for llms}.
\newblock \bibinfo{journal}{\emph{arXiv preprint arXiv:2402.04291}} (\bibinfo{year}{2024}).
\newblock


\bibitem[Huang et~al\mbox{.}(2024b)]%
        {huang2024good}
\bibfield{author}{\bibinfo{person}{Wei Huang}, \bibinfo{person}{Xudong Ma}, \bibinfo{person}{Haotong Qin}, \bibinfo{person}{Xingyu Zheng}, \bibinfo{person}{Chengtao Lv}, \bibinfo{person}{Hong Chen}, \bibinfo{person}{Jie Luo}, \bibinfo{person}{Xiaojuan Qi}, \bibinfo{person}{Xianglong Liu}, {and} \bibinfo{person}{Michele Magno}.} \bibinfo{year}{2024}\natexlab{b}.
\newblock \showarticletitle{How good are low-bit quantized llama3 models? an empirical study}.
\newblock \bibinfo{journal}{\emph{arXiv e-prints}} (\bibinfo{year}{2024}), \bibinfo{pages}{arXiv--2404}.
\newblock


\bibitem[Huang et~al\mbox{.}(2024c)]%
        {huang2024empirical}
\bibfield{author}{\bibinfo{person}{Wei Huang}, \bibinfo{person}{Xingyu Zheng}, \bibinfo{person}{Xudong Ma}, \bibinfo{person}{Haotong Qin}, \bibinfo{person}{Chengtao Lv}, \bibinfo{person}{Hong Chen}, \bibinfo{person}{Jie Luo}, \bibinfo{person}{Xiaojuan Qi}, \bibinfo{person}{Xianglong Liu}, {and} \bibinfo{person}{Michele Magno}.} \bibinfo{year}{2024}\natexlab{c}.
\newblock \showarticletitle{An empirical study of llama3 quantization: From llms to mllms}.
\newblock \bibinfo{journal}{\emph{Visual Intelligence}} \bibinfo{volume}{2}, \bibinfo{number}{1} (\bibinfo{year}{2024}), \bibinfo{pages}{36}.
\newblock


\bibitem[{HumanEval}(2024)]%
        {humaneval}
\bibfield{author}{\bibinfo{person}{{HumanEval}}.} \bibinfo{year}{Visited in 2024}\natexlab{}.
\newblock \bibinfo{howpublished}{https://github.com/openai/human-eval}.
\newblock


\bibitem[Husom et~al\mbox{.}(2024)]%
        {husom2024price}
\bibfield{author}{\bibinfo{person}{Erik~Johannes Husom}, \bibinfo{person}{Arda Goknil}, \bibinfo{person}{Lwin~Khin Shar}, {and} \bibinfo{person}{Sagar Sen}.} \bibinfo{year}{2024}\natexlab{}.
\newblock \showarticletitle{The Price of Prompting: Profiling Energy Use in Large Language Models Inference}.
\newblock \bibinfo{journal}{\emph{arXiv preprint arXiv:2407.16893}} (\bibinfo{year}{2024}).
\newblock


\bibitem[Ignatov et~al\mbox{.}(2019)]%
        {ignatov2019ai}
\bibfield{author}{\bibinfo{person}{Andrey Ignatov}, \bibinfo{person}{Radu Timofte}, \bibinfo{person}{Andrei Kulik}, \bibinfo{person}{Seungsoo Yang}, \bibinfo{person}{Ke Wang}, \bibinfo{person}{Felix Baum}, \bibinfo{person}{Max Wu}, \bibinfo{person}{Lirong Xu}, {and} \bibinfo{person}{Luc Van~Gool}.} \bibinfo{year}{2019}\natexlab{}.
\newblock \showarticletitle{Ai benchmark: All about deep learning on smartphones in 2019}. In \bibinfo{booktitle}{\emph{ICCVW'19}}. \bibinfo{pages}{3617--3635}.
\newblock


\bibitem[Jaiswal et~al\mbox{.}(2023)]%
        {jaiswal2023compressing}
\bibfield{author}{\bibinfo{person}{Ajay Jaiswal}, \bibinfo{person}{Zhe Gan}, \bibinfo{person}{Xianzhi Du}, \bibinfo{person}{Bowen Zhang}, \bibinfo{person}{Zhangyang Wang}, {and} \bibinfo{person}{Yinfei Yang}.} \bibinfo{year}{2023}\natexlab{}.
\newblock \showarticletitle{Compressing llms: The truth is rarely pure and never simple}.
\newblock \bibinfo{journal}{\emph{arXiv preprint arXiv:2310.01382}} (\bibinfo{year}{2023}).
\newblock


\bibitem[Jin et~al\mbox{.}(2024)]%
        {jin2024comprehensive}
\bibfield{author}{\bibinfo{person}{Renren Jin}, \bibinfo{person}{Jiangcun Du}, \bibinfo{person}{Wuwei Huang}, \bibinfo{person}{Wei Liu}, \bibinfo{person}{Jian Luan}, \bibinfo{person}{Bin Wang}, {and} \bibinfo{person}{Deyi Xiong}.} \bibinfo{year}{2024}\natexlab{}.
\newblock \showarticletitle{A comprehensive evaluation of quantization strategies for large language models}. In \bibinfo{booktitle}{\emph{Findings of the Association for Computational Linguistics ACL 2024}}. \bibinfo{pages}{12186--12215}.
\newblock


\bibitem[{Joulemeter}(2025)]%
        {joulemeter}
\bibfield{author}{\bibinfo{person}{{Joulemeter}}.} \bibinfo{year}{Visited in 2025}\natexlab{}.
\newblock \bibinfo{howpublished}{https://www.microsoft.com/en-us/research/project/joulemeter-computational-energy-measurement-and-optimization/}.
\newblock


\bibitem[{Joulescope}(2025)]%
        {joulescope}
\bibfield{author}{\bibinfo{person}{{Joulescope}}.} \bibinfo{year}{Visited in 2025}\natexlab{}.
\newblock \bibinfo{howpublished}{https://www.joulescope.com/}.
\newblock


\bibitem[Khoshsirat et~al\mbox{.}(2024)]%
        {khoshsirat2024decentralized}
\bibfield{author}{\bibinfo{person}{Aria Khoshsirat}, \bibinfo{person}{Giovanni Perin}, {and} \bibinfo{person}{Michele Rossi}.} \bibinfo{year}{2024}\natexlab{}.
\newblock \showarticletitle{Decentralized LLM Inference over Edge Networks with Energy Harvesting}.
\newblock \bibinfo{journal}{\emph{arXiv preprint arXiv:2408.15907}} (\bibinfo{year}{2024}).
\newblock


\bibitem[Lang et~al\mbox{.}(2024)]%
        {lang2024comprehensive}
\bibfield{author}{\bibinfo{person}{Jiedong Lang}, \bibinfo{person}{Zhehao Guo}, {and} \bibinfo{person}{Shuyu Huang}.} \bibinfo{year}{2024}\natexlab{}.
\newblock \showarticletitle{A Comprehensive Study on Quantization Techniques for Large Language Models}.
\newblock \bibinfo{journal}{\emph{arXiv preprint arXiv:2411.02530}} (\bibinfo{year}{2024}).
\newblock


\bibitem[Li et~al\mbox{.}(2024d)]%
        {li2024eliciting}
\bibfield{author}{\bibinfo{person}{Jiahuan Li}, \bibinfo{person}{Hao Zhou}, \bibinfo{person}{Shujian Huang}, \bibinfo{person}{Shanbo Cheng}, {and} \bibinfo{person}{Jiajun Chen}.} \bibinfo{year}{2024}\natexlab{d}.
\newblock \showarticletitle{Eliciting the translation ability of large language models via multilingual finetuning with translation instructions}.
\newblock \bibinfo{journal}{\emph{Transactions of the Association for Computational Linguistics}}  \bibinfo{volume}{12} (\bibinfo{year}{2024}), \bibinfo{pages}{576--592}.
\newblock


\bibitem[Li et~al\mbox{.}(2024c)]%
        {li2024evaluating}
\bibfield{author}{\bibinfo{person}{Shiyao Li}, \bibinfo{person}{Xuefei Ning}, \bibinfo{person}{Luning Wang}, \bibinfo{person}{Tengxuan Liu}, \bibinfo{person}{Xiangsheng Shi}, \bibinfo{person}{Shengen Yan}, \bibinfo{person}{Guohao Dai}, \bibinfo{person}{Huazhong Yang}, {and} \bibinfo{person}{Yu Wang}.} \bibinfo{year}{2024}\natexlab{c}.
\newblock \showarticletitle{Evaluating quantized large language models}.
\newblock \bibinfo{journal}{\emph{arXiv preprint arXiv:2402.18158}} (\bibinfo{year}{2024}).
\newblock


\bibitem[Li et~al\mbox{.}(2024b)]%
        {li2024palmbench}
\bibfield{author}{\bibinfo{person}{Yilong Li}, \bibinfo{person}{Jingyu Liu}, \bibinfo{person}{Hao Zhang}, \bibinfo{person}{M~Badri Narayanan}, \bibinfo{person}{Utkarsh Sharma}, \bibinfo{person}{Shuai Zhang}, \bibinfo{person}{Pan Hu}, \bibinfo{person}{Yijing Zeng}, \bibinfo{person}{Jayaram Raghuram}, {and} \bibinfo{person}{Suman Banerjee}.} \bibinfo{year}{2024}\natexlab{b}.
\newblock \showarticletitle{PalmBench: A Comprehensive Benchmark of Compressed Large Language Models on Mobile Platforms}.
\newblock \bibinfo{journal}{\emph{arXiv preprint arXiv:2410.05315}} (\bibinfo{year}{2024}).
\newblock


\bibitem[Li et~al\mbox{.}(2024a)]%
        {li2024tpi}
\bibfield{author}{\bibinfo{person}{Zonghang Li}, \bibinfo{person}{Wenjiao Feng}, \bibinfo{person}{Mohsen Guizani}, {and} \bibinfo{person}{Hongfang Yu}.} \bibinfo{year}{2024}\natexlab{a}.
\newblock \showarticletitle{TPI-LLM: Serving 70B-scale LLMs Efficiently on Low-resource Edge Devices}.
\newblock \bibinfo{journal}{\emph{arXiv preprint arXiv:2410.00531}} (\bibinfo{year}{2024}).
\newblock


\bibitem[Lin(2004)]%
        {lin2004rouge}
\bibfield{author}{\bibinfo{person}{Chin-Yew Lin}.} \bibinfo{year}{2004}\natexlab{}.
\newblock \showarticletitle{Rouge: A package for automatic evaluation of summaries}. In \bibinfo{booktitle}{\emph{Text summarization branches out}}. \bibinfo{pages}{74--81}.
\newblock


\bibitem[Lin et~al\mbox{.}(2024)]%
        {lin2024awq}
\bibfield{author}{\bibinfo{person}{Ji Lin}, \bibinfo{person}{Jiaming Tang}, \bibinfo{person}{Haotian Tang}, \bibinfo{person}{Shang Yang}, \bibinfo{person}{Wei-Ming Chen}, \bibinfo{person}{Wei-Chen Wang}, \bibinfo{person}{Guangxuan Xiao}, \bibinfo{person}{Xingyu Dang}, \bibinfo{person}{Chuang Gan}, {and} \bibinfo{person}{Song Han}.} \bibinfo{year}{2024}\natexlab{}.
\newblock \showarticletitle{AWQ: Activation-aware Weight Quantization for On-Device LLM Compression and Acceleration}.
\newblock \bibinfo{journal}{\emph{Proceedings of Machine Learning and Systems}}  \bibinfo{volume}{6} (\bibinfo{year}{2024}), \bibinfo{pages}{87--100}.
\newblock


\bibitem[Liu et~al\mbox{.}(2023)]%
        {liu2023emergent}
\bibfield{author}{\bibinfo{person}{Peiyu Liu}, \bibinfo{person}{Zikang Liu}, \bibinfo{person}{Ze-Feng Gao}, \bibinfo{person}{Dawei Gao}, \bibinfo{person}{Wayne~Xin Zhao}, \bibinfo{person}{Yaliang Li}, \bibinfo{person}{Bolin Ding}, {and} \bibinfo{person}{Ji-Rong Wen}.} \bibinfo{year}{2023}\natexlab{}.
\newblock \showarticletitle{Do emergent abilities exist in quantized large language models: An empirical study}.
\newblock \bibinfo{journal}{\emph{arXiv preprint arXiv:2307.08072}} (\bibinfo{year}{2023}).
\newblock


\bibitem[Liu et~al\mbox{.}(2024a)]%
        {liu2024edge}
\bibfield{author}{\bibinfo{person}{Shiwei Liu}, \bibinfo{person}{Kai Han}, \bibinfo{person}{Adriana Fernandez-Lopez}, \bibinfo{person}{Ajay~Kumar Jaiswal}, \bibinfo{person}{Zahra Atashgahi}, \bibinfo{person}{Boqian Wu}, \bibinfo{person}{Edoardo Ponti}, \bibinfo{person}{Callie Hao}, \bibinfo{person}{Rebekka Burkholz}, \bibinfo{person}{Olga Saukh}, {et~al\mbox{.}}} \bibinfo{year}{2024}\natexlab{a}.
\newblock \showarticletitle{Edge-LLMs: Edge-Device Large Language Model Competition}. In \bibinfo{booktitle}{\emph{NeurIPS 2024 Competition Track}}.
\newblock


\bibitem[Liu et~al\mbox{.}(2024b)]%
        {liu2024evaluating}
\bibfield{author}{\bibinfo{person}{Yijun Liu}, \bibinfo{person}{Yuan Meng}, \bibinfo{person}{Fang Wu}, \bibinfo{person}{Shenhao Peng}, \bibinfo{person}{Hang Yao}, \bibinfo{person}{Chaoyu Guan}, \bibinfo{person}{Chen Tang}, \bibinfo{person}{Xinzhu Ma}, \bibinfo{person}{Zhi Wang}, {and} \bibinfo{person}{Wenwu Zhu}.} \bibinfo{year}{2024}\natexlab{b}.
\newblock \showarticletitle{Evaluating the generalization ability of quantized llms: Benchmark, analysis, and toolbox}.
\newblock \bibinfo{journal}{\emph{arXiv preprint arXiv:2406.12928}} (\bibinfo{year}{2024}).
\newblock


\bibitem[{Llama.cpp}(2024)]%
        {llama.cpp}
\bibfield{author}{\bibinfo{person}{{Llama.cpp}}.} \bibinfo{year}{Visited in 2024}\natexlab{}.
\newblock \bibinfo{howpublished}{https://github.com/ggerganov/llama.cpp}.
\newblock


\bibitem[Mao et~al\mbox{.}(2024)]%
        {mao2024compressibility}
\bibfield{author}{\bibinfo{person}{Yu Mao}, \bibinfo{person}{Weilan Wang}, \bibinfo{person}{Hongchao Du}, \bibinfo{person}{Nan Guan}, {and} \bibinfo{person}{Chun~Jason Xue}.} \bibinfo{year}{2024}\natexlab{}.
\newblock \showarticletitle{On the compressibility of quantized large language models}.
\newblock \bibinfo{journal}{\emph{arXiv preprint arXiv:2403.01384}} (\bibinfo{year}{2024}).
\newblock


\bibitem[{Monsoon}(2025)]%
        {monsoon}
\bibfield{author}{\bibinfo{person}{{Monsoon}}.} \bibinfo{year}{Visited in 2025}\natexlab{}.
\newblock \bibinfo{howpublished}{https://www.msoon.com/high-voltage-power-monitor/}.
\newblock


\bibitem[{Ollama framework}(2025)]%
        {ollama}
\bibfield{author}{\bibinfo{person}{{Ollama framework}}.} \bibinfo{year}{Visited in 2025}\natexlab{}.
\newblock \bibinfo{howpublished}{https://github.com/ollama/ollama}.
\newblock


\bibitem[{Ollama Library}(2025)]%
        {ollama-library}
\bibfield{author}{\bibinfo{person}{{Ollama Library}}.} \bibinfo{year}{Visited in 2025}\natexlab{}.
\newblock \bibinfo{howpublished}{https://ollama.com/library}.
\newblock


\bibitem[Park et~al\mbox{.}(2022)]%
        {park2022lut}
\bibfield{author}{\bibinfo{person}{Gunho Park}, \bibinfo{person}{Baeseong Park}, \bibinfo{person}{Minsub Kim}, \bibinfo{person}{Sungjae Lee}, \bibinfo{person}{Jeonghoon Kim}, \bibinfo{person}{Beomseok Kwon}, \bibinfo{person}{Se~Jung Kwon}, \bibinfo{person}{Byeongwook Kim}, \bibinfo{person}{Youngjoo Lee}, {and} \bibinfo{person}{Dongsoo Lee}.} \bibinfo{year}{2022}\natexlab{}.
\newblock \showarticletitle{Lut-gemm: Quantized matrix multiplication based on luts for efficient inference in large-scale generative language models}.
\newblock \bibinfo{journal}{\emph{arXiv preprint arXiv:2206.09557}} (\bibinfo{year}{2022}).
\newblock


\bibitem[{PowerAPI}(2025)]%
        {PyJoules}
\bibfield{author}{\bibinfo{person}{{PowerAPI}}.} \bibinfo{year}{Visited in 2025}\natexlab{}.
\newblock \bibinfo{title}{PyJoules: Python-based energy measurement library}.
\newblock \bibinfo{howpublished}{https://github.com/powerapi-ng/pyJoules}.
\newblock


\bibitem[Qin et~al\mbox{.}(2024)]%
        {qin2024empirical}
\bibfield{author}{\bibinfo{person}{Ruiyang Qin}, \bibinfo{person}{Dancheng Liu}, \bibinfo{person}{Zheyu Yan}, \bibinfo{person}{Zhaoxuan Tan}, \bibinfo{person}{Zixuan Pan}, \bibinfo{person}{Zhenge Jia}, \bibinfo{person}{Meng Jiang}, \bibinfo{person}{Ahmed Abbasi}, \bibinfo{person}{Jinjun Xiong}, {and} \bibinfo{person}{Yiyu Shi}.} \bibinfo{year}{2024}\natexlab{}.
\newblock \showarticletitle{Empirical Guidelines for Deploying LLMs onto Resource-constrained Edge Devices}.
\newblock \bibinfo{journal}{\emph{arXiv preprint arXiv:2406.03777}} (\bibinfo{year}{2024}).
\newblock


\bibitem[Qu et~al\mbox{.}(2024)]%
        {qu2024mobile}
\bibfield{author}{\bibinfo{person}{Guanqiao Qu}, \bibinfo{person}{Qiyuan Chen}, \bibinfo{person}{Wei Wei}, \bibinfo{person}{Zheng Lin}, \bibinfo{person}{Xianhao Chen}, {and} \bibinfo{person}{Kaibin Huang}.} \bibinfo{year}{2024}\natexlab{}.
\newblock \showarticletitle{Mobile edge intelligence for large language models: A contemporary survey}.
\newblock \bibinfo{journal}{\emph{arXiv preprint arXiv:2407.18921}} (\bibinfo{year}{2024}).
\newblock


\bibitem[Radford et~al\mbox{.}(2018)]%
        {radford2018improving}
\bibfield{author}{\bibinfo{person}{Alec Radford}, \bibinfo{person}{Karthik Narasimhan}, \bibinfo{person}{Tim Salimans}, \bibinfo{person}{Ilya Sutskever}, {et~al\mbox{.}}} \bibinfo{year}{2018}\natexlab{}.
\newblock \showarticletitle{Improving language understanding by generative pre-training}.
\newblock  (\bibinfo{year}{2018}).
\newblock


\bibitem[Rahman et~al\mbox{.}(2023)]%
        {rahman2023quantized}
\bibfield{author}{\bibinfo{person}{Mohammad Wali~Ur Rahman}, \bibinfo{person}{Murad~Mehrab Abrar}, \bibinfo{person}{Hunter~Gibbons Copening}, \bibinfo{person}{Salim Hariri}, \bibinfo{person}{Sicong Shao}, \bibinfo{person}{Pratik Satam}, {and} \bibinfo{person}{Soheil Salehi}.} \bibinfo{year}{2023}\natexlab{}.
\newblock \showarticletitle{Quantized transformer language model implementations on edge devices}. In \bibinfo{booktitle}{\emph{ICMLA'23}}. \bibinfo{pages}{709--716}.
\newblock


\bibitem[Reddi et~al\mbox{.}(2020)]%
        {reddi2020mlperf}
\bibfield{author}{\bibinfo{person}{Vijay~Janapa Reddi}, \bibinfo{person}{Christine Cheng}, \bibinfo{person}{David Kanter}, \bibinfo{person}{Peter Mattson}, \bibinfo{person}{Guenther Schmuelling}, \bibinfo{person}{Carole-Jean Wu}, \bibinfo{person}{Brian Anderson}, \bibinfo{person}{Maximilien Breughe}, \bibinfo{person}{Mark Charlebois}, \bibinfo{person}{William Chou}, {et~al\mbox{.}}} \bibinfo{year}{2020}\natexlab{}.
\newblock \showarticletitle{Mlperf inference benchmark}. In \bibinfo{booktitle}{\emph{ISCA'20}}. \bibinfo{pages}{446--459}.
\newblock


\bibitem[{Scaphandre}(2025)]%
        {scaphandre}
\bibfield{author}{\bibinfo{person}{{Scaphandre}}.} \bibinfo{year}{Visited in 2025}\natexlab{}.
\newblock \bibinfo{howpublished}{https://github.com/hubblo-org/scaphandre}.
\newblock


\bibitem[Shen et~al\mbox{.}(2024a)]%
        {shen2024agile}
\bibfield{author}{\bibinfo{person}{Xuan Shen}, \bibinfo{person}{Peiyan Dong}, \bibinfo{person}{Lei Lu}, \bibinfo{person}{Zhenglun Kong}, \bibinfo{person}{Zhengang Li}, \bibinfo{person}{Ming Lin}, \bibinfo{person}{Chao Wu}, {and} \bibinfo{person}{Yanzhi Wang}.} \bibinfo{year}{2024}\natexlab{a}.
\newblock \showarticletitle{Agile-quant: Activation-guided quantization for faster inference of LLMs on the edge}. In \bibinfo{booktitle}{\emph{Proceedings of the AAAI Conference on Artificial Intelligence}}, Vol.~\bibinfo{volume}{38}. \bibinfo{pages}{18944--18951}.
\newblock


\bibitem[Shen et~al\mbox{.}(2024b)]%
        {shen2024edgeqat}
\bibfield{author}{\bibinfo{person}{Xuan Shen}, \bibinfo{person}{Zhenglun Kong}, \bibinfo{person}{Changdi Yang}, \bibinfo{person}{Zhaoyang Han}, \bibinfo{person}{Lei Lu}, \bibinfo{person}{Peiyan Dong}, \bibinfo{person}{Cheng Lyu}, \bibinfo{person}{Chih-hsiang Li}, \bibinfo{person}{Xuehang Guo}, \bibinfo{person}{Zhihao Shu}, {et~al\mbox{.}}} \bibinfo{year}{2024}\natexlab{b}.
\newblock \showarticletitle{Edgeqat: Entropy and distribution guided quantization-aware training for the acceleration of lightweight llms on the edge}.
\newblock \bibinfo{journal}{\emph{arXiv preprint arXiv:2402.10787}} (\bibinfo{year}{2024}).
\newblock


\bibitem[Sheng et~al\mbox{.}(2023)]%
        {sheng2023flexgen}
\bibfield{author}{\bibinfo{person}{Ying Sheng}, \bibinfo{person}{Lianmin Zheng}, \bibinfo{person}{Binhang Yuan}, \bibinfo{person}{Zhuohan Li}, \bibinfo{person}{Max Ryabinin}, \bibinfo{person}{Beidi Chen}, \bibinfo{person}{Percy Liang}, \bibinfo{person}{Christopher R{\'e}}, \bibinfo{person}{Ion Stoica}, {and} \bibinfo{person}{Ce Zhang}.} \bibinfo{year}{2023}\natexlab{}.
\newblock \showarticletitle{Flexgen: High-throughput generative inference of large language models with a single gpu}. In \bibinfo{booktitle}{\emph{ICML'23}}. \bibinfo{pages}{31094--31116}.
\newblock


\bibitem[Shi et~al\mbox{.}(2024)]%
        {shi2024greening}
\bibfield{author}{\bibinfo{person}{Jieke Shi}, \bibinfo{person}{Zhou Yang}, \bibinfo{person}{Hong~Jin Kang}, \bibinfo{person}{Bowen Xu}, \bibinfo{person}{Junda He}, {and} \bibinfo{person}{David Lo}.} \bibinfo{year}{2024}\natexlab{}.
\newblock \showarticletitle{Greening large language models of code}. In \bibinfo{booktitle}{\emph{ICSE'24}}. \bibinfo{pages}{142--153}.
\newblock


\bibitem[Sudhakaran et~al\mbox{.}(2024)]%
        {sudhakaran2024mariogpt}
\bibfield{author}{\bibinfo{person}{Shyam Sudhakaran}, \bibinfo{person}{Miguel Gonz{\'a}lez-Duque}, \bibinfo{person}{Matthias Freiberger}, \bibinfo{person}{Claire Glanois}, \bibinfo{person}{Elias Najarro}, {and} \bibinfo{person}{Sebastian Risi}.} \bibinfo{year}{2024}\natexlab{}.
\newblock \showarticletitle{Mariogpt: Open-ended text2level generation through large language models}.
\newblock \bibinfo{journal}{\emph{Advances in Neural Information Processing Systems}}  \bibinfo{volume}{36} (\bibinfo{year}{2024}).
\newblock


\bibitem[Sun et~al\mbox{.}(2020)]%
        {sun2020mobilebert}
\bibfield{author}{\bibinfo{person}{Zhiqing Sun}, \bibinfo{person}{Hongkun Yu}, \bibinfo{person}{Xiaodan Song}, \bibinfo{person}{Renjie Liu}, \bibinfo{person}{Yiming Yang}, {and} \bibinfo{person}{Denny Zhou}.} \bibinfo{year}{2020}\natexlab{}.
\newblock \showarticletitle{Mobilebert: a compact task-agnostic bert for resource-limited devices}.
\newblock \bibinfo{journal}{\emph{arXiv preprint arXiv:2004.02984}} (\bibinfo{year}{2020}).
\newblock


\bibitem[Tan et~al\mbox{.}(2024)]%
        {tan2024mobilequant}
\bibfield{author}{\bibinfo{person}{Fuwen Tan}, \bibinfo{person}{Royson Lee}, \bibinfo{person}{{\L}ukasz Dudziak}, \bibinfo{person}{Shell~Xu Hu}, \bibinfo{person}{Sourav Bhattacharya}, \bibinfo{person}{Timothy Hospedales}, \bibinfo{person}{Georgios Tzimiropoulos}, {and} \bibinfo{person}{Brais Martinez}.} \bibinfo{year}{2024}\natexlab{}.
\newblock \showarticletitle{Mobilequant: Mobile-friendly quantization for on-device language models}.
\newblock \bibinfo{journal}{\emph{arXiv preprint arXiv:2408.13933}} (\bibinfo{year}{2024}).
\newblock


\bibitem[Tian et~al\mbox{.}(2024)]%
        {tian2024greenllm}
\bibfield{author}{\bibinfo{person}{Chunlin Tian}, \bibinfo{person}{Xinpeng Qin}, {and} \bibinfo{person}{Li Li}.} \bibinfo{year}{2024}\natexlab{}.
\newblock \showarticletitle{Greenllm: Towards efficient large language model via energy-aware pruning}. In \bibinfo{booktitle}{\emph{IWQoS'24}}. \bibinfo{pages}{1--2}.
\newblock


\bibitem[Touvron et~al\mbox{.}(2023)]%
        {touvron2023llama}
\bibfield{author}{\bibinfo{person}{Hugo Touvron}, \bibinfo{person}{Thibaut Lavril}, \bibinfo{person}{Gautier Izacard}, \bibinfo{person}{Xavier Martinet}, \bibinfo{person}{Marie-Anne Lachaux}, \bibinfo{person}{Timoth{\'e}e Lacroix}, \bibinfo{person}{Baptiste Rozi{\`e}re}, \bibinfo{person}{Naman Goyal}, \bibinfo{person}{Eric Hambro}, \bibinfo{person}{Faisal Azhar}, {et~al\mbox{.}}} \bibinfo{year}{2023}\natexlab{}.
\newblock \showarticletitle{Llama: Open and efficient foundation language models}.
\newblock \bibinfo{journal}{\emph{arXiv preprint arXiv:2302.13971}} (\bibinfo{year}{2023}).
\newblock


\bibitem[{TruthfulQA}(2024)]%
        {truthfulqa}
\bibfield{author}{\bibinfo{person}{{TruthfulQA}}.} \bibinfo{year}{Visited in 2024}\natexlab{}.
\newblock \bibinfo{howpublished}{https://github.com/sylinrl/TruthfulQA}.
\newblock


\bibitem[Tschand et~al\mbox{.}(2024)]%
        {tschand2024mlperf}
\bibfield{author}{\bibinfo{person}{Arya Tschand}, \bibinfo{person}{Arun Tejusve~Raghunath Rajan}, \bibinfo{person}{Sachin Idgunji}, \bibinfo{person}{Anirban Ghosh}, \bibinfo{person}{Jeremy Holleman}, \bibinfo{person}{Csaba Kiraly}, \bibinfo{person}{Pawan Ambalkar}, \bibinfo{person}{Ritika Borkar}, \bibinfo{person}{Ramesh Chukka}, \bibinfo{person}{Trevor Cockrell}, {et~al\mbox{.}}} \bibinfo{year}{2024}\natexlab{}.
\newblock \showarticletitle{MLPerf Power: Benchmarking the Energy Efficiency of Machine Learning Systems from Microwatts to Megawatts for Sustainable AI}.
\newblock \bibinfo{journal}{\emph{arXiv preprint arXiv:2410.12032}} (\bibinfo{year}{2024}).
\newblock


\bibitem[van Schaik and Pugh(2024)]%
        {schaik2024field}
\bibfield{author}{\bibinfo{person}{Tempest~A. van Schaik} {and} \bibinfo{person}{Brittany Pugh}.} \bibinfo{year}{2024}\natexlab{}.
\newblock \showarticletitle{A Field Guide to Automatic Evaluation of LLM-Generated Summaries}. In \bibinfo{booktitle}{\emph{SIGIR '24}}. \bibinfo{publisher}{ACM}, \bibinfo{pages}{2832–2836}.
\newblock


\bibitem[Wang et~al\mbox{.}(2024)]%
        {wang2024model}
\bibfield{author}{\bibinfo{person}{Wenxiao Wang}, \bibinfo{person}{Wei Chen}, \bibinfo{person}{Yicong Luo}, \bibinfo{person}{Yongliu Long}, \bibinfo{person}{Zhengkai Lin}, \bibinfo{person}{Liye Zhang}, \bibinfo{person}{Binbin Lin}, \bibinfo{person}{Deng Cai}, {and} \bibinfo{person}{Xiaofei He}.} \bibinfo{year}{2024}\natexlab{}.
\newblock \showarticletitle{Model compression and efficient inference for large language models: A survey}.
\newblock \bibinfo{journal}{\emph{arXiv preprint arXiv:2402.09748}} (\bibinfo{year}{2024}).
\newblock


\bibitem[Wei et~al\mbox{.}(2024)]%
        {wei2024t}
\bibfield{author}{\bibinfo{person}{Jianyu Wei}, \bibinfo{person}{Shijie Cao}, \bibinfo{person}{Ting Cao}, \bibinfo{person}{Lingxiao Ma}, \bibinfo{person}{Lei Wang}, \bibinfo{person}{Yanyong Zhang}, {and} \bibinfo{person}{Mao Yang}.} \bibinfo{year}{2024}\natexlab{}.
\newblock \showarticletitle{T-mac: Cpu renaissance via table lookup for low-bit llm deployment on edge}.
\newblock \bibinfo{journal}{\emph{arXiv preprint arXiv:2407.00088}} (\bibinfo{year}{2024}).
\newblock


\bibitem[Wei et~al\mbox{.}(2022)]%
        {wei2022outlier}
\bibfield{author}{\bibinfo{person}{Xiuying Wei}, \bibinfo{person}{Yunchen Zhang}, \bibinfo{person}{Xiangguo Zhang}, \bibinfo{person}{Ruihao Gong}, \bibinfo{person}{Shanghang Zhang}, \bibinfo{person}{Qi Zhang}, \bibinfo{person}{Fengwei Yu}, {and} \bibinfo{person}{Xianglong Liu}.} \bibinfo{year}{2022}\natexlab{}.
\newblock \showarticletitle{Outlier suppression: Pushing the limit of low-bit transformer language models}.
\newblock \bibinfo{journal}{\emph{Advances in Neural Information Processing Systems}}  \bibinfo{volume}{35} (\bibinfo{year}{2022}), \bibinfo{pages}{17402--17414}.
\newblock


\bibitem[Wulf and Meierhofer(2024)]%
        {wulf2024exploring}
\bibfield{author}{\bibinfo{person}{Jochen Wulf} {and} \bibinfo{person}{Juerg Meierhofer}.} \bibinfo{year}{2024}\natexlab{}.
\newblock \showarticletitle{Exploring the potential of large language models for automation in technical customer service}.
\newblock \bibinfo{journal}{\emph{arXiv preprint arXiv:2405.09161}} (\bibinfo{year}{2024}).
\newblock


\bibitem[Xiao et~al\mbox{.}(2023)]%
        {xiao2023smoothquant}
\bibfield{author}{\bibinfo{person}{Guangxuan Xiao}, \bibinfo{person}{Ji Lin}, \bibinfo{person}{Mickael Seznec}, \bibinfo{person}{Hao Wu}, \bibinfo{person}{Julien Demouth}, {and} \bibinfo{person}{Song Han}.} \bibinfo{year}{2023}\natexlab{}.
\newblock \showarticletitle{Smoothquant: Accurate and efficient post-training quantization for large language models}. In \bibinfo{booktitle}{\emph{ICML'23}}. \bibinfo{pages}{38087--38099}.
\newblock


\bibitem[Xu et~al\mbox{.}(2024)]%
        {xu2024unleashing}
\bibfield{author}{\bibinfo{person}{Minrui Xu}, \bibinfo{person}{Hongyang Du}, \bibinfo{person}{Dusit Niyato}, \bibinfo{person}{Jiawen Kang}, \bibinfo{person}{Zehui Xiong}, \bibinfo{person}{Shiwen Mao}, \bibinfo{person}{Zhu Han}, \bibinfo{person}{Abbas Jamalipour}, \bibinfo{person}{Dong~In Kim}, \bibinfo{person}{Xuemin Shen}, {et~al\mbox{.}}} \bibinfo{year}{2024}\natexlab{}.
\newblock \showarticletitle{Unleashing the power of edge-cloud generative ai in mobile networks: A survey of aigc services}.
\newblock \bibinfo{journal}{\emph{IEEE Communications Surveys \& Tutorials}} (\bibinfo{year}{2024}).
\newblock


\bibitem[Yao et~al\mbox{.}(2022)]%
        {yao2022zeroquant}
\bibfield{author}{\bibinfo{person}{Zhewei Yao}, \bibinfo{person}{Reza Yazdani~Aminabadi}, \bibinfo{person}{Minjia Zhang}, \bibinfo{person}{Xiaoxia Wu}, \bibinfo{person}{Conglong Li}, {and} \bibinfo{person}{Yuxiong He}.} \bibinfo{year}{2022}\natexlab{}.
\newblock \showarticletitle{Zeroquant: Efficient and affordable post-training quantization for large-scale transformers}.
\newblock \bibinfo{journal}{\emph{Advances in Neural Information Processing Systems}}  \bibinfo{volume}{35} (\bibinfo{year}{2022}), \bibinfo{pages}{27168--27183}.
\newblock


\bibitem[Yin et~al\mbox{.}(2024)]%
        {yin2024llm}
\bibfield{author}{\bibinfo{person}{Wangsong Yin}, \bibinfo{person}{Mengwei Xu}, \bibinfo{person}{Yuanchun Li}, {and} \bibinfo{person}{Xuanzhe Liu}.} \bibinfo{year}{2024}\natexlab{}.
\newblock \showarticletitle{Llm as a system service on mobile devices}.
\newblock \bibinfo{journal}{\emph{arXiv preprint arXiv:2403.11805}} (\bibinfo{year}{2024}).
\newblock


\bibitem[Yu et~al\mbox{.}(2024)]%
        {yu2024edge}
\bibfield{author}{\bibinfo{person}{Zhongzhi Yu}, \bibinfo{person}{Zheng Wang}, \bibinfo{person}{Yuhan Li}, \bibinfo{person}{Ruijie Gao}, \bibinfo{person}{Xiaoya Zhou}, \bibinfo{person}{Sreenidhi~Reddy Bommu}, \bibinfo{person}{Yang Zhao}, {and} \bibinfo{person}{Yingyan Lin}.} \bibinfo{year}{2024}\natexlab{}.
\newblock \showarticletitle{Edge-llm: Enabling efficient large language model adaptation on edge devices via unified compression and adaptive layer voting}. In \bibinfo{booktitle}{\emph{Proceedings of the 61st ACM/IEEE Design Automation Conference}}. \bibinfo{pages}{1--6}.
\newblock


\bibitem[Yuan et~al\mbox{.}(2024)]%
        {yuan2024generative}
\bibfield{author}{\bibinfo{person}{Xingyu Yuan}, \bibinfo{person}{He Li}, \bibinfo{person}{Kaoru Ota}, {and} \bibinfo{person}{Mianxiong Dong}.} \bibinfo{year}{2024}\natexlab{}.
\newblock \showarticletitle{Generative inference of large language models in edge computing: An energy efficient approach}. In \bibinfo{booktitle}{\emph{IWCMC'24}}. \bibinfo{pages}{244--249}.
\newblock


\bibitem[Zhang et~al\mbox{.}(2024)]%
        {zhang2024edgeshard}
\bibfield{author}{\bibinfo{person}{Mingjin Zhang}, \bibinfo{person}{Jiannong Cao}, \bibinfo{person}{Xiaoming Shen}, {and} \bibinfo{person}{Zeyang Cui}.} \bibinfo{year}{2024}\natexlab{}.
\newblock \showarticletitle{EdgeShard: Efficient LLM Inference via Collaborative Edge Computing}.
\newblock \bibinfo{journal}{\emph{arXiv preprint arXiv:2405.14371}} (\bibinfo{year}{2024}).
\newblock


\bibitem[Zhong et~al\mbox{.}(2023)]%
        {zhong2023can}
\bibfield{author}{\bibinfo{person}{Qihuang Zhong}, \bibinfo{person}{Liang Ding}, \bibinfo{person}{Juhua Liu}, \bibinfo{person}{Bo Du}, {and} \bibinfo{person}{Dacheng Tao}.} \bibinfo{year}{2023}\natexlab{}.
\newblock \showarticletitle{Can chatgpt understand too? a comparative study on chatgpt and fine-tuned bert}.
\newblock \bibinfo{journal}{\emph{arXiv preprint arXiv:2302.10198}} (\bibinfo{year}{2023}).
\newblock


\bibitem[Zhou et~al\mbox{.}(2024)]%
        {zhou2024survey}
\bibfield{author}{\bibinfo{person}{Zixuan Zhou}, \bibinfo{person}{Xuefei Ning}, \bibinfo{person}{Ke Hong}, \bibinfo{person}{Tianyu Fu}, \bibinfo{person}{Jiaming Xu}, \bibinfo{person}{Shiyao Li}, \bibinfo{person}{Yuming Lou}, \bibinfo{person}{Luning Wang}, \bibinfo{person}{Zhihang Yuan}, \bibinfo{person}{Xiuhong Li}, {et~al\mbox{.}}} \bibinfo{year}{2024}\natexlab{}.
\newblock \showarticletitle{A survey on efficient inference for large language models}.
\newblock \bibinfo{journal}{\emph{arXiv preprint arXiv:2404.14294}} (\bibinfo{year}{2024}).
\newblock


\end{thebibliography}
\bibliographystyle{ACM-Reference-Format}

\end{document}